\documentclass[preprint,12pt,eqsecnum,nofootinbib,amsmath,amssymb]{revtex4}

\newcommand{\datechange}{625March 2018}

\newcommand{\laurnumber}{LA-UR-19-22641}

\newcommand{\mytitle}{Multimaterial Heat Flow Verification}

\usepackage{graphicx} 
\usepackage{dcolumn}  
\usepackage{bm}          
\usepackage{latexsym} 
\usepackage{fancyhdr} 
\usepackage{wrapfig}

\newcommand{\smA}{{\scriptscriptstyle \rm A}}
\newcommand{\smB}{{\rm\scriptscriptstyle B}}

\newcommand{\F}{{\rm F}}

\newcommand{\sm}[1]{{\scriptstyle #1}}

\newcommand{\bodyskip}{\baselineskip 18pt plus 1pt minus 1pt}
\newcommand{\bibskip}{\baselineskip16pt plus 1pt minus 1pt}
\newcommand{\tableofcontentsskip}{\baselineskip 14pt plus 1pt minus 1pt}
\newcommand{\footnoteskip}{\baselineskip 12pt plus 1pt minus 1pt}
\newcommand{\abstractskip}{\baselineskip 13pt plus 1pt minus 1pt}
\newcommand{\titleskip}{\baselineskip 18pt plus 1pt minus 1pt}
\newcommand{\affiliationskip}{\baselineskip 15pt plus 1pt minus 1pt}

\def\frame#1#2#3#4{\vbox{\hrule height #1pt    
  \hbox{\vrule width #1pt\kern #2pt                     
  \vbox{\kern #2pt                                               
  \vbox{\hsize #3\noindent #4}                            
  \kern #2pt}                                                        
  \kern #2pt\vrule width #1pt}                              
  \hrule height0pt depth #1pt}                            
}
\def\myframe#1{\vbox{\hrule height 0.1pt    
  \hbox{\vrule width 0.1pt\kern 2pt                     
  \vbox{\kern 2pt                                               
  \vbox{\hsize 16.5cm\noindent #1}                            
  \kern 2pt}                                                        
  \kern 2pt\vrule width 0.1pt}                              
  \hrule height0pt depth 0.1pt}                            
}

\def\fitframe #1#2#3{\vbox{\hrule height#1pt  
  \hbox{\vrule width#1pt\kern #2pt             
  \vbox{\kern #2pt\hbox{#3}\kern #2pt}         
  \kern #2pt\vrule width#1pt}                  
  \hrule height0pt depth#1pt}                  
}
\def\shframe #1#2#3#4{\vbox{\hrule height 0pt 
 \hbox{\vrule width #1pt\kern 0pt             
 \vbox{\kern-#1pt\frame{.3}{#2}{#3}{#4}       
 \kern-.3pt}                                  
 \kern-#2pt\vrule width 0pt}                  
 \hrule height #1pt}                          
}

\begin{document}

\thispagestyle{empty}
\pagestyle{empty}
\setcounter{page}{1}

%
%
%
\baselineskip 20pt plus 1pt minus 1pt


\preprint{\laurnumber}

\title{\titleskip
  \mytitle
}

\author{$^1$Robert L Singleton Jr, $^1$Christopher M Malone, and $^2$Cora L Brown}
\vskip 0.2cm 
\affiliation{\affiliationskip
     $^1$Los Alamos National Laboratory\\
     Los Alamos, New Mexico 87545, USA
     \\ 
     $^2$The University of Minnesota\\
     Minneapolis, Minnesota 55455, USA
}

\date{\datechange}

\begin{abstract}
\abstractskip

\noindent
Multimaterial heat diffusion can be a challenging numerical problem when the 
material boundaries are misaligned with the numerical grid. Even when the boundaries 
start out aligned, they typically become misaligned  through hydrodynamic motion.
There are usually a number of methods for handling multimaterial cells in any given 
hydro code. One of the simplest methods is to replace the multimaterial cell by an 
{\em average} single-material cell whose heat capacity and conductivity are averages 
over the constituent materials. One can further refine this model by using either the 
arithmetic or harmonic averages, thereby providing two distinct (albeit naive) 
multimaterial models for the arithmetic and harmonic averages.  More sophisticated 
models typically involve a surrogate mesh of some kind, as with the thin mesh and 
static condensation methods. In this paper, we perform rigorous code verification of 
the multiphysics hydrocode FLAG, including grid resolution studies. We employ a number 
of newly constructed 2D heat flow solutions that generalize the standard {\em planar 
sandwich} solution, and this paper offers a smorgasbord of exact solutions for heat flow 
verification. To perform the analyses and to produce the corresponding convergence plots, 
we employ the code verification tool ExactPack. 

\end{abstract}
\thispagestyle{empty}
\maketitle

\pagebreak
\tableofcontentsskip
\tableofcontents
\pagebreak
\bodyskip
\newpage
\bodyskip
\pagestyle{fancy}

\lhead{}
\chead{}
\rhead{}
\lfoot{}
\cfoot{\thepage}
\rfoot{}

\section{Introduction}

This paper is concerned with {\em verifying} a number of 2D multimaterial
heat-flow algorithms in the multi-physics computational hydrodynamics 
code FLAG\,\cite{flag_ref}.  As a general principle, {\em code verification} is 
the process of comparing and analyzing the differences between {\em numerical} 
results and {\em exact} analytic results. Technically, the term {\em exact solution} 
means a solution that can be expressed solely in terms of known analytic 
functions.\footnote{\footnoteskip
By a {\em known analytic function}, we mean a function defined in terms of 
the conical analytic functions from classical $19^{\rm th}$ century mathematics.
These functions have been exhaustively studied, and have been implemented in 
hosts of numerical packages. 
}
These solutions are exceedingly rare, with the Noh problem providing the 
quintessential example of an exact solution.  A more common form of solution 
is the semi-exact or semi-analytic solution. These solutions can be expressed 
in terms known analytic functions, supplemented by simple numerical operations, 
such as 1D quadrature, root finding, numerical ODE solves, or summing infinite 
series.  The planar sandwich solutions are of the latter category.  We will use
the term {\em exact solution} for both cases. The relevance 
of exact solutions is that their errors can be systematically controlled.
Exact solutions usually exploit the symmetry of the problem, such 
as spherical or planar symmetry, scale invariance, or more general 
Lie Group symmetries. 

We concentrate on an exact 2D solution of the heat flow equation called the 
{\em planar sandwich}\,\cite{Shashkov}, performing a series of rigorous convergence
analyses.  This solution has been analyzed by Dawes,  Shashkov, and Malone\,\cite{DMS} 
in the context of multimaterial heat diffusion, although these authors do not present convergence 
analyses. A number of generalizations of the planar sandwich test problem have 
been presented in Ref.~\cite{PlanarSandwichExactPackDoc}, and we explore these solutions 
as well. To perform the analyses and to produce the corresponding convergence plots, we 
employ the code verification tool ExactPack\,\cite{exactpack}. This paper provides a summary 
of the planar sandwich solution and its implementation in ExactPack, including the basic 
Python source code used to produce the various figures and to perform the convergence analyses.

\begin{figure}[t!]
\includegraphics[scale=0.40]{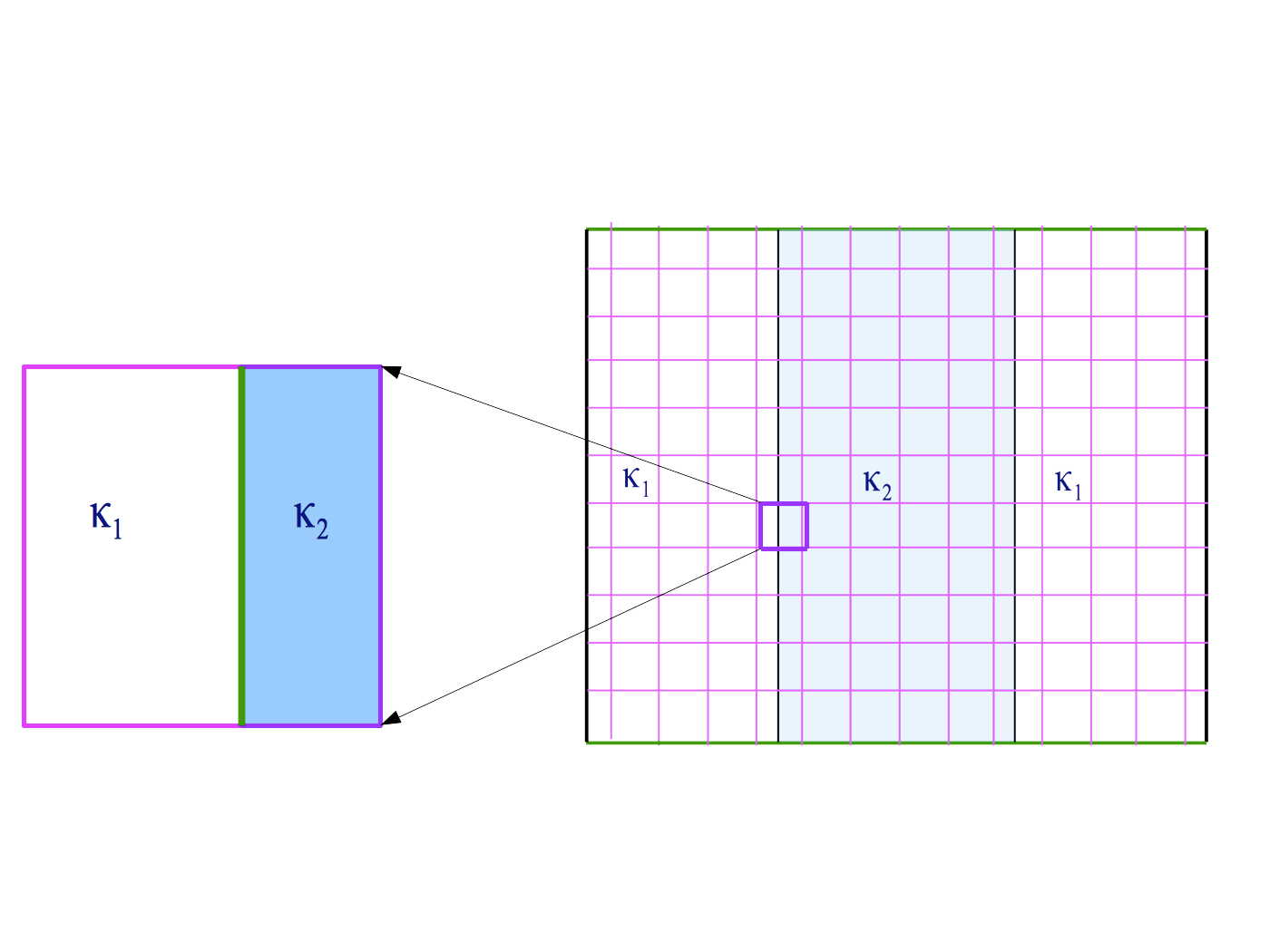}
\vskip-1.0cm
\caption{\footnoteskip  
An example of a multimaterial grid. The rightmost figure shows a 
square numerical grid overlaid upon a rectangular physical geometry 
consisting of three parallel material regions with differing diffusion 
coefficients. The numerical grid partitions the physical geometry into 
a number of corresponding numerical cells, none of which need align 
with the material regions. The outer two regions are composed of the 
same material with diffusion coefficient $\kappa_1$, while the inner 
region has diffusion coefficient $\kappa_2$, forming a sandwich-like 
configuration. The planar sandwich test problem takes the outer two 
regions (the bread) to be nonconducting, with $\kappa_1=0$, while 
the inner region (the meat of the sandwich) conducts heat with 
$\kappa_2 \equiv \kappa > 0$. In the Figure, the numerical grid 
is misaligned relative to the material boundaries of the inner 
conducting region. This is illustrated in left panel figure.
}
\label{fig_sub_grid}
\end{figure}

The focus of this paper is the multimaterial heat flow algorithms in FLAG.
By a {\em multimaterial} cell, we mean a computational cell containing multiple 
materials, each with their own distinct physical properties, and with a clear 
interface between the separate materials. Figure~\ref{fig_sub_grid} 
illustrates a numerical mesh with multimaterial cells for a square grid based 
on the planar sandwich geometry. In the context of heat flow, a multimaterial 
cell contains a number of individual materials labeled by an index $m$ with 
heat conductivity $\kappa_m$. Subgrid models must be employed to resolve 
such physics in a hydrocode.  The simplest subgrid model is obtained by replacing 
a multimaterial cell by a uniform single-material cell with an {\em  average} 
conductivity $\bar\kappa$. An average material is meant to reproduce the 
collective effects of the individual sub-materials with differing values of 
$\kappa_m$, and FLAG utilizes both the {\em arithmetic} and {\em harmonic} 
averages,
\begin{eqnarray}
  \bar\kappa_a &=& \sum_m V_m \, \kappa_m
  \\[5pt] 
  \bar\kappa^{-1}_h &=& \sum_m V_m \, \kappa_m^{-1}
  \ ,
\label{eq:averages}
\end{eqnarray}
where $V_m$ is the corresponding volume fraction  of the cell associated 
with material $m$ and diffusion coefficient $\kappa_m$.\,\footnote{
\footnoteskip
If appropriate, one can also use the mass faction $M_m$, rather 
than the volume fraction $V_m$, to define a mass-weighted average.
} 
In Fig.~\ref{fig_sub_grid}, the multimaterial index runs over $m=1,2$ along 
the cells containing the material boundary, and for the  arithmetic average, 
the heat diffusion coefficient is \hbox{$\bar\kappa_a= V_1\, \kappa_1 + V_2\, \kappa_2$};  
for the harmonic average, the heat diffusion coefficient along the boundary 
is determined by $\bar\kappa^{-1}_h = V_1 \,\kappa_1^{-1} + V_2 \, \kappa_2^{-1}$.
As we shall see, the arithmetic average overestimates the heat flow along
the boundary, while the harmonic average underestimates the heat flow.

Averaging techniques cannot always faithfully represent the physics of multimaterial 
cells. Consequently, FLAG employs  more sophisticated multimaterial diffusion 
options, namely the {\em thin mesh}\,\cite{tm}  and {\em static condensation}\,\cite{sc}
algorithms. The thin mesh method starts with the volume fractions of each 
material region, and reconstructs the material interfaces using interface reconstruction 
methods. The mesh is then subdivided along the interfaces, making sure that the final 
polyhedral mesh conforms with the numerical mesh. This new unstructured mesh is 
constructed with full connectivity each cycle. The heat diffusion equation is solved on 
the subdivided mesh containing only single material cells.

The static condensation approach also makes use of the reconstructed material 
interfaces, but does not require all the details regarding connectivity across material 
interfaces within a cell. Instead, the global system for the diffusion equation is 
rewritten in terms of {\em unknown} face-centered temperature values. Total flux
continuity is enforced at each cell face by ensuring that the sum of the fluxes from 
all materials on either side of the face are the same. There is an approximation here 
in that each cell face is assigned a {\em single} temperature associated with it;
however, the fluxes contain the material-based diffusion coefficient $\kappa_m$,
which are allowed to vary.  The material-centered temperatures are eliminated 
from the system (via the Schur Complement), thus {\em condensing} the number 
of degrees of freedom.  The global system is then solved for the unknown face 
temperatures using standard mimetic techniques.  The result is that each cell now 
has a known solution on its boundary (faces), which becomes a local Dirichlet problem 
that can be solved independently to recover updated material-centered temperatures.  
For more details on the algorithm, see Ref.~\cite{asc}; this method is reported to 
be second order accurate in Ref.~\cite{sc}.

\section{The Planar Sandwich Test Problem}
\begin{figure}[t!]
\includegraphics[scale=0.35]{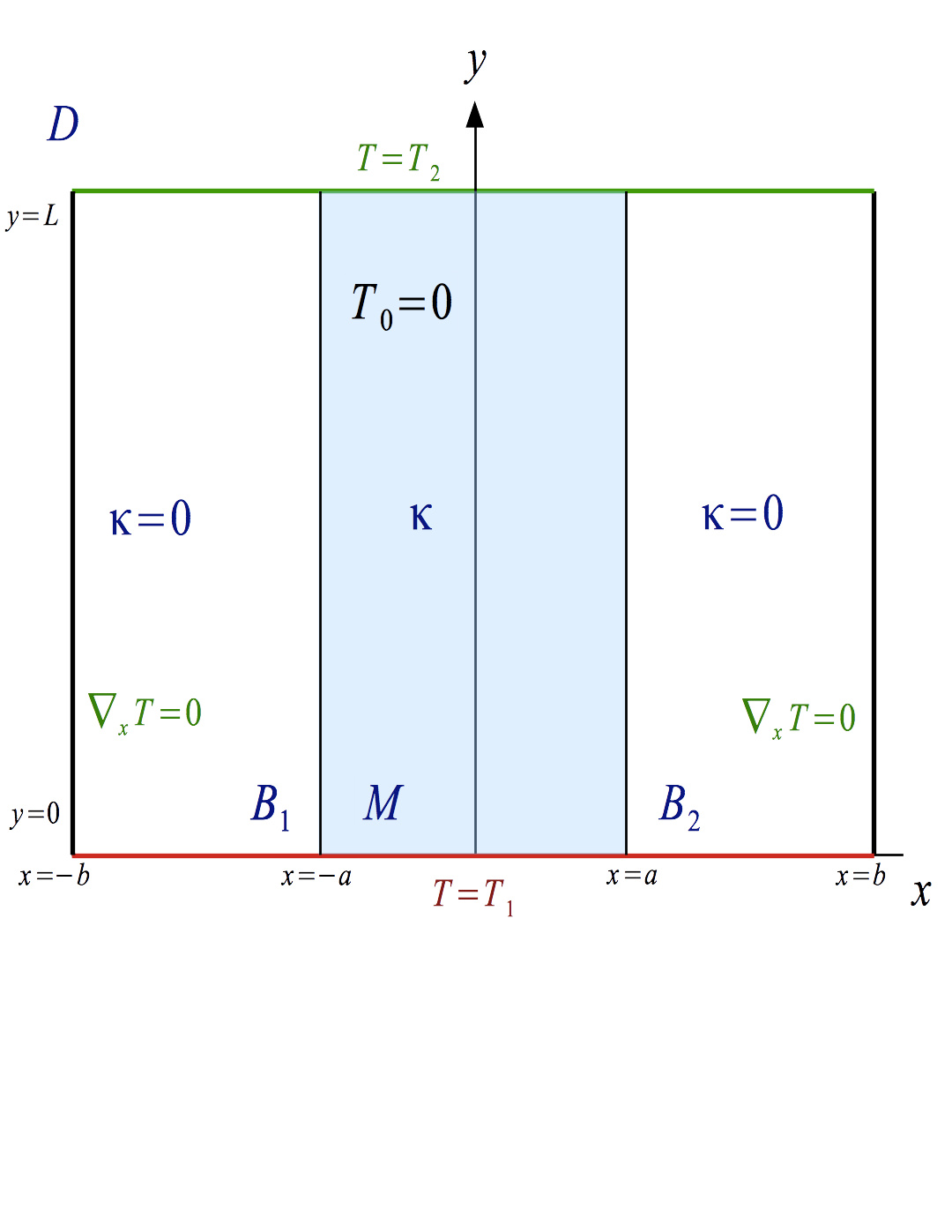} 
\vskip-2.5cm
\caption{\footnoteskip  
The Planar Sandwich is a 2D-Cartesian heat flow problem defined 
on the square domain $D = [0,L] \times [0, L]$. The temperature 
at location $(x,y) \in D$ at time $t$ is denoted by $T(x,y,t)$. The inner 
heat-conducing region $M = \{ (x,y) \in D \,\vert\, a_1 \le x \le a_2 \}$,
for which $\kappa > 0$, is called the meat of the sandwich. The outer 
non-conducting materials, the bread of the sandwich, are located within 
$B_1 = \{ 0 \le x < a_1 \}$ or $B_2=\{a_2 < x \le L\}$, and are insulating,
with $\kappa=0$. In numerical work, we take $\kappa=10^{-12}$ rather 
than zero. In setting the boundary conditions, we take the upper and 
lower boundary temperatures to be constant and uniform along $y=0$ 
and $y=L$ with values $T_1$ and $T_2$, respectively, {\em i.e.} we impose 
the Dirichlet boundary conditions \hbox{$T(x, y = 0,t)=T_1$} and  
\hbox{$T(x, y = L,t)=T_2$}.  On the left and right boundaries, we 
take the temperature flux in the $x$-direction to vanish, {\em i.e.} we use 
the Neumann boundary conditions \hbox{$\partial_x T(x  = 0,y,t)=0$} 
and \hbox{$\partial_x T(x = L,y,t)=0$}.   We must also impose an initial 
condition: we set the temperature in the interior of the sandwich to zero, 
{\em i.e.} \hbox{$T(x,y,t = 0)=0$} for $(x,y) \in (0,L)\times (0,L)$. When 
$T_1 > T_2$, heat flows in the $y$-direction from the lower to the upper 
boundary. This allows us to describe the planar sandwich in terms of 1D 
profiles $T(y,t)$.  For numerical work, we take $a_1=0.75$ and 
$a_2=1.25$, or a shifted variant with $a_1=0.77$ and $a_2=1.27$.
}
\label{fig_planar_sandwich}
\end{figure}

This section is devoted to the planar sandwich test problem of Ref.~\cite{DMS}, 
as illustrated in Fig.~\ref{fig_planar_sandwich}. It is a 2D-Cartesian heat flow 
problem on the  square domain $D = [0,L] \times [0,L]$, where the $x$-axis 
runs horizontally and the $y$-axis is vertical. We take $L=2$ (in arbitrary 
units) in all figures and examples that follow. The domain $D$ is partitioned 
into three vertical sandwich-like regions delimited by $x=a_1$ and $x=a_2$,  
with $0 < a_1 < a_2 < L$. The inner region $a_1 \le x \le a_2$ is composed of 
a heat-conduction material with heat diffusion coefficient $\kappa > 0$, and 
is called the {\em meat} of the sandwich. The meat is surrounded by two non-heat 
conducting materials, called the {\em bread} of the sandwich, for which $\kappa=0$ 
in $0 \le x < a_1$ and $a_2 < x \le L$. Working in arbitrary temperature units, we 
wish to solve the 2D heat equation 
\begin{eqnarray}
  \frac{\partial T}{\partial t}
  &=&
  {\bm\nabla}\cdot \Big[\kappa\, {\bm \nabla} T \,\Big]
  ~~~{\rm for}~~(x,y) \in (0,L) \times (0,L)
  \ ,
\end{eqnarray}
where $T=T(x,y,t)$ is the temperature field at position $(x,y)$ and time $t$, with 
the diffusion coefficient taking the form  
\begin{eqnarray}
  \kappa(x, y)
  =
 \left\{
 \begin{array}{ll}
  0 & ; ~0 \le x < a_1 \\[-5pt]
  \kappa & ; ~a_1 \le x \le a_2 \\[-5pt]
  0 & ; ~a_2 < x \le L \ .
  \end{array}
  \right.
\end{eqnarray}
The initial condition (IC) is chosen so that the temperature vanishes in the interior 
of the domain $D$ at time $t=0$,
\begin{eqnarray}
  T(x,y,0) = 0
  ~~~{\rm for}~~ (x,y) \in (0,L)\times (0,L) 
  \ ,
\end{eqnarray}
while the boundary conditions (BCs) are taken  to be 
\begin{eqnarray}
  T(x, 0, t) &=& T_1    \hskip1cm {\rm for}~~ x \in [0,L]
  \label{TBCa}
  \\
  T(x, L, t) &=& T_2
  \label{TBCb}
  \\
  \partial_x T(0, y, t) &=& 0  \hskip1.2cm {\rm for}~~ y \in [0,L]
  \label{TBCc}
  \\
  \partial_x T(L, y, t) &=& 0
   \label{TBCd}
  \ .
\end{eqnarray}
Note that the left- and right-hand sides of the rectangle are insulating, {\em i.e.} 
the temperature flux along the $x$-direction at the far left- and right-ends of $D$ 
vanishes, \hbox{$\partial_x T(x = 0, y, t)=0$} and $\partial_x T(x = L, y, t)=0$ for 
$0 \le y \le L$.  Along the lower and upper boundaries \hbox{$y=0,L$}, the temperature 
profiles are uniform in $x$ with $T(x,y = 0,t)=T_1$ and \hbox{$T(x,y = L,t)=T_2$}, 
where $T_1$ and $T_2$ are constant temperature values over the length of the region 
\hbox{$0 \le x \le L$.} When $T_1 > T_2$, the Second Law of Thermodynamics ensures 
that heat flows upward from $y=0$ to $y=L$.  We shall generalize the planar sandwich 
problem in Section~\ref{sec_generalized} by considering nonhomogeneous boundary 
condition with nonzero heat flux and non-trivial initial conditions\,\cite{PlanarSandwichExactPackDoc}.

\begin{figure}[t!]
\includegraphics[scale=0.40]{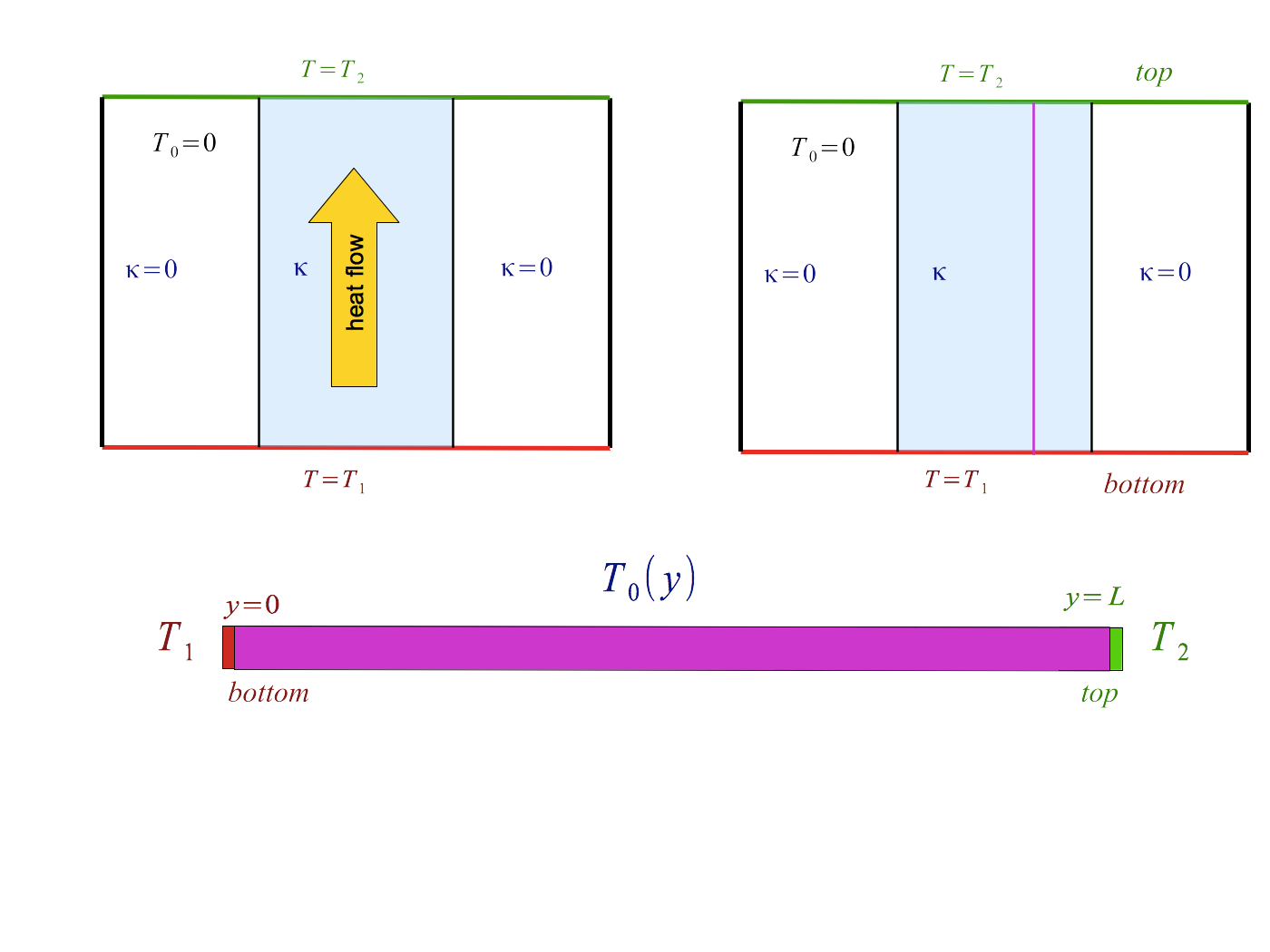}
\vskip-1.5cm
\caption{\footnoteskip  
The 2D planar sandwich reduces to a 1D heat flow problem along a vertical
rod of length $L$. In the top panels, the 2D temperature profile is a function 
of the $x$ and $y$ coordinates, $T= T(x, y,t)$. Since there is no $x$ dependence 
in the boundary conditions $T_1$ and $T_2$, the problem reduces to 1D 
temperature flow along the vertical direction, in which case $T= T(y,t)$ in
the central heat-conducting region. To be consistent with the 2D geometrical setup, 
we will refer to the 1D boundary conditions $T(0,t)=T_1$ and  $T(L,t)=T_2$ as
the  {\em bottom} and  {\em top} boundary condition, respectively.
}
\label{fig_planarSandwichDeto1Drod}
\end{figure}

As illustrated in Fig.~\ref{fig_planarSandwichDeto1Drod}, the planar symmetry of  
the problem allows us to express the 2D solution in terms of a 1D profile along the 
\hbox{$y$-direction}, independent of the $x$-position within the central heat-conducting 
region \hbox{$a_1 \le x \le a_2$}. This is because the upper and lower boundary 
conditions are uniform along the \hbox{$x$-direction} over the whole range 
\hbox{$0 \le x \le L$},  and therefore heat flows only along the vertical direction. 
Thus, the corresponding 1D heat flow equation for $y$-the profile $T=T(y,t)$ is
\begin{eqnarray}
  \frac{\partial T}{\partial t}
  &=&
  \kappa\, \frac{\partial^2 T}{\partial y^2}
  \ ,
\end{eqnarray}
where the 1D initial condition (IC) is
\begin{eqnarray}
  T(y, 0) = T_0 = 0 
  \label{OneDIC}
  \ ,
\end{eqnarray}
and the corresponding 1D boundary conditions (BCs) are 
\begin{eqnarray}
  T(0, t) &=& T_1  \hskip 1.0cm {\rm (bottom)}
  \label{OneDBCa}
  \\
  T(L, t) &=& T_2 \hskip 1.0cm {\rm (top)}
  \label{OneDBCb}  
    \ .
\end{eqnarray}
We refer to the BCs as the {\em bottom} and {\em top} boundary conditions, 
respectively, as suggested by the 1D rod in Fig.~\ref{fig_planarSandwichDeto1Drod}. 
\begin{figure}[t!]
\includegraphics[scale=0.45]{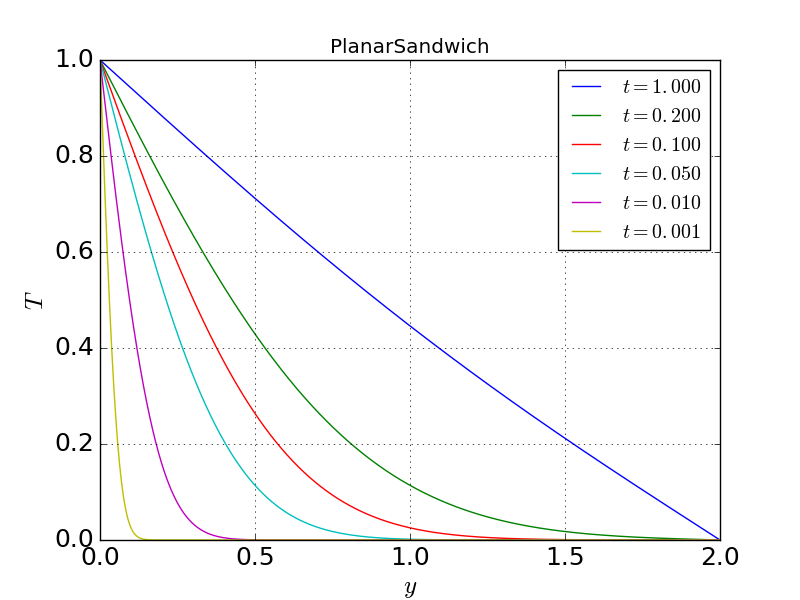}
\caption{\footnoteskip  
Temperature profiles $T(y,t)$ for the planar sandwich solution at 
six representative time slices, \hbox{$t = 0.001, 0.01$}, $0.05, 0.1, 0.2, 1$, 
ranging from early to late times. The heat diffusion coefficient is taken to 
be $\kappa=1$, the length of the 1D heat-conducting rod is set to $L=2$ 
(the abscissa of the graph), and the the first 1000 terms of the series 
have been summed. We have employed the boundary conditions 
$T_1=1$, $T_2=0$ at $y=0$, $y=L$, with the initial condition $T_0=0$.  
Note that the solution at the earliest time $t=0.001$ has negligible 
temperature for all $y$ outside a small neighborhood about $y=0$, as 
required by the initial condition $T_0=0$. Also note that the late-time 
solution at $t=1$ is very close to the static equilibrium solution 
$\bar T(y) = T_1 + (T_2 - T_1)\,y/L$. We shall perform all future verification 
analyses and convergence plots at time $t=0.1$. This time is early enough 
to capture the dynamics of the diffusive heat flow across a large range of 
$y$, and it is late enough to have a contribution from the static equilibrium 
solution. 
}
\label{figs1_planar_sandwich}
\end{figure}
The exact analytic solution for IC (\ref{OneDIC}) and the BCs 
(\ref{OneDBCa})--(\ref{OneDBCb}) was presented in Ref.~\cite{DMS}, 
and takes the form
\begin{eqnarray}
  T(y,t)
  &=&
  T_1 + \frac{(T_2 - T_1) \, y}{L}
  + 
  \sum_{n=1}^\infty B_n \, \sin(k_n \, y) \, e^{-\kappa\, k_n^2 t}
\label{eq_planar_sandwichA}
  \\[5pt]
  k_n &=& \frac{n \pi}{L} 
  \hskip0.5cm {\rm and}\hskip0.5cm
  B_n 
  =
  \frac{2 T_2 \, (-1)^n - 2 T_1}{n\pi}
  \ .
  \label{eq_planar_sandwichB}
\end{eqnarray}
%
The solution profiles $T(y,t)$ are plotted in Fig.~\ref{figs1_planar_sandwich} for six 
time slices \hbox{$t = 0.001, 0.01, 0.05$}, $0.1, 0.2, 1$. The boundary conditions 
are $T_1=1$ and $T_2=0$, and the initial condition is $T_0=0$. We also take 
the diffusion constant to be $\kappa=1$, the length of the domain to be $L=2$, 
and we sum over the first 1000 terms of the series. As illustrated in 
Fig.~\ref{figs1_planar_sandwich}, the solution at the earliest time $t=0.001$ has 
a negligible temperature for all values of $y$ outside a small neighborhood about 
$y=0$. Also note that the late-time solution at $t=1$ is very close to the static 
equilibrium solution 
\begin{eqnarray}
 \bar T(y)
  &=&
  T_1 +  \frac{( T_2 - T_1) \, y}{L}
  \ .
\end{eqnarray}
We shall preform the verification analyses and convergence plots at time $t=0.1$. The
choice  $t=0.1$ is an intermediate time that is sensitive to both the dynamics of the 
heat flow and to the late-time equilibrium solution. This choice tests the static boundary 
conditions and the dynamics of the heat flow solver. 

\begin{figure}[t!]
\includegraphics[scale=0.40]{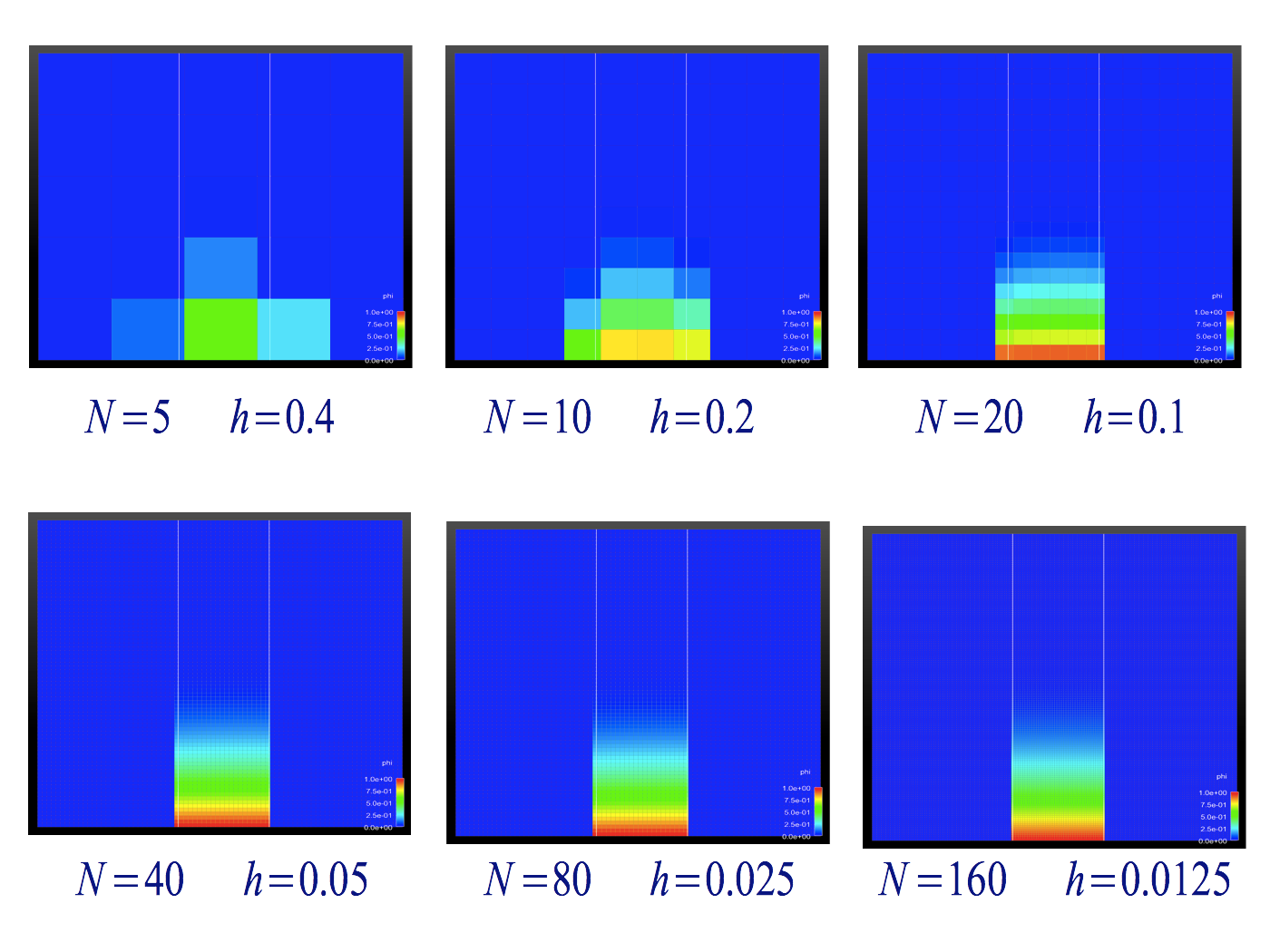}
\caption{\footnoteskip  
Illustration of the 2D planar sandwich solution employing the arithmetic 
average multimaterial algorithm for six resolutions. The $x$- and $y$-axes 
are divided into $N$ segments, with $N=5, 10, 20, 40, 80, 160$. For $L=2$ 
this gives the six resolutions $h = 0.5, 0.2, 0.1, 0.05, 0.025, 0.0125$. Note 
that heat flows outside the central boundary region 
$a_1 = 0.77 \le x  \le 1.27 = a_2$. 
This is particularly noticeable at low resolutions. 
}
\label{fig_planar_sandwich_c1_ensight}
\end{figure}

\begin{figure}[t!]
\includegraphics[scale=0.40]{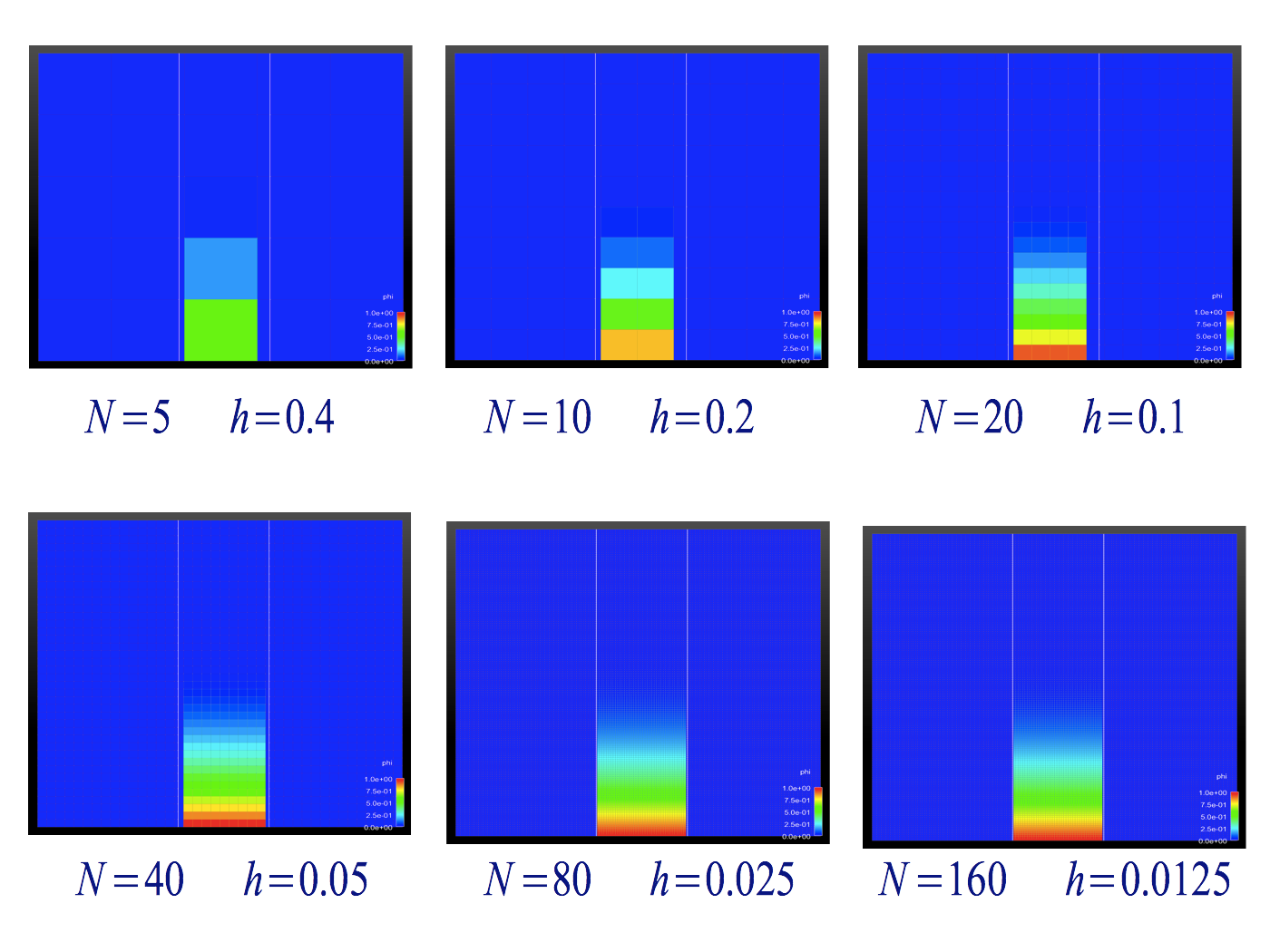}
\caption{\footnoteskip  
Illustration of the 2D planar sandwich solution employing the harmonic 
average multimaterial algorithm for six resolutions. The $x$- and $y$-axes 
are divided into $N$ segments, with $N=5, 10, 20, 40, 80, 160$. For $L=2$ 
this gives the six resolutions $h = 0.5, 0.2, 0.1, 0.05, 0.025, 0.0125$. 
Note that the heat is underestimated in the multimaterial regions.
}
\label{fig_planar_sandwich_c2_ensight}
\end{figure}

Even though we perform  the FLAG simulations in a 2D cartesian geometry,  we shall 
express the numerical solution in terms of its 1D profile. However, before proceeding, 
it is convenient to visualize the solutions in 2D. The $x$- and $y$-axes are divided 
into $N$ segments, with $N=5, 10, 20, 40, 80, 160$. This gives the six grid spacings 
$h = 0.5, 0.2, 0.1, 0.05, 0.025, 0.0125$. Note that heat flows outside the boundary 
region, particularly at low resolutions. Figures~\ref{fig_planar_sandwich_c1_ensight} 
and \ref{fig_planar_sandwich_c2_ensight} illustrate the arithmetic and harmonic 
average models for the 2D numerical FLAG solutions, for six levels of increasing 
resolution. To better understand these Figures, we examine the arithmetic technique
in more detail. We shall see that the arithmetic average 
has the effect of extending the inner material beyond the boundary, while
the harmonic average decreases the inner material within the boundary. 
For the planar sandwich geometry illustrated in Fig.~\ref{fig_sub_grid}, the
material boundaries are static with $x=a_1$ and $x=a_2$. The
arithmetic average of the diffusion coefficient on a boundary cell is 
\begin{eqnarray}
  \bar\kappa_a 
  &=& 
  V_1\, \kappa_1 + V_2\, \kappa_2 
  =
  V_1\, \epsilon + V_2\, \kappa   
  \to V_2 \kappa > 0
  \ ,
\end{eqnarray}
and we should expect the arithmetic average to overestimate the effects of multimaterial 
heat flow along the boundary cells.
\noindent
Conversely, the harmonic average gives
\begin{eqnarray}
  \bar\kappa_h^{-1}
  &=& 
  V_1\, \kappa_1^{-1} + V_2\, \kappa_2^{-1} 
  =
  V_1\, \epsilon^{-1} + V_2\, \kappa^{-1}
  \to V_1 \epsilon^{-1}
  ~~{\rm or}~~
  \bar\kappa_h \to 0
  \ ,
\end{eqnarray}
which underestimates the effects of multimaterial heat flow along the boundary cells.  
The numerical FLAG results of Ref.~\cite{DMS} indeed show that the arithmetic mean emphasizes 
larger values of the diffusion coefficient, while the harmonic mean emphasizes smaller values.

\section{Using ExactPack Heat Solvers}
\label{sec:sample_ep_script}

In this section we present an example of how to use ExactPack\,\cite{exactpack} 
to perform code verification for the exact solution of the planar 
sandwich.\,\footnote{\footnoteskip ExactPack is an open source project  
available at \url{https://github.com/lanl/ExactPack}. 
}  
One starts by importing the planar sandwich solution module into Python,

\vskip0.2cm
\begin{verbatim}
from exactpack.solvers.heat import PlanarSandwich 
\end{verbatim}
\noindent
As discussed in the last section, the exact solution for the 2D planar sandwich 
can be described in terms of a 1D rod of length $L$ with a uniform heat diffusion 
coefficient $\kappa$. The 1D solution is implemented in ExactPack by the solver 
\verb+PlanarSandwich+. The planar sandwich class comes with a number of 
default settings for the input parameters, such as the length of the rod, the 
value of the heat diffusion coefficient, settings for the boundary and initial conditions, 
and the number of terms to be summed in the series. To instantiate and use the 
\verb+PlanarSandwich+ class with default values, one invokes
\vskip0.2cm
\footnoteskip
\begin{verbatim}
solver = PlanarSandwich() 
\end{verbatim}
\bodyskip
\noindent
The parameter values can be explicitly set by
\vskip0.2cm
\footnoteskip
\begin{verbatim}
solver = PlanarSandwich(T1=1, T2=0, L=2, kappa=1, Nsum=1000)
\end{verbatim}
\bodyskip
\noindent
This creates an ExactPack object  called \verb+solver+  with boundary 
conditions $T_1=1$ and $T_2=0$, the length of the 1D rod  set to $L=2$, 
the diffusion coefficient set to $\kappa=1$, and we have specified that we 
want to retain the first 1000 terms in the summation. The default initial 
condition is $T_0=0$; this can be changed by setting the variables
\verb+TA+ and \verb+TB+.
Access to all the properties of the planar sandwich definition can be controlled 
through the class \verb+PlanarSandwich+. 

The planar sandwich object encapsulates the definition of the problem, but it is 
unaware of the spatial grid necessary for a specific realization of the 
problem.  Therefore, a {\em solver} object must be used to  produce a 
{\em solution} object on a spatial array at a given time \verb+t=0.1+. The  
corresponding Python code is
\vskip0.2cm
\footnoteskip
\begin{verbatim}
solver = PlanarSandwich()

L = 2
y = numpy.linspace(0, L, 1000)
t = 0.1

soln = solver(y, t)
soln.plot('temperature') 
\end{verbatim}
\bodyskip

\begin{figure}[t!]
\includegraphics[scale=0.45]{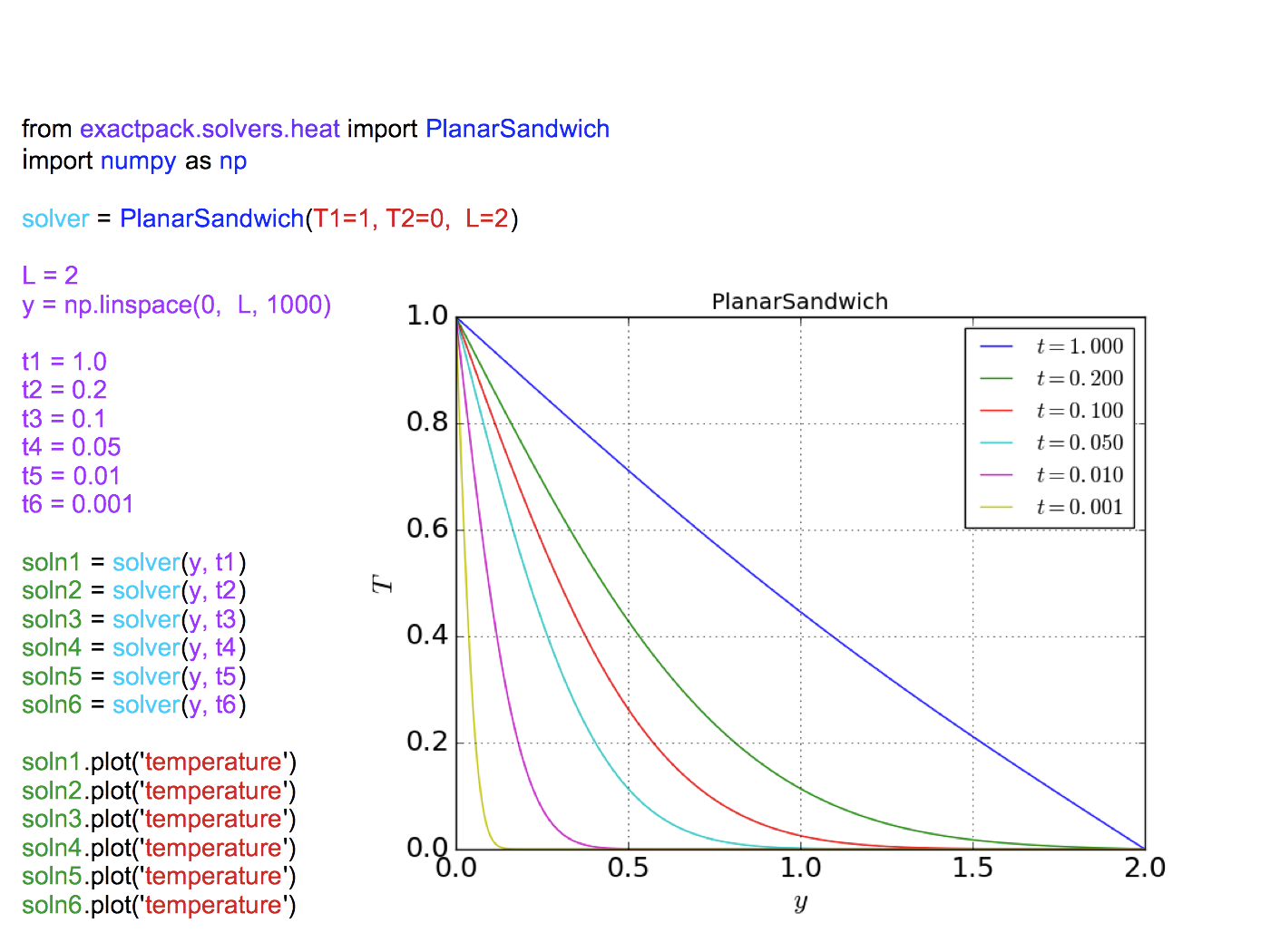}
\caption{\footnoteskip  
An illustration of the Python source code used to plot the planar sandwich 
contour solutions of Fig.~\ref{figs1_planar_sandwich}. The Python script
to produce this figure is provided in Appendix~\ref{sec_python_script}. 
The planar sandwich solver has been instantiated by 
\texttt{solver = PlanarSandwich(T1=1, T2=0, L=2)}, 
which sets the boundary conditions to $T(0,t)=1$, $T(L,t)=0$, the initial condition 
to $T(y,0)=0$, and the length of the rod (along the $y$-direction) to $L=2$. 
The values $\kappa=1$ and $N_{\rm sum}=1000$ are the default settings, 
and are not specified in the above code segment. 
The solver is used to form a solution object from a spatial array \texttt{y} and
a time \texttt{t} by  \texttt{soln = solver(y, t)}. 
}
\label{fig_planar_sandwich_exactpack}
\end{figure}

\noindent
The solver object \verb+solver+ takes a spatial array \verb+y+ and a
time variable \verb+t+, and produces a solution object called
\verb+soln+ with the exact solution evaluated on the spatial array at
the given time. As illustrated by the last line above, a solution
object is equipped with a plotting method, in addition to various
analysis methods not shown here. The Python script that produces
Fig.~\ref{figs1_planar_sandwich} is given in
Appendix~\ref{sec_python_script}, and is summarized in
Fig.~\ref{fig_planar_sandwich_exactpack}.

%

\section{Grid Resolution Studies of the Planar Sandwich}

In the previous sections, we examined the planar sandwich test problem in some 
detail, in particular, we provided the exact solution in a semi-analytic form in
Eq.~(\ref{eq_planar_sandwichA}). The geometry of the planar sandwich is 
illustrated in Fig.~\ref{fig_sub_grid}, and the material interfaces and their heat 
conduction properties are defined in Fig.~\ref{fig_planar_sandwich}. In this section, 
we build on these results by performing rigorous convergence analyses for the four 
primary multimaterial heat flow algorithms in FLAG, namely,  (i) the arithmetic average,  
(ii) the harmonic average,  (iii) thin mesh, and  (iv) static condensation. As we have 
already emphasized, the analyses are performed at time $t=0.1$, and the domain of 
the planar sandwich is the two dimensional region $D=[0,L] \times [0,L]$.  In numerical 
simulations we take $L=2$, partitioning the domain $D$ into $N \times N$ square cells 
with sides of length $h = L/N$.  In other words, the 2D computational grid 
is formed by dividing the $x$- and $y$-grids into $N$ equal segments of length 
$h$, thereby creating $N^2$ square cells with sides of length 
$\Delta x = \Delta y = h$. It should be noted that the algorithms in FLAG do not require 
the cells to be square, and all multimaterial methods work on general polytopal meshes. 
We only use square cells to provide a unique length scale $h$ with which to plot the 
norms. In all numerical simulations, we take the number of segments to increase by a factor 
of two, starting with five segments for the lowest resolution and ending with 640 for the 
highest resolution, 
\begin{eqnarray}
  N=5, 10, 20, 40, 80, 160, 320, 640 \ .
\label{eq:N}
\end{eqnarray}
For $L=2$, the square cells have sides of length
\begin{eqnarray}
h \equiv L/N = 0.4, 0.2, 0.1, 0.05, 0025, 0.0125, 0.00625, 0.003125.
\label{eq:dxdy}  
\end{eqnarray}
It is important to note that in all numerical simulations, we halve the maximum 
time step for every doubling in $N$. The central heat-conducting material, the 
meat of the sandwich, is the 2D region \hbox{$M = \big\{ (x,y) \in D \,\vert\,  
a_1 \le x \le a_2 ~{\rm and}~ 0 \le y \le L \big\}$}, within which the heat diffusion 
coefficient takes the value $\kappa=1$. The outer two materials $B = \big\{  (x,y) 
\in D \,\vert\, 0 \le x < a_1 ~{\rm or}~ a_2 < x \le L ~{\rm with}~ 0 \le y \le L \big\}$
are called the bread of the sandwich, and are composed of an insulated material 
for which $\kappa =0$. In our numerical simulations, 
we do not actually take the heat diffusion coefficient to vanish inside $B$, but rather, 
we set $\kappa=\epsilon \equiv 10^{-12}$.  We should therefore think of the condition 
$\kappa=10^{-12}$ as a limiting procedure in which $\kappa = \epsilon$ with 
$\epsilon \to 0^+$.

\subsection{The Arithmetic Average}

\begin{figure}[t!]
\includegraphics[scale=0.40]{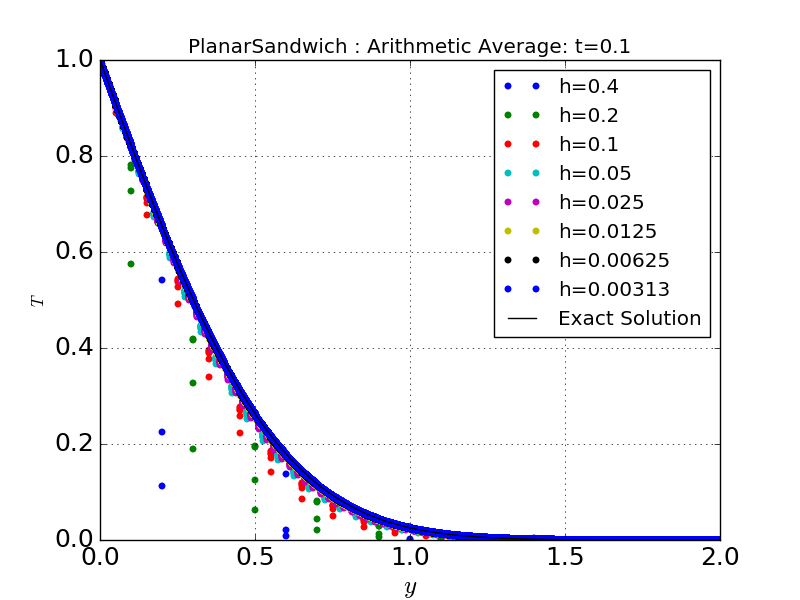}
\caption{\footnoteskip  
The numerical FLAG runs for the planar sandwich at time $t=0.1$ for the 
arithmetic average technique. The exact  solution is plotted in black,  although 
it is difficult to resolve against the dense numerical background. We have 
performed numerical runs for \hbox{$N=5, 10$}, $20, 40, 80, 160, 320, 640$. 
Choosing length $L=2$, the corresponding resolutions are 
\hbox{$h  \equiv N/L = 0.4$}, $0.2, 0.1, 0.05, 0025, 0.0125, 0.00625, 0.003125$.  
For every resolution $h$, the numerical solution can be expressed by the triplets 
$(x_i, y_j, T_{ij})$ for $i, j = 1, 2 \cdots N$. We have projected out the $x$-coordinates, 
plotting $(y_j, T_{ij})$ in the Figure.  At high resolutions, the solution becomes 
independent of $x_i$, and indeed, as the grid resolution increases, the points become 
more closely spaced, and lie closer to the exact solution. For these numerical runs, we 
have taken the conducting region (the meat of the sandwich) to be 
$a_1=0.75 \le x \le 1.25=a_2$. The numerical results for the shifted region 
$a_1=0.77 \le x \le 1.27=a_2$ are qualitatively similar in appearance. 
}
\label{fig_planar_sandwich_c1_run8}
\end{figure}

In this section we perform the convergence analysis for the arithmetic average 
multimaterial algorithm in FLAG. We first consider the case in which $a_1=0.75$ 
and $a_2 = 1.25$ (with $L=2$).  At grid point $(x_i, y_j) \in D$ and time $t=0.1$, 
the numerical algorithm returns a temperature $T_{ij}$, and therefore, 
the numerical results can be expressed by the $N^2$ triplets $(x_i, y_j, T_{ij})$ 
for $i, j = 1, 2 \cdots N$.   In Fig.~\ref{fig_planar_sandwich_c1_run8}, we have 
projected out the $x$-coordinate and plotted the points $(y_j, T_{ij})$  at time 
$t$. Note that the numerical solutions becomes more finely spaced in $y$ with 
increasing $x$-resolution, until the points $(y_j, T_{ij})$ lie on top of the exact 
1D profile $T(y,t)$. In the Fig.~\ref{fig_planar_sandwich_c1_run8}, the exact 
solution is the solid black line, although it is difficult to resolve against the 
dense numerical background.  

Let us examine the ExactPack script used to produce Fig.~\ref{fig_planar_sandwich_c1_run8}.
ExactPack contains an object called \verb+Study+ that, among other things, can be used 
to plot the numerical data alongside the exact solution:
\vskip0.2cm
\vbox{
\footnoteskip
\begin{verbatim}
study = Study(datasets=dumpfiles,
    reference=PlanarSandwich(),
    study_parameters=[0.4, 0.2, 0.1, 0.05, 0025, 0.0125, 0.00625, 0.003125],
    time=0.1,
    reader=FlagVarDump(),
    abscissa='y_position'
    )
    
study.plot('temperature')
\end{verbatim}
}
\bodyskip

\begin{figure}[t!]
\includegraphics[scale=0.45]{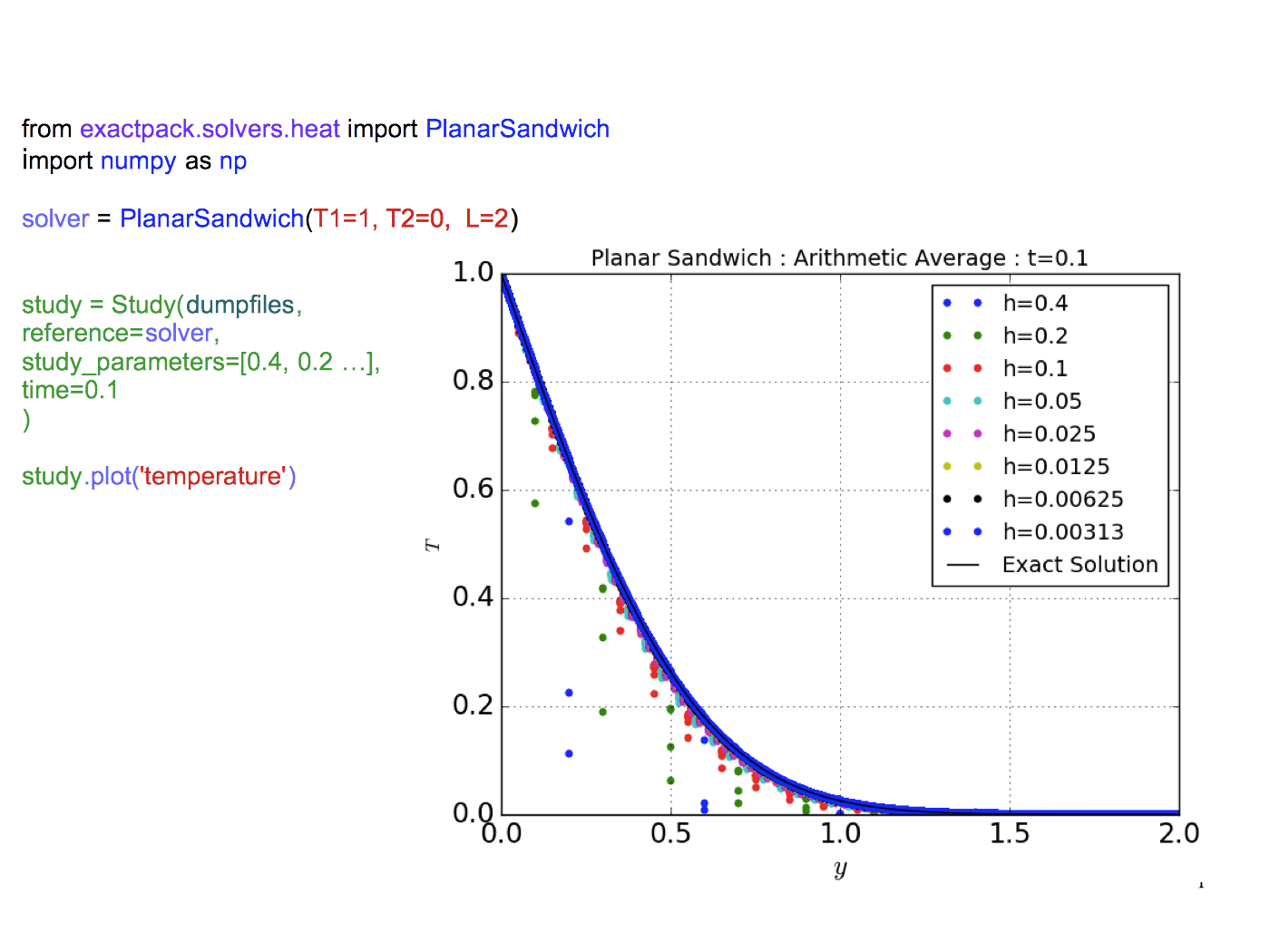}
\caption{\footnoteskip  
A summary of the Python source code used to produce
Fig.~\ref{fig_planar_sandwich_c1_run8}.  The exact solution is plotted 
in black,  although it is hard to resolve against the dense numerical background. 
The study object contains a plotting method that plots the exact solution 
alongside the numerical results. 
}
\label{fig_planar_sandwich_exactpack_data}
\end{figure}

\noindent
The statement \verb+study=Study(+$\cdots$\verb+)+ instantiates the object \verb+Study+ 
by the instance \verb+study+, where the latter inherits it properties and methods from the 
former.  The study object contains a plot method, \verb+study.plot('temperature')+,
which instructs the object to plot itself, thereby producing Fig.~\ref{fig_planar_sandwich_c1_run8}.
The object  \verb+Study()+ takes a  number of arguments. The first argument \verb+datasets+
has been assigned the value \verb+dumpfiles+, which is a regular expression for the path of
the code output. The output consists of separate code runs at the resolutions 
specified by \verb+study_parameters+ and at the time specified by \verb+time+. The argument  
\verb+reference+ is used to select the reference solver in ExactPack, which in this case is 
\verb+PlanarSandwich+.  The argument \verb+reader+ provides the interface between the code 
output and ExactPack. In this case, the code reader is specific to FLAG; however, ExactPack will 
soon use the VTK format by default. In general, the numerical output can always be expressed 
in the form $(x_i, y_j, T_{ij})$ with $i,j, = 1, \cdots, N$, for every resolution $N$, and the final 
argument \verb+abscissa=`y_position`+ specifies that only data only along the $y$-direction 
be used in the analyses. With this setting, the analyses and figures are performed using the 1D 
profile representation. The ExactPack script and corresponding Figure are summarized in 
Fig.~\ref{fig_planar_sandwich_exactpack_data}, with the upper left margin specifying the 
ExactPack script.

\vskip0.2cm

The $L_1$ error norm is defined by
\begin{eqnarray}
  L_1 
  =
  \sum_{i,j =1}^N \,\bigg\vert T_{ij}^{\,\rm num} - T_{ij}^{\, \rm ex} \bigg\vert
  \ ,
\end{eqnarray}
where $T_{ij}^{\rm num}=T_{ij}$ is the numerical solution at position $(x_i, y_j)$ 
and time $t$, and $T_{ij}^{\rm ex}=T(x_i,y_j,t)$ is the corresponding exact solution. 
We remind the reader that we take $t=0.1$ in all numerical analyses, simulations, and 
figures. We can define a restricted metric $L_1(R)$ over a subset $R \subset D$, such 
that
\begin{eqnarray}
  L_1(R)
  = \!
  \sum_{ij ,\, (x_i, y_j) \in R} \, 
  \bigg\vert T_{ij}^{\,\rm num} - T_{ij}^{\, \rm ex} \bigg\vert
  \ .
\end{eqnarray}
ExactPack currently only supports the data format of the 1D profile along 
$y$; therefore, the norms are only calculated over 1D regions $R \subset (0,L)$ 
along the $y$-axis, 
\begin{eqnarray}
  L_1(R)
  =
  \sum_{ij ,\, y_j \in R} \bigg\vert T_{ij}^{\,\rm num} - T_{ij}^{\, \rm ex} \bigg\vert
  \ .
\end{eqnarray}
Since the temperature vanishes inside region $B$, the only sizable contributions 
to $L_1$ occur for $(x_i, y_j) \in M$. 

\begin{figure}[t!]
\includegraphics[scale=0.40]{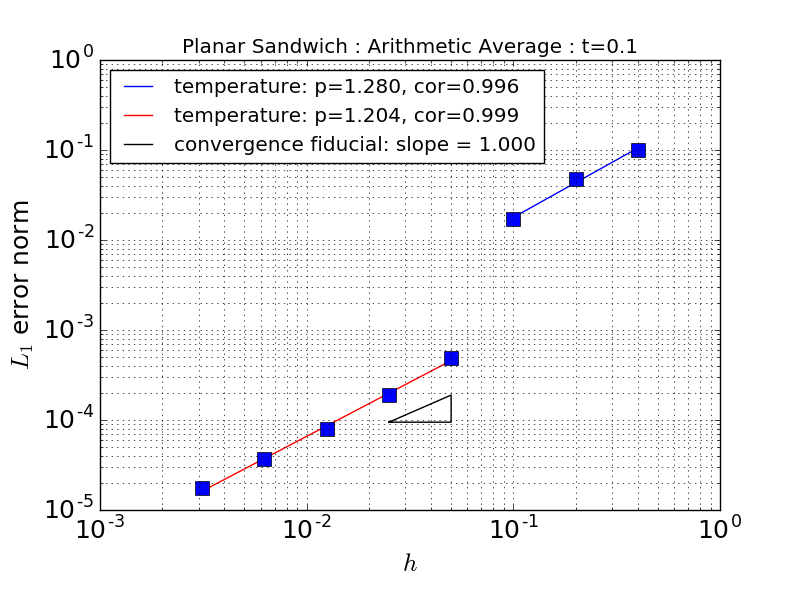}
\caption{\footnoteskip  
Convergence analysis at time $t  = 0.1$ for the planar sandwich 
using the arithmetic average treatment of multimaterial cells. The length
of the domain is $L=2$ with eight uniform regions $N=5, 10, 20, 40, 80, 
160, 320, 640$, corresponding to resolutions \hbox{$h=0.4, 0.2, 0.1, 0.05,
0025$}, $0.0125, 0.00625, 0.003125$.  Since $a_1=0.75 \le x \le 1.25=a_2$ 
with  $L=2$, the numerical grid and the material interface align on the the 
$4^{\rm th}$ iteration at $h=0.05$. When the mesh aligns with the material 
boundary, the accuracy increases by an order of magnitude, producing
a discontinuity in the convergence graph. However, both the upper and 
lower branches converge at approximately the same rate.
}
\label{fig_planar_sandwich_c1_res_run8}
\end{figure}

\begin{figure}[h!]
\includegraphics[scale=0.37]{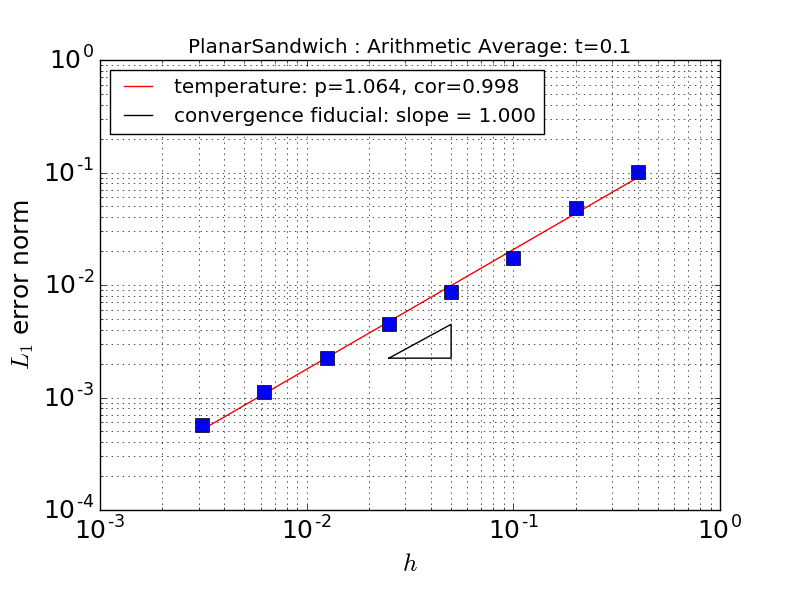}
\caption{\footnoteskip
The convergence analysis in FLAG at time $t = 0.1$ for the planar sandwich 
using the arithmetic average treatment of multimaterial cells gives $p=1.0$. 
The length of the domain is $L=2$ with eight uniform regions $N=5, 10, 20, 
40, 80, 160, 320, 640$, corresponding to resolutions \hbox{$h=0.4, 0.2, 0.1, 
0.05, 0025$}, $0.0125, 0.00625, 0.003125$.  The central region has been 
shifted to $a_1=0.77 \le x \le 1.27=a_2$, so that the mesh refinements 
are never commensurate with this region. The convergence plot is then
rendered continuous.
}
\label{fig_planar_sandwich_c1_res_run9}
\end{figure}

Figure~\ref{fig_planar_sandwich_c1_res_run8} illustrates the convergence analysis 
for the planar sandwich test problem in FLAG, where we have plotted the $L_1$ 
norm against the resolution $h$ on a \hbox{log-log} scale.  For the heat-conducting 
region delimited by \hbox{$a_1=0.75 \le x \le 1.25=a_2$} with $L=2$ and $N=5$,  
the $4^{\rm th}$ mesh refinement has resolution $h=0.05$, at which point the 
numerical grid aligns with the material boundaries at $x=a_1$ and $x=a_2$.
At the alignment, the accuracy 
of the code increases by over an order of magnitude, thereby giving rise to a 
discontinuity in the convergence plot. Note, however, that the upper and lower 
branches converge at approximately the same rate $p=1.2$. As shown in
Fig.~\ref{fig_planar_sandwich_c1_res_run9}, we can eliminate the discontinuity 
by slightly shifting the conduction region to $a_1=0.77 \le x \le 1.27=a_2$,
in which case the grid points of the shifted region never align with a material 
interfaces. As we see in the lower panel of the Figure,  the convergence rate is 
$p=1.0$. In everything that follows, we use the shifted region by default. Before 
continuing, we provide a brief summary of the ExactPack code used to create the 
convergence study of Fig.~\ref{fig_planar_sandwich_c1_res_run8}:

\vbox{
\footnoteskip
\begin{verbatim}
domain = (0, L)
fiducials = {'temperature': 1}
fit = RegressionConvergenceRate(study, domain=domain fiducials=fiducials)
fit[0:3].plot_fit('temperature', "-", c='b')
fit[3:8].plot_fit('temperature', "-", c='r')
fit.norms.plot('temperature')
fit.plot_fiducial('temperature')
fit.plot('temperature')
\end{verbatim}
\bodyskip
}


\begin{figure}[h!]
\includegraphics[scale=0.37]{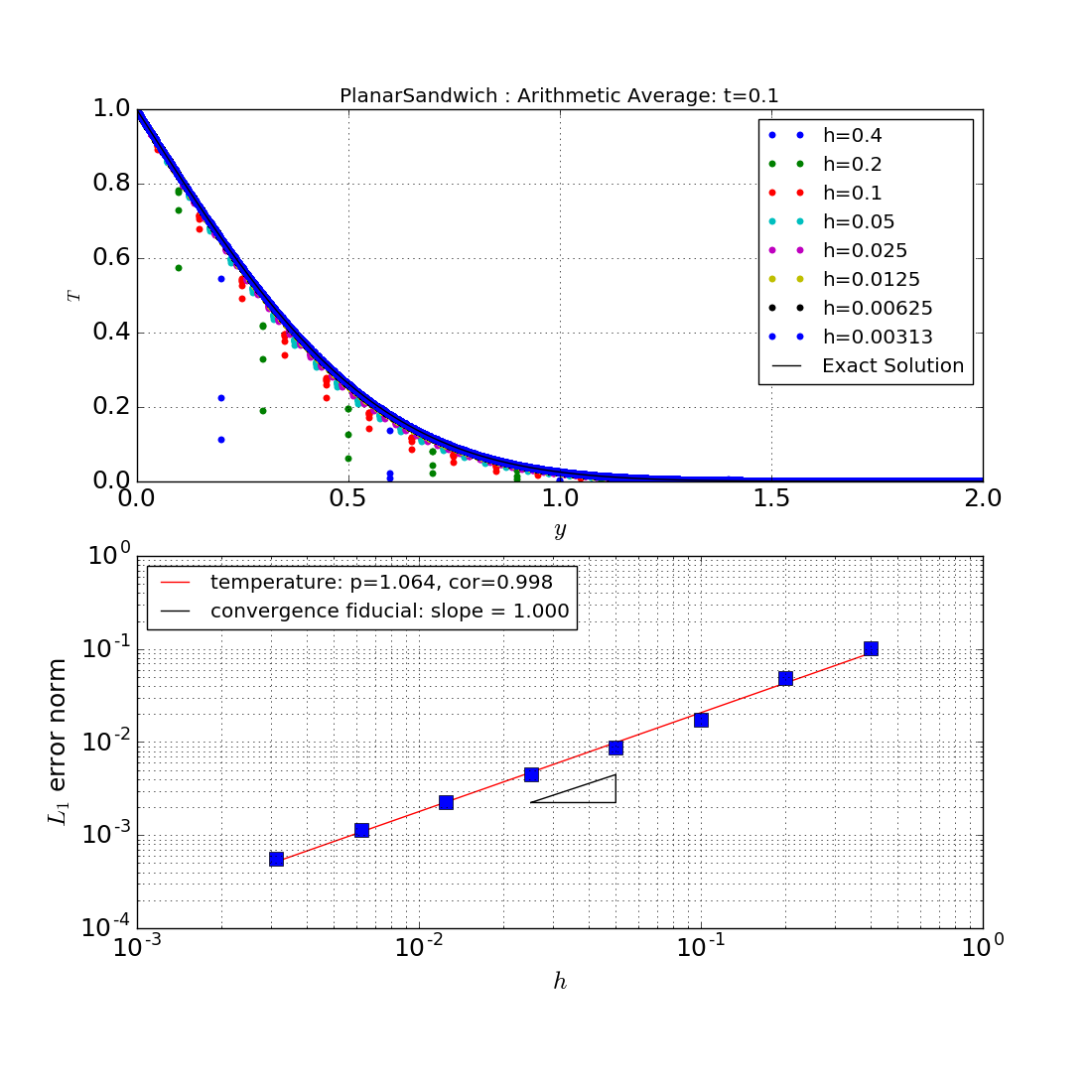}
\vskip-1.3cm
\caption{\footnoteskip  
The arithmetic average for multimaterial cells. Numerical profiles at time \hbox{$t = 0.1$} 
are show in the top panel, and the convergence analysis in the bottom panel, corresponding 
to Figs.~\ref{fig_planar_sandwich_c1_res_run8} and \ref{fig_planar_sandwich_c1_res_run9}. 
The length of the domain is $L=2$ for eight uniform segments with $N=5, 10, 20,  40, 80, 
160, 320, 640$, corresponding to resolutions \hbox{$h=0.4, 0.2, 0.1, 0.05, 0025, 0.0125$}, 
$0.00625, 0.003125$.  The central heat-conducting region has been shifted to $a_1=0.77 
\le x \le 1.27=a_2$, so that the mesh refinements never align with a material boundary. 
The convergence rate for the arithmetic average is approximately $p=1$.
}
\label{fig_planar_sandwich_c1_res_run9_soln}
\end{figure}

\noindent
The above script starts with the assignment statement \verb+domain = (0,L)+, 
which specifies the region along $y$ over which to calculate the norm, where, 
as usual, we take \hbox{$L = 2$}.  The command 
\hbox{\verb+fiducials = {'temperature': 1}+} defines a Python dictionary that 
sets the slope of the fiducial triangle to unity. The next line, 
\verb+fit = RegressionConvergenceRate(+$\cdots$\verb+)+, instantiates the 
convergence rate object \verb+RegressionConvergenceRate+ by the instance 
\verb+fit+. The object \verb+RegressionConvergenceRate+ performs a convergence 
analysis based on an error {\em Ansatz} $L_1(h) = A \, h^p$, and it takes  the 
following arguments: the object \verb+study+ discussed above, the domain over 
which the norm is calculated, and the fiducial 
dictionary. The command \verb+fit.plot_fit+ exercises the method \verb+plot_fit+, 
which uses the matplotlib.pyplot.plot routine to plot the error norms and the best 
fit convergence rate for the error {\em Ansatz}. Other error {\em Ansatze} can be 
used if desired. Since the convergence rate is discontinuous at the $4^{\rm th}$ 
iteration in resolution, we use the nomenclature \verb+fit[0,3]+ and \verb+fit[3,8]+ 
to perform the fits on the first three and last five data sets independently, thereby 
giving convergence rates for the upper and lower branches. It is necessary to 
do an explicit matplotlib.pyplot.show to display the plot, \verb+fit.plot(`temperature`)+.
Figure~\ref{fig_planar_sandwich_c1_res_run9_soln} summarizes the results of this
section. 

\begin{figure}[t!]
\includegraphics[scale=0.37]{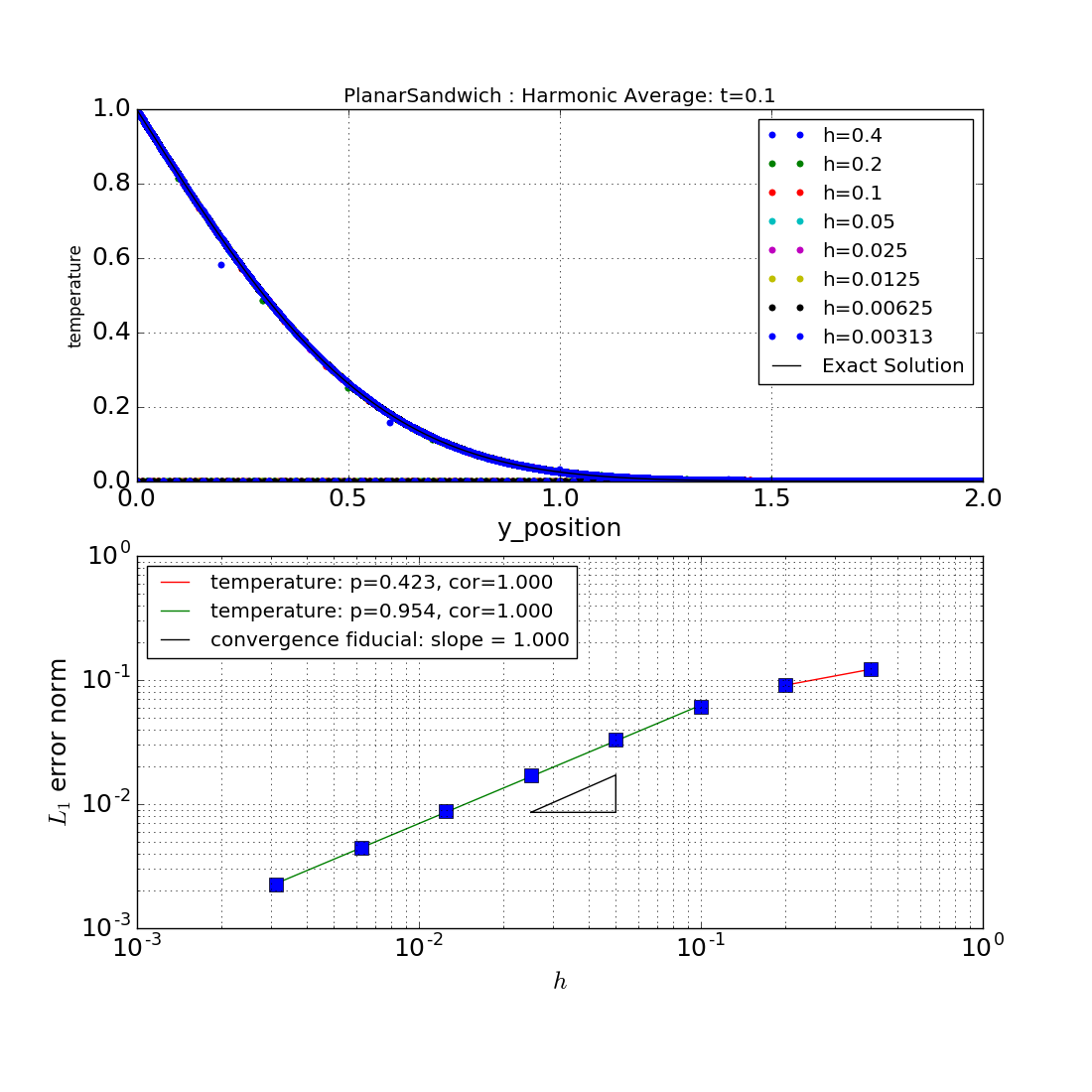}
\vskip-1.3cm
\caption{\footnoteskip  
The harmonic average for multimaterial cells. The parameter setting are the same as in Fig.~\ref{fig_planar_sandwich_c1_res_run9_soln}.
}
\label{fig_planar_sandwich_c2_res_run9_soln}
\end{figure}

\subsection{The Harmonic Average}

We now perform a convergence analysis for the harmonic average multimaterial model 
in FLAG. As discussed in the previous section, it is convenient to use the shifted 
heat-conducting region delimited by $a_1=0.77 \le x \le 1.27=a_2$, where the 
length of the domain is take to be $L=2$. In 
this way, the numerical grid never aligns with  the material interfaces at $x=a_1$ and 
$x=a_2$, thereby rendering the convergence plot continuous. 
Figure~\ref{fig_planar_sandwich_c2_res_run9_soln} illustrates the convergence analysis 
for the harmonic average. The upper panel of the Figure plots the numerical solutions 
$(y_j, T_{ij})$ alongside the exact 1D profile, and the lower panel gives the corresponding
convergence analysis. Note that the first two points, $N=5, 10$, converges at a rate 
$p=0.4$, while $N=20, 40, \cdots, 640$ converges approximately linearly with $p=0.95$, 
indicating that the first two points lie outside the asymptotic range of convergence. 
Comparing the upper panels of Figs.~\ref{fig_planar_sandwich_c1_res_run9_soln} and
\ref{fig_planar_sandwich_c2_res_run9_soln},  we see that the harmonic average solutions 
do not have the same degree of scattering as the arithmetic average, and they lie much 
closer to the exact analytic profile. Note, however, that the harmonic average solutions 
contain a spurious branch along the horizontal axis. This arises because of a bug in FLAG, 
which fails to propagate the heat flow on a small number of discrete grid points. 


\subsection{Thin Mesh and Static Condensation}
\begin{figure}[b!]
\includegraphics[scale=0.37]{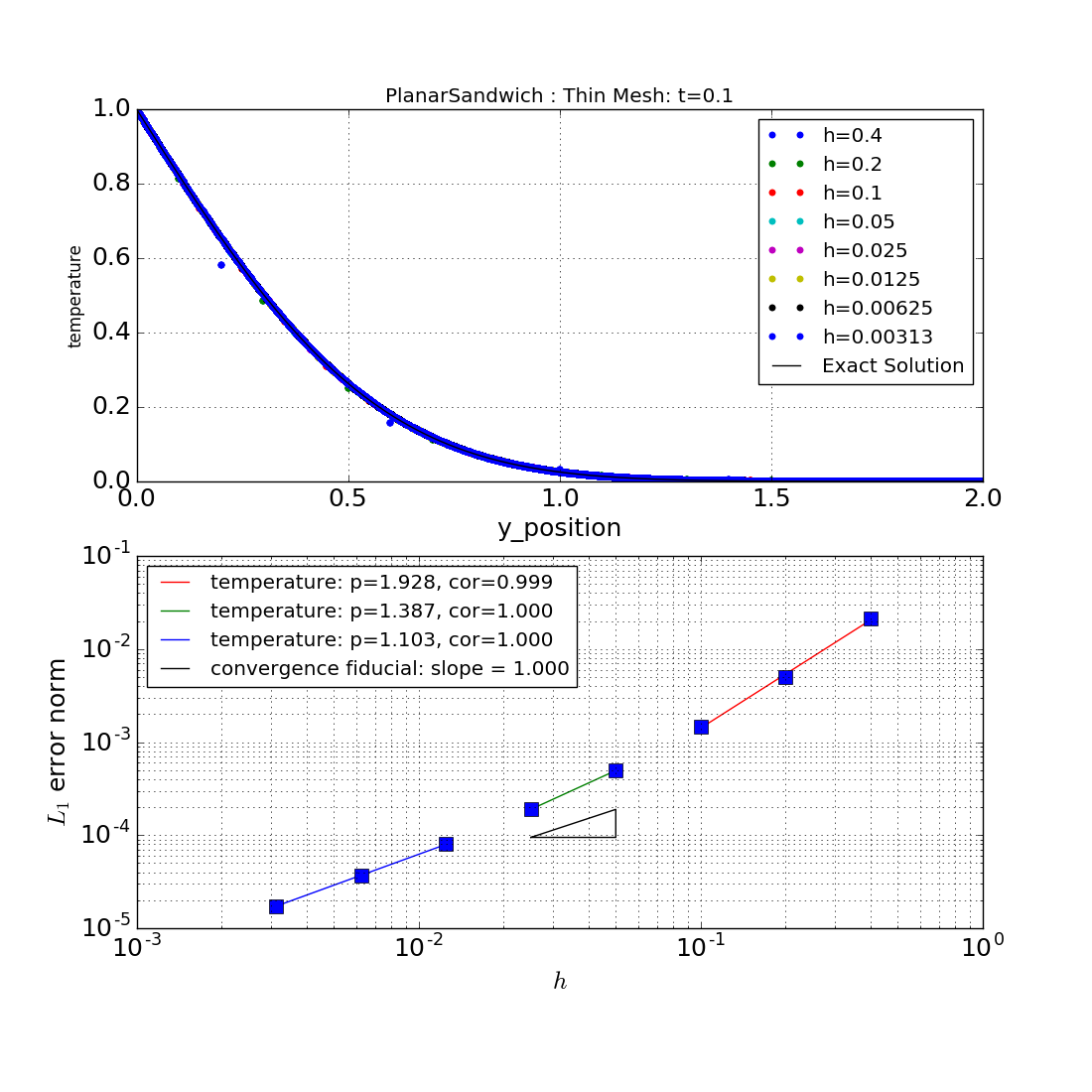}
\vskip-1.0cm
\caption{\footnoteskip  
The thin mesh option for multimaterial cells. The parameter setting are the same as in Fig.~\ref{fig_planar_sandwich_c1_res_run9_soln}.
}
\label{fig_planar_sandwich_tm_run8}
\end{figure}

We now turn to the more advanced multimaterial algorithms of FLAG, 
namely, {\em thin mesh}\,\cite{tm} and {\em static condensation}\,\cite{sc}. 
As discussed in the Introduction, the thin mesh algorithm uses the volume
fractions of each material to reconstruct the material interfaces by employing 
interface reconstruction methods. The mesh is then subdivided along the 
interfaces, making sure that the final polyhedral mesh conforms with the 
numerical mesh. This method is quite accurate but very time consuming.
Figure~\ref{fig_planar_sandwich_tm_run8} illustrates the analysis, with 
the upper panel showing the numerical solutions plotted alongside the 
exact solution, and the lower panel giving the convergence plot.  Note
that thin mesh starts out with $2^{\rm nd}$ order convergence, $p=2$, 
and becomes lower order order as the mesh resolution is refined, eventually
giving $p=1$ at the smallest resolutions. There are a number of possible 
reasons why the convergence rate levels off as the grid is refined, and we 
are currently investigating this. One possibility is that the order of accuracy 
is limited by the interface reconstruction algorithm critical to both of these 
methods.

\begin{figure}[t!]
\includegraphics[scale=0.37]{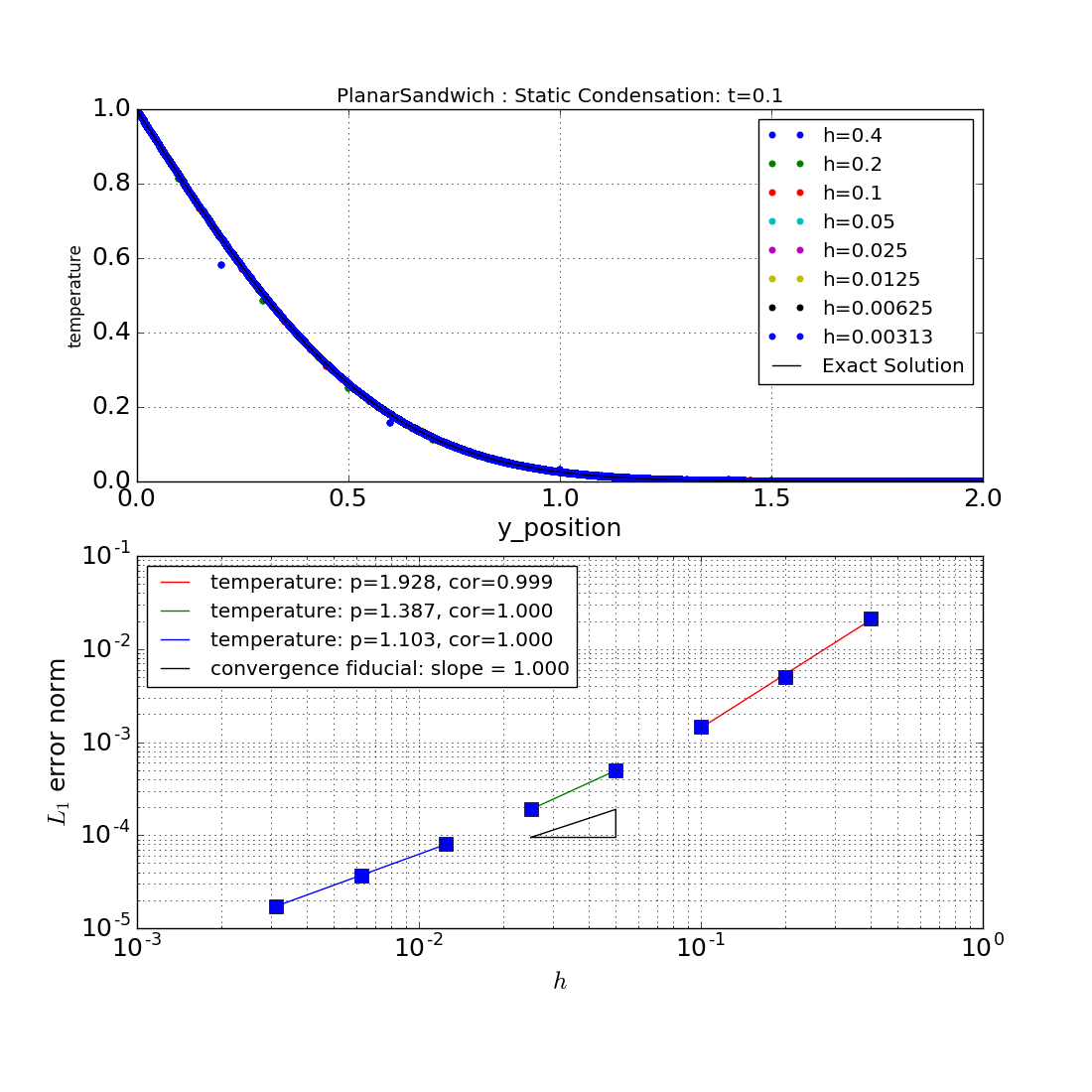}
\vskip-1.3cm
\caption{\footnoteskip  
The static condensation option for multimaterial cells. The parameter
setting are the same as in Fig.~\ref{fig_planar_sandwich_c1_res_run9_soln}.
}
\label{fig_planar_sandwich_c_run9}
\end{figure}

The static condensation approach also makes use of the reconstructed material 
interfaces, but does not require the connectivity information across material 
interfaces within a cell. The global system for the diffusion equation is rewritten 
in terms of unknown face-centered temperature values, and flux continuity is 
enforced at each cell face by ensuring that the sum of the fluxes from all materials 
on either side of the face are the same. The global system is then solved for the 
unknown face temperatures.  The result is that each cell now has a known solution 
on its boundary (faces), which becomes a local Dirichlet problem that can be solved
independently to recover updated material-centered temperatures.  For more details 
on the algorithm, see Ref.~\cite{asc}; this method is reported to be second order 
accurate in Ref.~\cite{sc}. Figure~\ref{fig_planar_sandwich_c_run9} illustrates the 
analysis for static condensation, which exhibits quantitatively similar scaling to
thin mesh. In fact, for the planar sandwich test problem, static condensation agrees 
with thin mesh almost exactly.

%

\pagebreak
\clearpage
\section{Generalizations of the Planar Sandwich}
\label{sec_generalized}

As we have seen, the 2D planar sandwich test problem can be expressed 
in terms of heat flow along a 1D rod oriented in the horizontal direction, 
as illustrated in Fig.~\ref{fig_planarSandwichDeto1Drod}.  The planar 
sandwich is a special case of the more general heat flow problem\,\cite{Berg}, 
\begin{eqnarray}
  \frac{\partial T(y,t)}{\partial t}
  &=&
  \kappa\, \frac{\partial^2 T(y,t)}{\partial y^2}
\label{eq_oneDrodAnh}
\\[5pt]
  \alpha_1 T(0,t) + \beta_1 \partial_y T(0,t) &=& \gamma_1
\label{eq_oneDrodBnhA}    
\\[-3pt]
  \alpha_2 T(L,t) + \beta_2 \partial_y T(L,t) &=& \gamma_2
\label{eq_oneDrodBnhB}    
\\[5pt]
  T(y,0) &=& T_0(y)   
  \label{eq_IC_a}
  \ .
\label{eq_oneDrodCnh}
\end{eqnarray}
This problem is implemented by the ExactPack class \verb+Rod1D+. We use an arbitrary 
but consistent set of temperature units. Equation~(\ref{eq_oneDrodAnh}) is a diffusion 
equation describing the temperature response to heat flow in  a material with constant 
diffusivity. The next two equations, Eqs.~(\ref{eq_oneDrodBnhA}) and (\ref{eq_oneDrodBnhB}) 
are the boundary conditions (BCs), which we take to be nonhomogeneous linear combinations 
of Dirichlet and Neumann conditions.
The initial condition (IC) is given by Eq.~(\ref{eq_oneDrodCnh}), and specifies the 
$t=0$ temperature profile along the rod.  When  the right-hand sides of the BCs vanish, 
$\gamma_1= \gamma_2=0$, the problem is called {\em homogeneous}, otherwise the 
problem is called {\em nonhomogeneous}. The distinction between homogenous and
nonhomogeneous solutions has far-reaching implications for how one goes about
solving the heat problem. The special property of homogeneous solutions is that the 
sum of any two homogeneous solutions is another homogeneous solution. However, 
such linearity is not true for nonhomogeneous solutions, as adding two nonzero values
of $\gamma_i$ violates the BCs.

Finding a solution to the general problem (\ref{eq_oneDrodAnh})--(\ref{eq_oneDrodCnh}) 
involves solving both the homogeneous and nonhomogeneous problems. The general 
{\em homogenous} solution, for which $\gamma_1=\gamma_2=0$, will be denoted by 
$\tilde T(y,t)$. To construct the exact solution, we must also find a {\em specific} static 
solution to the nonhomogeneous problem, which we denote by $\bar T(y)$.\footnote{
\footnoteskip
We choose the strategy of finding a static $\bar T(y)$ because this is usually easier than
finding a time-dependent specific solution. 
}
The general solution therefore takes the form
\begin{eqnarray}
  T(y,t) = \tilde T(y,t) + \bar T(y)
  \ .
  \label{Tgeneral}
\end{eqnarray}
The homogeneous solution
$\tilde T(y,t)$ will be expressed as a Fourier series, and its coefficients 
will be chosen so that the initial condition (\ref{eq_oneDrodCnh}) is satisfied
by $T(y,t)$. This means that we choose the Fourier coefficients of $\tilde T$ 
such that
\begin{eqnarray}
  \tilde T(y,0) = T_0(y) - \bar T(y) 
  \ .
  \label{TBC}
\end{eqnarray}
The static nonhomogeneous solution $\bar T(y)$ is itself linear in $y$,
and takes the form
\begin{eqnarray}
  \bar T(y; T_1, T_2)
  &=& 
    T_1 + \frac{T_2 - T_1}{L}\, y
    \ ,
  \label{barTform}
\end{eqnarray}
where the temperatures $T_1$ and $T_2$ are defined in terms of the parameters
$\alpha_i$, $\beta_i$, and $\gamma_i$ for $i=1,2$. Although there exists a solution
to the heat flow equations for any continuous initial condition $T_0(y)$, it is convenient
for our purposes to consider only the linear initial condition

\begin{eqnarray}
  T_0(y; T_\smA, T_\smB) 
  =
  T_\smA + \frac{T_\smB - T_\smA}{L}\, y
  \ ,
 \label{IClinearA}
\end{eqnarray}
for independent temperature parameters $T_\smA$ and $T_\smB$. 
This form can also be used for the flux boundary condition  $\partial_y T(L)=F_\smB$ 
by setting $T_\smB=F_\smB L$. Therefore, (\ref{TBC}) can be written as
\begin{eqnarray}
  \tilde T(y,0) 
  &=& 
  T_0(y) - \bar T(y) 
  \\[5pt]
  &=&
  T_a + \frac{T_b - T_a}{L}\, y
  \ ,
  \label{barTT0}
 \end{eqnarray}
where
\begin{eqnarray}
  T_a &=& T_\smA- T_1
  \label{Ta_def}
  \\
  T_b &=& T_\smB - T_2
  \label{Tb_def}
    \ .
 \end{eqnarray}
By assuming a linear IC, we find that the boundary conditions and initial conditions 
can be interchanged according to (\ref{Ta_def}) and (\ref{Tb_def}). It would be an
interesting verification exercise to observe the extent to which this holds true in a 
code, since code algorithms handle boundary and initial conditions quite differently.

Let us briefly discuss some basic issues involving {\em uniform convergence}\,\cite{Rubin}.
This is related to order-of-limits questions in classical mathematics, which have direct 
practical implications for many mathematical systems of interest. The parameters $T_\smA$ 
and $T_\smB$ specify the initial temperatures 
at the bottom and top of the rod, or the bottom and top of the 1D slice in
Fig.~\ref{fig_planarSandwichDeto1Drod}. By {\em bottom} and {\em top}, we 
really mean the limits $y=0^+$ and $y=L^-$, respectively. In other words, 
the linear initial condition $T_0(y)$ is defined only on the open interval $(0,L)$, 
with $T_\smA$ being the value of $T_0(y)$ as $y \to 0^+$ and $T_\smB$ being 
the value of $T_0(y)$ as $y \to L^-$. The temperatures $T_\smA$ and $T_\smB$ 
need not be equal to the boundary conditions $T_1$ and $T_2$. If the BCs 
and the IC do not agree,  then the solution is nonuniformly convergent for $t \to 
0$. As an example, consider BC1 with $T_1 \ne T_\smA$ and $T_2 \ne T_\smB$. 
The profile $T(y,t)$ converges point-wise to the initial profile $T_0(y)$ as $t$ goes 
to zero over the domain $(0,L)$, that is to say, the temperature $T(y,t) \to 
T_0(y)$ as $t \to 0$ for all $y \in (0,L)$.  However, this point-wise convergence 
in $y$ is {\em nonuniform} on the closed interval $[0,L]$, in that $T(0,t)=T_1
\ne T_\smA$, although $\lim_{y \to 0^+} T(y,t)= T_\smA \ne  T_\sm1 = T(0,t)$. 
Similarly,  $\lim_{y \to L^-} T(y,t) \ne T(L,t)$. These conditions place a limit on how 
close to $t=0$ one can set the time $t$ in convergence plots. See Ref.~\cite{Rubin} 
for an introductory but solid treatment of real analysis and nonuniform convergence. 

The boundary conditions (\ref{eq_oneDrodBnhA}) and (\ref{eq_oneDrodBnhB}) are 
specified by the coefficients $\alpha_i$, $\beta_i$, and $\gamma_i$ for $i=1,2$. 
These parameters are not all independent, and various combinations will produce 
the same temperatures $T_i$ and fluxes \hbox{$F_i = \partial_y T_i$}; therefore, 
it is often more convenient to specify the boundary conditions directly in terms of 
$T_i$ and $F_i$. For example, if $\beta_1=0$ in (\ref{eq_oneDrodBnhA}), then the 
boundary condition becomes $\alpha_1 T(0,t) = \gamma_1$, which we can rewrite 
in the form $T(0,t)=T_1$ with $T_1 = \gamma_1/\alpha_1$. There are four special 
boundary conditions that provide particularly simple solutions. The first class is specified 
by  $\beta_1=\beta_2=0$ with $\alpha_i \ne 0$, and gives the Dirichlet boundary 
conditions
\begin{eqnarray}
  && {\rm BC1} 
  \nonumber \\
  &&
  T(0,t) =T_1 \hskip0.2cm :~
  \hskip0.15cm 
  \alpha_1 \ne 0 \hskip0.5cm \beta_1 = 0 
  \hskip1.0cm T_1  = \frac{\gamma_1}{\alpha_1} 
  \label{BCTone}
  \\[5pt]
  &&
  T(L,t) = T_2 ~:~
  \hskip0.15cm 
  \alpha_2 \ne 0 \hskip0.5cm \beta_2 = 0 
  \hskip1.0cm  T_2 = \frac{\gamma_2}{\alpha_2}
  \label{BCTtwo}  
  \ .
\end{eqnarray}
The planar sandwich of the previous sections is a subclass of BC1. The next class 
of boundary conditions is obtained by by setting $\alpha_1=\alpha_2=0$ with 
$\beta_i \ne 0$, and this gives the Neumann boundary conditions
\begin{eqnarray}
  &&{\rm BC2} 
  \nonumber \\
  &&
  \partial_y T(0,t) = F_1 \hskip0.2cm :~
  \hskip0.15cm
  \alpha_1 = 0 \hskip0.5cm \beta_1 \ne 0 
  \hskip1.0cm  F_1 = \frac{\gamma_1}{\beta_1}
  \label{eq_BC2A_non}
  \\[5pt]
  &&
  \partial_y T(L,t) = F_2 ~:~
  \hskip0.15cm
  \alpha_2 = 0 \hskip0.5cm \beta_2 \ne 0   
  \hskip1.0cm  F_2 = \frac{\gamma_2}{\beta_2}  
  \label{eq_BC2B_non}
  \ .
\end{eqnarray}
Since the differential equation does not contain sources or sinks of heat, energy conservation 
requires that we must further constrain the heat fluxes to be equal, $F_1=F_2 \equiv F$. 
In  other words, the heat flowing into the system must equal the heat flowing out of the 
system.  We will refer to such solutions as the {\em hot and warm planar sandwiches}, 
depending on whether $F=0$ or $F \ne 0$. 
The final two classes of BCs are mixed Dirichlet and Neumann conditions, 
\begin{eqnarray}
  && {\rm BC3} 
  \nonumber \\
  &&
  T(0,t) = T_1 \hskip0.65cm : \hskip0.25cm
  \alpha_1 \ne 0 \hskip0.5cm \beta_1 = 0
  \hskip1.0cm T_1 = \frac{\gamma_1}{\alpha_1}
  \label{BC3nonhomoA}
  \\[5pt]
  &&
  \partial_y T(L,t) = F_2 ~:~
  \hskip0.15cm 
  \alpha_2 = 0 \hskip0.5cm \beta_2 \ne 0 
  \hskip1.0cm F_2 = \frac{\gamma_2}{\alpha_2}  
  \label{BC3nonhomoB}
  \ ,
\end{eqnarray}
and
\begin{eqnarray}
  && {\rm BC4} 
  \nonumber \\
  &&
  \partial_y T(0,t) = F_1 ~:~ 
  \alpha_1 = 0 \hskip0.55cm \beta_1 \ne 0 
  \hskip1.0cm   F_1 = \frac{\gamma_1}{\beta_1}
  \label{BC4nonhomoA}
  \\[5pt]
  &&
  T(L,t) =T_2 \hskip0.45cm :~
  \hskip0.08cm 
  \alpha_2 \ne 0 \hskip0.5cm \beta_2 = 0   
  \hskip1.0cm   T_2 = \frac{\gamma_2}{\alpha_2}  
  \label{BC4nonhomoB}  
  \ .
\end{eqnarray}
The boundary conditions BC3 and BC4 define the {\em half planar sandwich}. Note 
that BC3 and BC4 are physically equivalent, and represent a rod that has been flipped 
about its midpoint. Boundary conditions BC1--BC4 can be instantiated by 

\footnoteskip
\begin{verbatim}

solver1 = Rod1D(alpha1=1, beta1=0, gamma1=T1,
                alpha2=1, beta2=0, gamma2=T2, TA=Ta, TB=Tb)
                
solver2 = Rod1D(alpha1=0, beta1=1, gamma1=F, 
                alpha2=0, beta2=1, gamma2=F, TA=Ta, TB=Tb)

solver3 = Rod1D(alpha1=1, beta1=0, gamma1=T1, 
                alpha2=0, beta2=1, gamma2=F2, TA=Ta, TB=Tb)

solver4 = Rod1D(alpha1=0, beta1=1, gamma1=F1, 
                alpha2=1, beta2=0, gamma2=T2, TA=Ta, TB=Tb)
\end{verbatim}
\bodyskip
We have specified the boundary conditions of \verb+Rod1D+ by setting $\gamma_i$ 
to the appropriate temperature or flux, and by taking the corresponding coefficients 
$\alpha_i$ and $\beta_i$ to unity or zero, as in  \verb+alpha1=0+, \verb+beta1=0+
and \verb+gamma1=T1+ for BC1. The linear initial condition $T_0(y)$ is specified by 
\verb+TA=Ta+ and \verb+TB=Tb+, which sets the temperature values $T_\smA$ 
and $T_\smB$ at the bottom and top of the 1D profile. 

\begin{figure}[h!]
\includegraphics[scale=0.40]{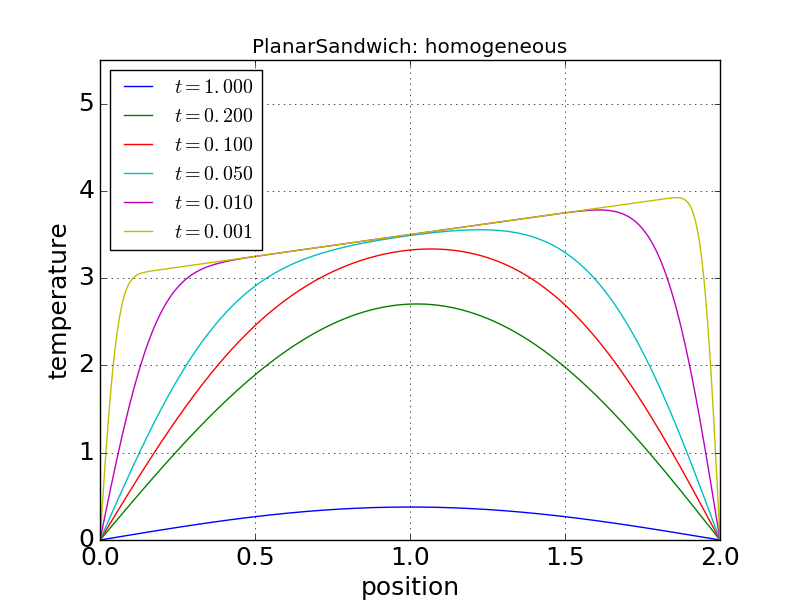} 
\caption{\footnoteskip  
Temperature profiles for the {\em homogeneous planar sandwich} 
at times $t=1, 0.2, 0.1, 0.01$, and $0.001$. The ExactPack class
is \texttt{PlanarSandwich(T1=0, T2=0, TA=3, TB=4, L=2,  Nsum=1000)},
where we have set $\kappa=1$, $L=2$,  $T_\smA=3$, $T_\smB=4$,
with $T_1=T_2=0$.  The boundary conditions $T_1=0$ and $T_2=0$ 
render the solution homogenous, while the linear initial condition $T_0(y)$
is specified by $T_\smA$ and $T_\smB$ via $T_0(y)  =  T_\smA + 
\big(T_\smB - T_\smA\big)y/L$. 
}
\label{fig_planar_sandwich_homo_ep}
\end{figure}

As noted above, the planar sandwich is of type BC1. As another example of BC1, we
impose the homogeneous BCs
\begin{eqnarray}
  T_1 &=& 0
  \\
  T_2 &=& 0
  \ ,
\end{eqnarray}
and we choose the linear initial condition (\ref{IClinearA}) specified by the values $T_\smA$ 
and $T_\smB$ at the endpoints. The specific nonhomogenous solution $\bar T(y)$ vanishes, 
since the boundary conditions are zero, and the solution takes the form\,\cite{PlanarSandwichExactPackDoc},
\begin{eqnarray}
 T(y,t)
 &=&
 \sum_{n=1}^\infty 
 B_n \, \sin(k_n y)\, e^{-\kappa \, k_n^2 t}
 \\[5pt]
  k_n &=& 
  \frac{n \pi}{L}
  \hskip0.5cm {\rm with}\hskip0.5cm
  B_n 
  =
  \frac{2T_\smA - 2T_\smB (-1)^n}{n\pi}
  \ .
  \label{eq_variant_two}
\end{eqnarray}
We are tempted to call this solution the {\em homogenous planar sandwich}, and it
is illustrated in Fig.~\ref{fig_planar_sandwich_homo_ep} for $T_\smA=3$ and $T_\smB=4$.
We have plotted the temperature profiles at times $t=1, 0.2, 0.1, 0.05, 0.01, 0.001$ with 
$\kappa=1$ and $L=2$. We could have taken $T_\smA = T_\smB$, but we chose to plot 
the case for which the initial profile $T_0(y)$ has a nonzero slope. This case is a bit subtle 
to implement in a code. For example, if the temperature lives on the midpoint of cells, we 
must initialize the problem with the values of $T_0(y)$ at these points. For nonuniform 
initial conditions, one will always encounter the problem of sampling the profile $T_0(y)$ 
at specific points. While we have derived the solutions for a general linear profile specified 
by independent values of $T_\smA$ and $T_\smB$, the numerical work has been performed 
only for constant initial contions, $T_\smA = T_\smB = T_0$. We intend to use the solutions 
with $T_\smA \ne T_\smB$ for verification problems involving nonuniform initial setups. 

We close this discussion with a few comments on the general BCs for 
the class \verb+Rod1D+. As an example let us consider the case 

\footnoteskip
\begin{verbatim}
solver = Rod1D(alpha1=3, beta1=-1, gamma1=1, 
               alpha2=1, beta2=2,  gamma2=1, L=2, TL=3, TR=3) ,
\end{verbatim}
\bodyskip

\begin{figure}[h!]
\includegraphics[scale=0.45]{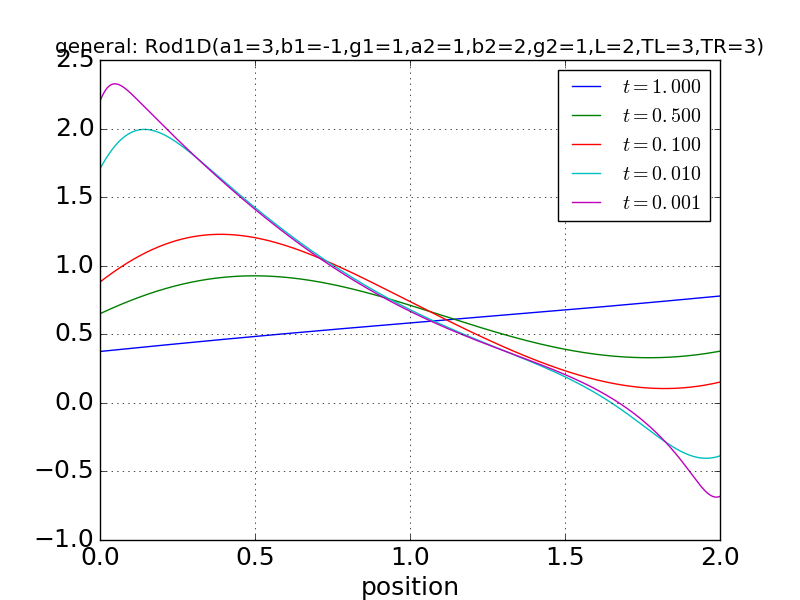} 
\caption{\footnoteskip  
General boundary conditions: \texttt{Rod1D(alpha1=3, beta1=-1, gamma1=1, alpha2=1, 
beta2=2, gamma2=1,  TL=3, TR=3)}.
}
\label{figBCgen} 
\end{figure}

\noindent
This corresponds to the BCs
\begin{eqnarray}
  3 \,T(0,t) -  \partial_y T(0,t) &=& 1
\\[5pt]
  T(L,t) + 2\, \partial_y T(L,t) &=& 1
  \ ,
\end{eqnarray}
and the solution is plotted in Fig.~\ref{figBCgen}. Unlike the previous case, the IC is uniform,
with $T_\smA=T_\smB=3$. The boundary conditions are mixed, and require numerically solving 
the equation $\mu \tan \mu = 1$. In the next section, we perform 
rigorous convergence analyses for the half, hot, and warm planar sandwich variants. In fact, 
\verb+Rod1D+ is the parent class of \verb+PlanarSandwich+ and all other specialized planar 
sandwich classes, such as \verb+PlanarSandwichHalf+ and \verb+PlanarSandwichHot+ of the 
next two sections. 
%

\subsection{BC2: The Hot and Warm Planar Sandwiches}

\begin{figure}[h!]
\includegraphics[scale=0.40]{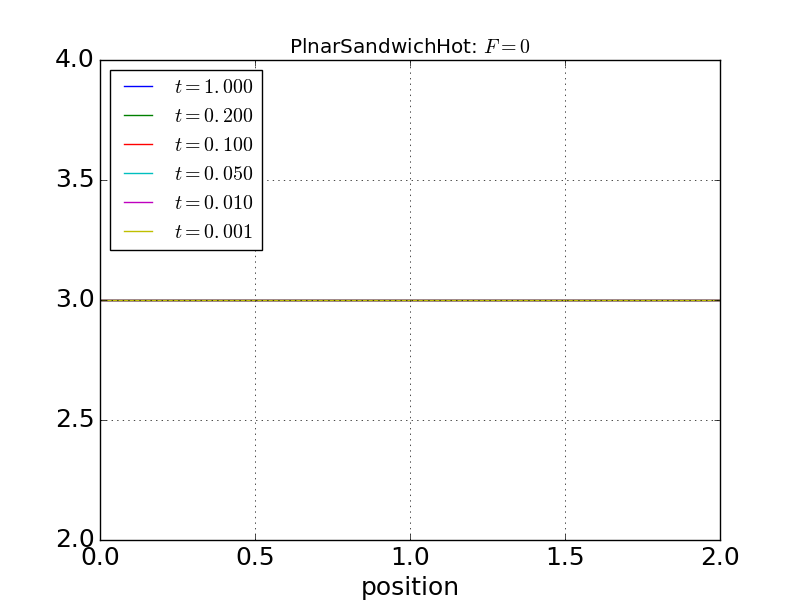} 
\caption{\footnoteskip  
The {\em hot planar sandwich} in ExactPack: 
\texttt{PlanarSandwichHot(F=0, TA=3, TB=3, L=2, Nsum=1000)}. Since 
the heat flux on the boundaries vanishes, heat cannot escape from the 
material, and the temperature must remain constant in time. The 
temperature profile has been plotted for the times $t=1, 0.2, 0.1,
0.01$, and $0.001$, and is indeed constant.
}
\label{fig_planar_sandwich_hot_ep}
\end{figure}

We now perform convergence analyses for several solutions of the boundary condition 
class BC2. These solutions are specified by the flux $F$ at the boundary points $y=0,L$,
and the linear IC specified by $T_\smA$ and $T_\smB$. As shown in 
Ref.~\cite{PlanarSandwichExactPackDoc}, the corresponding exact solutions are of the 
form
\begin{eqnarray}
 T(y,t)
 &=&
 F y + 
 A_0 + \sum_{n=1}^\infty 
 A_n \, \cos(k_n y)\, e^{-\kappa \, k_n^2 t}
 \label{hotA}
\\[5pt]
  k_n &=& \frac{n \pi}{L}
  \hskip0.1cm , \hskip0.1cm
  A_0 = \frac{T_\smA + T_\smB}{2}
  \hskip0.1cm , \hskip0.1cm
  A_n 
  =
  2 \Big(T_\smA - T_\smB \Big)\frac{1 - (-1)^n}{n^2\pi^2}
  \hskip0.35cm  
  n \ge 1 
 \label{hotB}
  \ .
\end{eqnarray}
The {\em hot planar sandwich} of Ref.~\cite{DMS} is a special case in which the BC is $F=0$ 
and the IC is $T(y,0)=T_0$, and on physical grounds, the exact solution is given by the trivial 
solution
\begin{eqnarray}
  T(y, t ) = T_0
  \ .
  \label{T0sol}
\end{eqnarray}
This solution is constant in time and uniform in space with value $T_0$. This also follow 
from (\ref{hotA}) and  (\ref{hotB}) by setting $F_\smA=\F_\smB=T_0$ and $F=0$.
The hot planar sandwich is therefore given by $T_\smA = T_\smB= T_0$.  
In the analysis that follows, we take $T_0=3$, and consequently, $A_0 = T_0$ 
and $A_n =0$ for $n \ge 1$, which reduces to the constant solution 
$T(y,t)=T_0$. This is illustrated in Fig.~\ref{fig_planar_sandwich_hot_ep}.
This new variant of the planar sandwich can be instantiated by

\vskip0.2cm
\noindent
\verb+solver = PlanarSandwichHot(F=0, TA=3, TB=3, L=2, Nsum=1000)+ \ .
\vskip0.2cm

\begin{figure}[h!]
\includegraphics[scale=0.40]{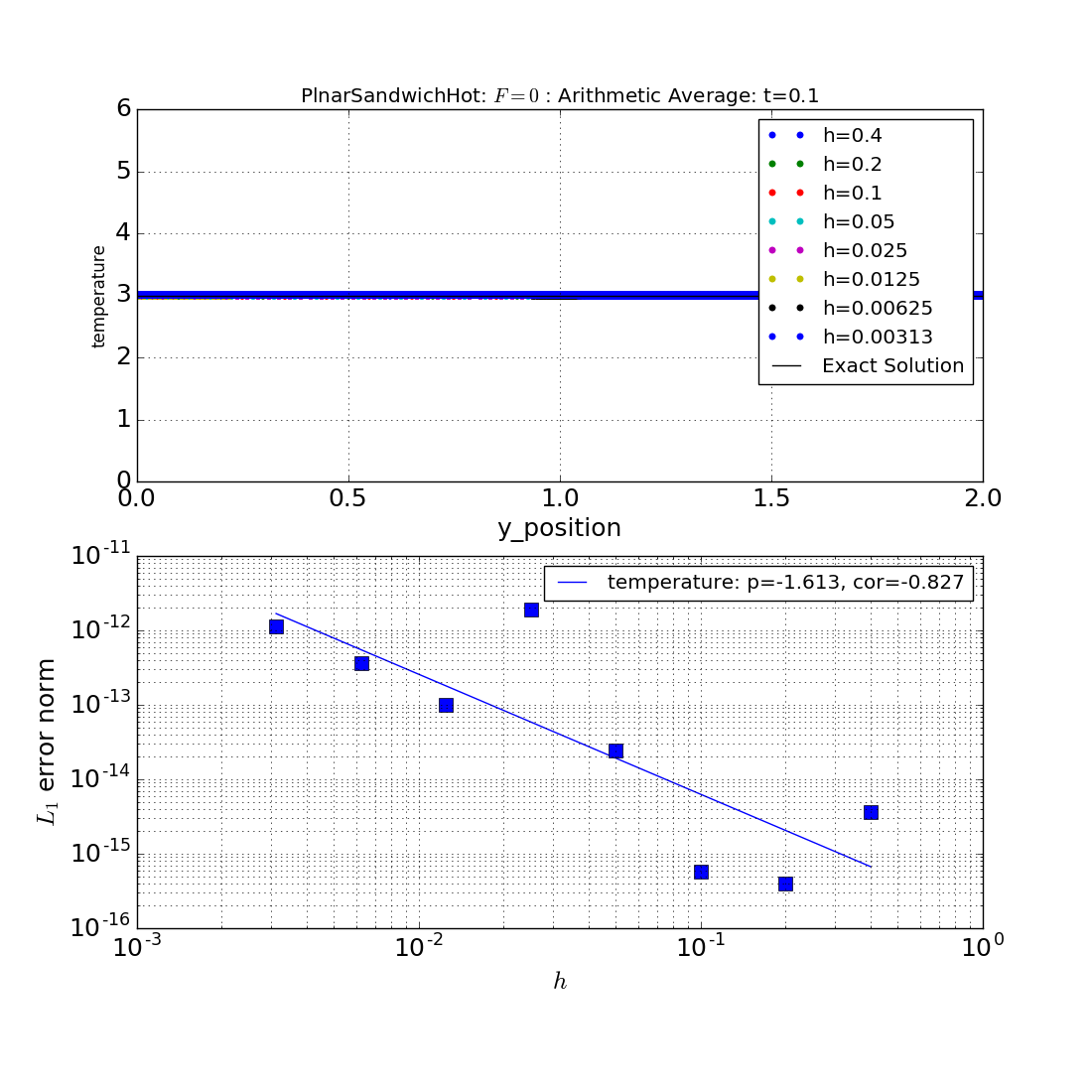} 
\vskip-1.0cm
\caption{\footnoteskip  
The hot planar sandwich. 
The parameter setting are the same as in Fig.~\ref{fig_planar_sandwich_c1_res_run9_soln},
except the arithmetic average is used for multimaterial cells.
}
\label{fig_planar_sandwich_hot_flat}
\end{figure}

\noindent
The heat flux $F$ on the boundaries has been set to zero, and a constant initial 
condition $T_0=3$, which has been specified  $T_0=T_\smA=T_\smB=3$ in
the ExactPack solution interface.  On physical grounds, heat cannot escape from 
the material,  and the temperature must remain constant,  $T(y,t) = T_0$, as 
illustrated by the exact solution plotted in Fig.~\ref{fig_planar_sandwich_hot_ep}.  
The numerical results are give in Fig.~\ref{fig_planar_sandwich_hot_flat} at time 
$t=0.1$, which indeed shows that the temperature remains constant. The lower 
panel of this Figure gives the convergence analysis. Note that the error is of order 
machine precision, and the points are scattered somewhat randomly, with a systematic
linear increase at the higher precisions.


\begin{figure}[h!]
\includegraphics[scale=0.37]{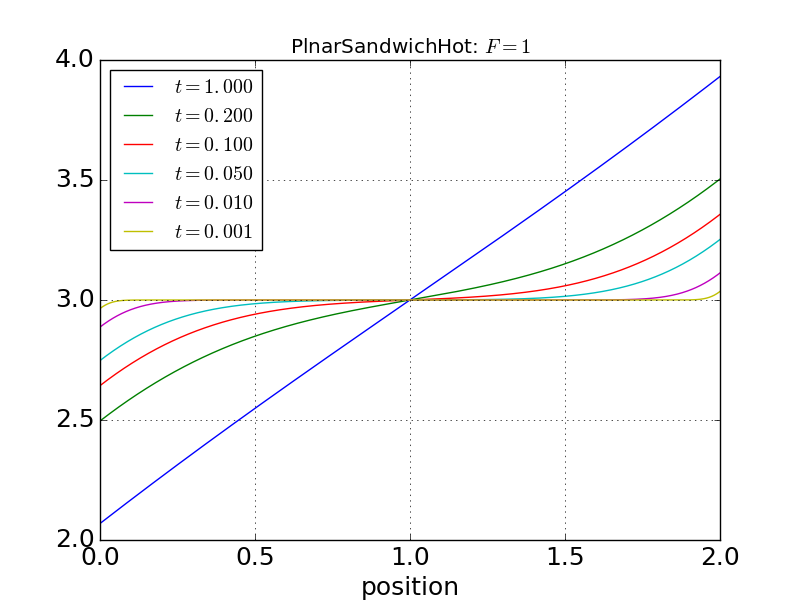}
\caption{\footnoteskip  
The {\em warm planar sandwich} in ExactPack:
\texttt{PlanarSandwichHot(F=1, TA=3, TB=3, L=2, Nsum=1000)}. The profiles 
are plotted for times $t=1, 0.2, 0.1, 0.01$, and $0.001$. The heat flux at the 
boundaries is $F=1$, and we see that the temperature profile changes as heat 
flows out of the rod. In contrast to Fig.~\ref{fig_planar_sandwich_hot_flat}, when 
the heat flux is nonzero, heat is free to flow from the sandwich to the environment, 
and the temperature need not remain constant. 
}
\label{fig_planar_sandwich_hot_ep_FOne}
\end{figure}

\begin{figure}[h!]
\includegraphics[scale=0.37]{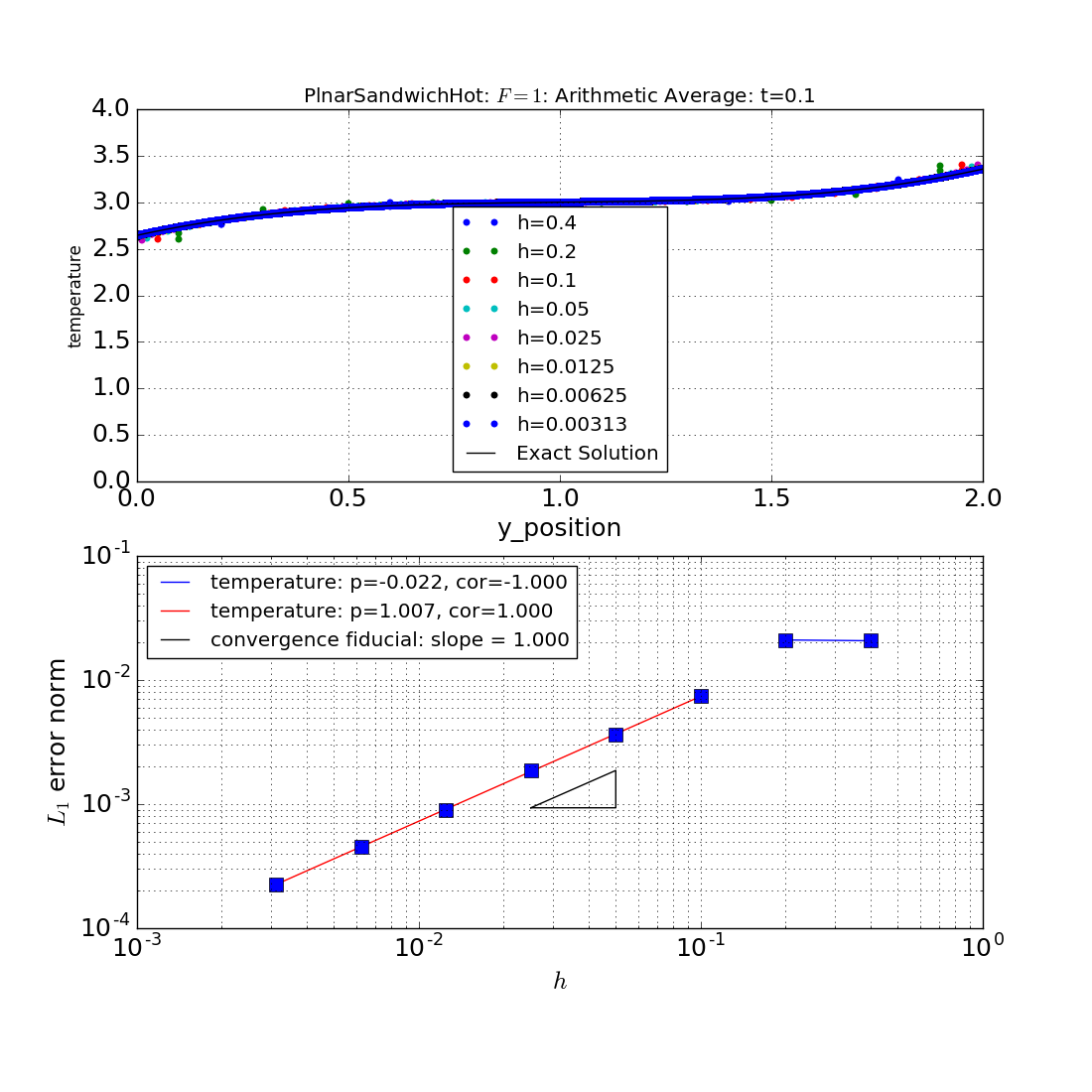}
\vskip-1.0cm
\caption{\footnoteskip  
The parameter setting are the same as in Fig.~\ref{fig_planar_sandwich_c1_res_run9_soln},
except the arithmetic average is used for multimaterial cells.
}
\label{planar_sandwich_hotF1_c1_run16_temperature_soln_conv}
\end{figure}

\begin{figure}[h!]
\includegraphics[scale=0.37]{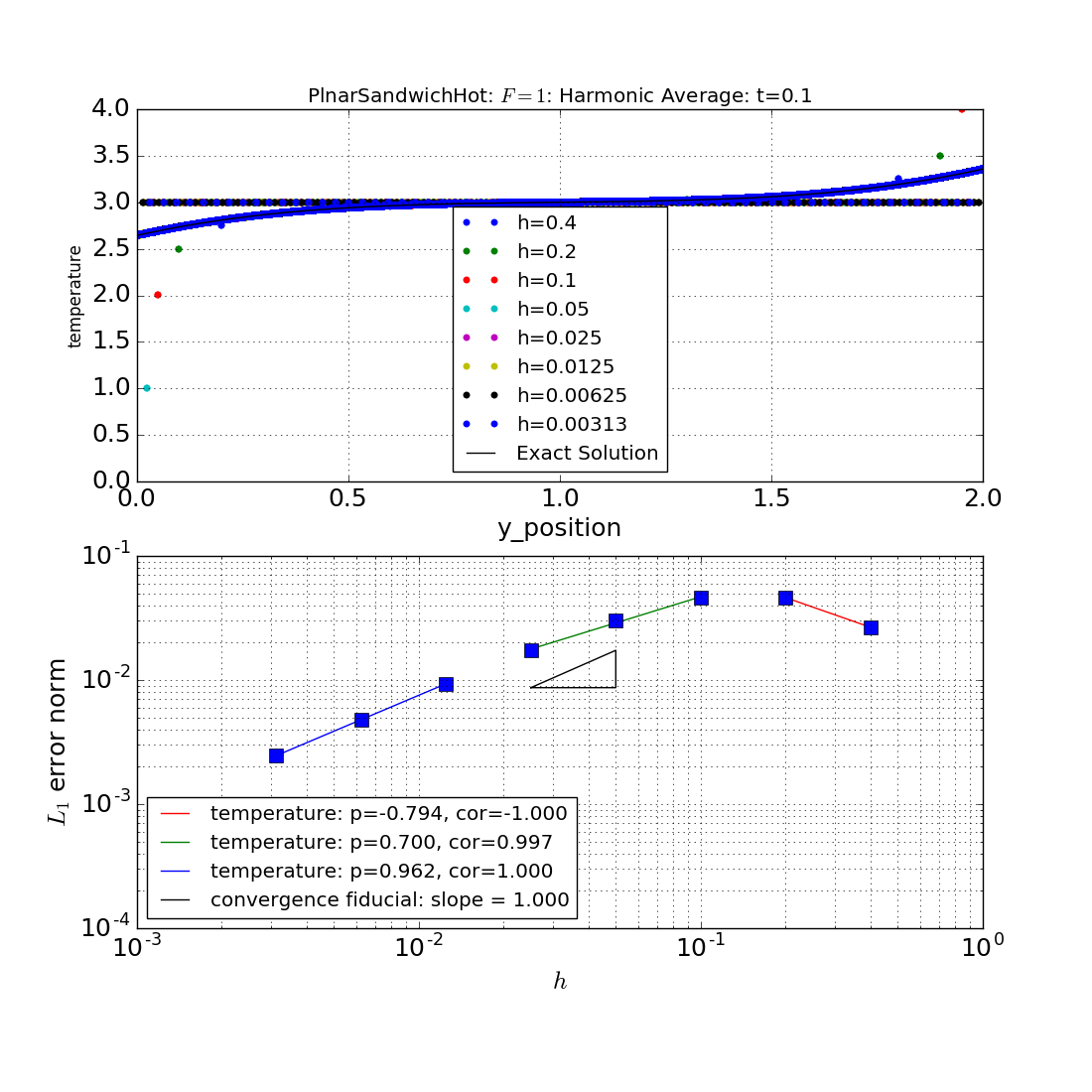}
\vskip-1.0cm
\caption{\footnoteskip  
The parameter setting are the same as in Fig.~\ref{fig_planar_sandwich_c1_res_run9_soln},
except the harmonic average is used for multimaterial cells.
}
\label{planar_sandwich_hotF1_c2_run16_temperature_soln_conv}
\end{figure}

\begin{figure}[h!]
\includegraphics[scale=0.37]{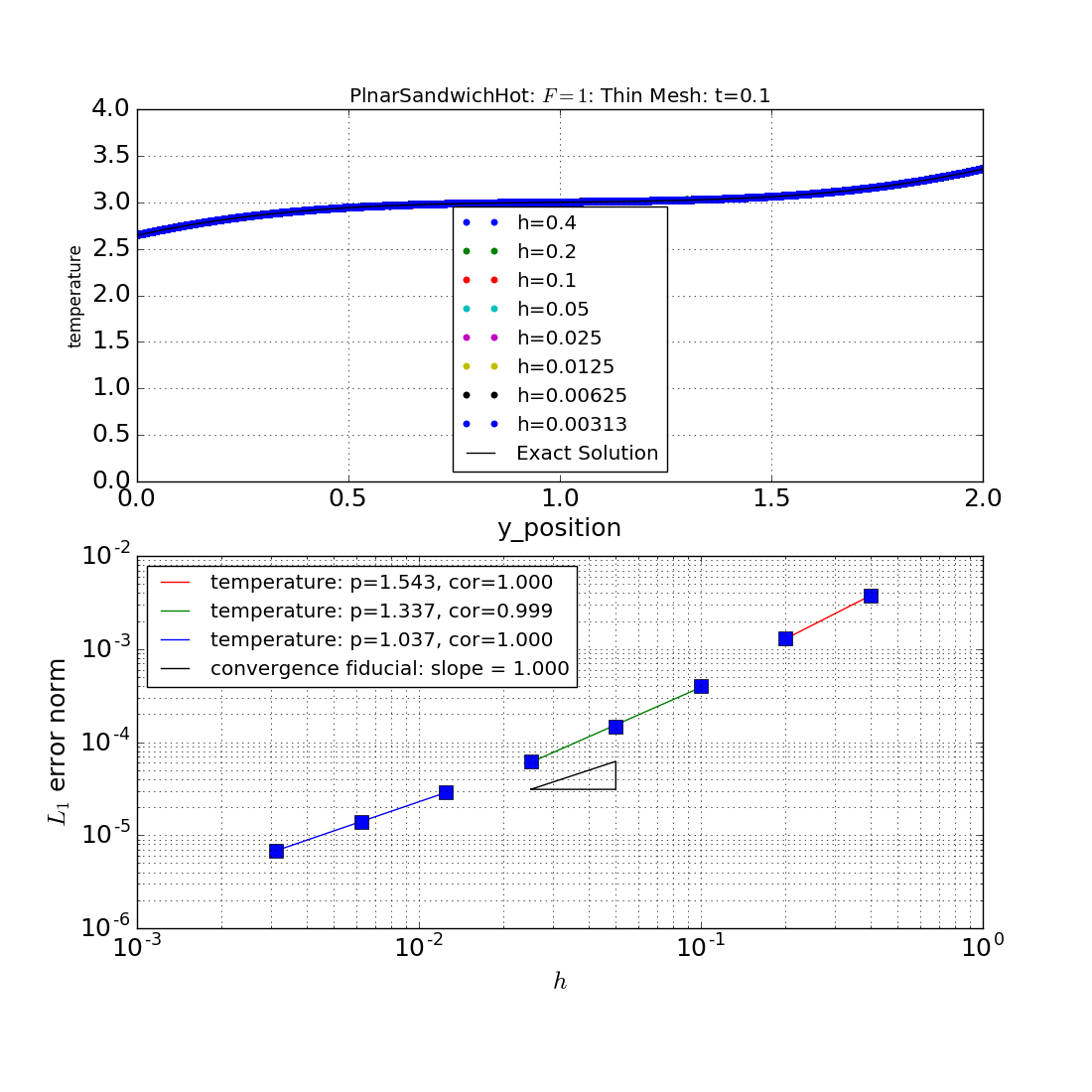}
\vskip-1.0cm
\caption{\footnoteskip  
The parameter setting are the same as in Fig.~\ref{fig_planar_sandwich_c1_res_run9_soln},
except the thin mesh option is used for multimaterial cells.
}
\label{planar_sandwich_hotF1_tm_run16_temperature_soln_conv}
\end{figure}

\begin{figure}[h!]
\includegraphics[scale=0.37]{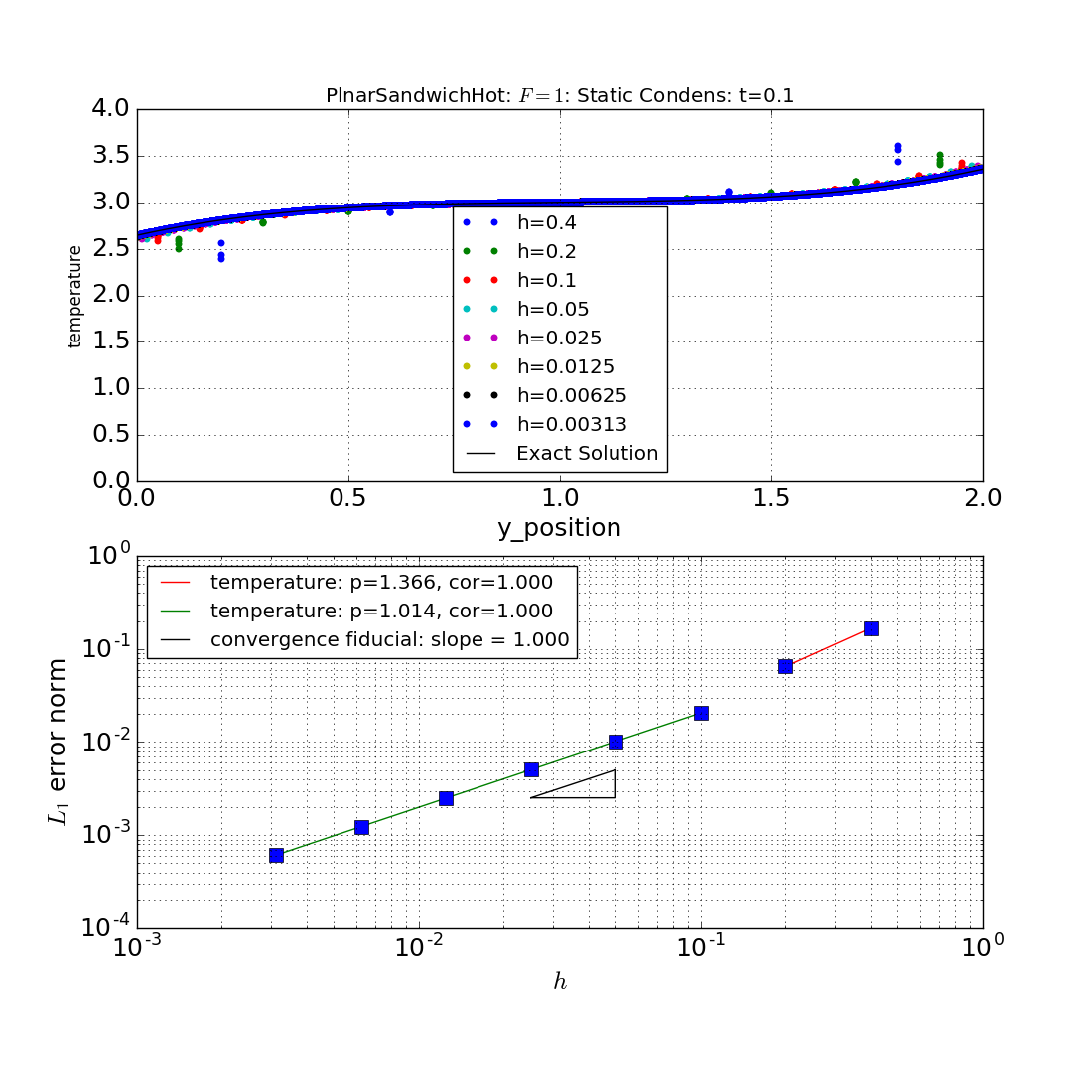}
\vskip-1.0cm
\caption{\footnoteskip  
The parameter setting are the same as in Fig.~\ref{fig_planar_sandwich_c1_res_run9_soln},
except the static condensation option is used for multimaterial cells.
}
\label{planar_sandwich_hotF1_sc_run16_temperature_soln_conv}
\end{figure}

\pagebreak
\clearpage
The next test problem will be called the {\em warm planar sandwich}. This problem 
allows heat to escape along the boundaries with flux $F=1$.
The temperature profiles are illustrated in  
Fig.~\ref{fig_planar_sandwich_hot_ep_FOne}, and the numerical results in
Figs.~\ref{planar_sandwich_hotF1_c1_run16_temperature_soln_conv} -- \ref{planar_sandwich_hotF1_sc_run16_temperature_soln_conv}.

%

%

\subsection{The Half Planar Sandwich}

The final variant we shall consider is given by choosing a vanishing heat flux on 
the upper boundary, $F_2=\partial_y T(L)=0$, and zero temperature on the lower 
boundary, $T_1 = T(0)=0$. This is an example of boundary condition BC3, and 
we call the solution the {\em half planar sandwich}. For initial condition (\ref{IClinearA}), 
Ref.~\cite{PlanarSandwichExactPackDoc} shows that the solution takes the form

\begin{eqnarray}
 T(y,t)
 &=&
 \sum_{n=0}^\infty 
 B_n \, \sin(k_n y)\, e^{-\kappa \, k_n^2 t}
 \\[5pt]
  k_n &=& 
  \frac{(2 n + 1) \pi}{2L}
  \hskip0.5cm {\rm with}\hskip0.5cm
  B_n 
  =
  \frac{4 T_\smB}{(2 n + 1) \pi}
  -
  \frac{8\big(T_\smB - T_\smA\big) }{(2n+1)^2 \pi^2} 
  \ .
\end{eqnarray}
\begin{figure}[h!]
\includegraphics[scale=0.40]{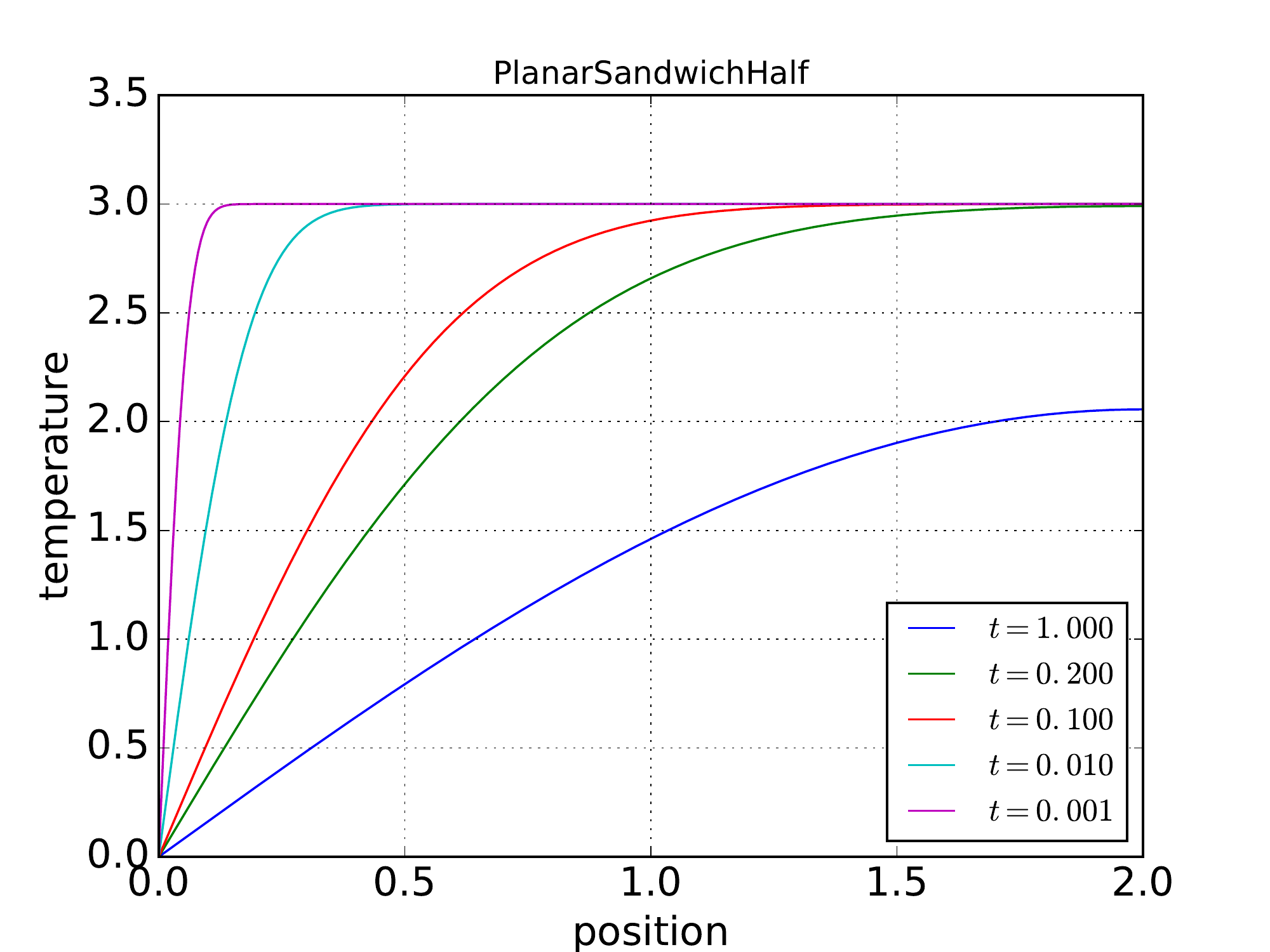} 
\caption{\footnoteskip  
The {\em half planar sandwich} in ExactPack: 
PlanarSandwichHalf(T1=0, F2=0, TA=3, TB=3, L=2, Nsum=1000). 
The profiles are plotted for times $t=1, 0.2, 0.1, 0.01$, and $0.001$. 
Note that the temperature profile vanishes on the bottom and the derivative
of the temperature vanishes on the top.
}
\label{fig_planar_sandwich_half_ep}
\end{figure}

\noindent
Taking the initial condition $T_0=3$ ($T_\smA=T_\smB=3)$ gives 
Fig.~\ref{fig_planar_sandwich_half_ep}, which is instantiated by
\vskip0.2cm
\noindent
\verb+solver = PlanarSandwichHalf(T1=0, F2=0, TA=3, TB=3, L=2, Nsum=1000)+ \ .
\vskip0.2cm

\begin{figure}[h!]
\includegraphics[scale=0.45]{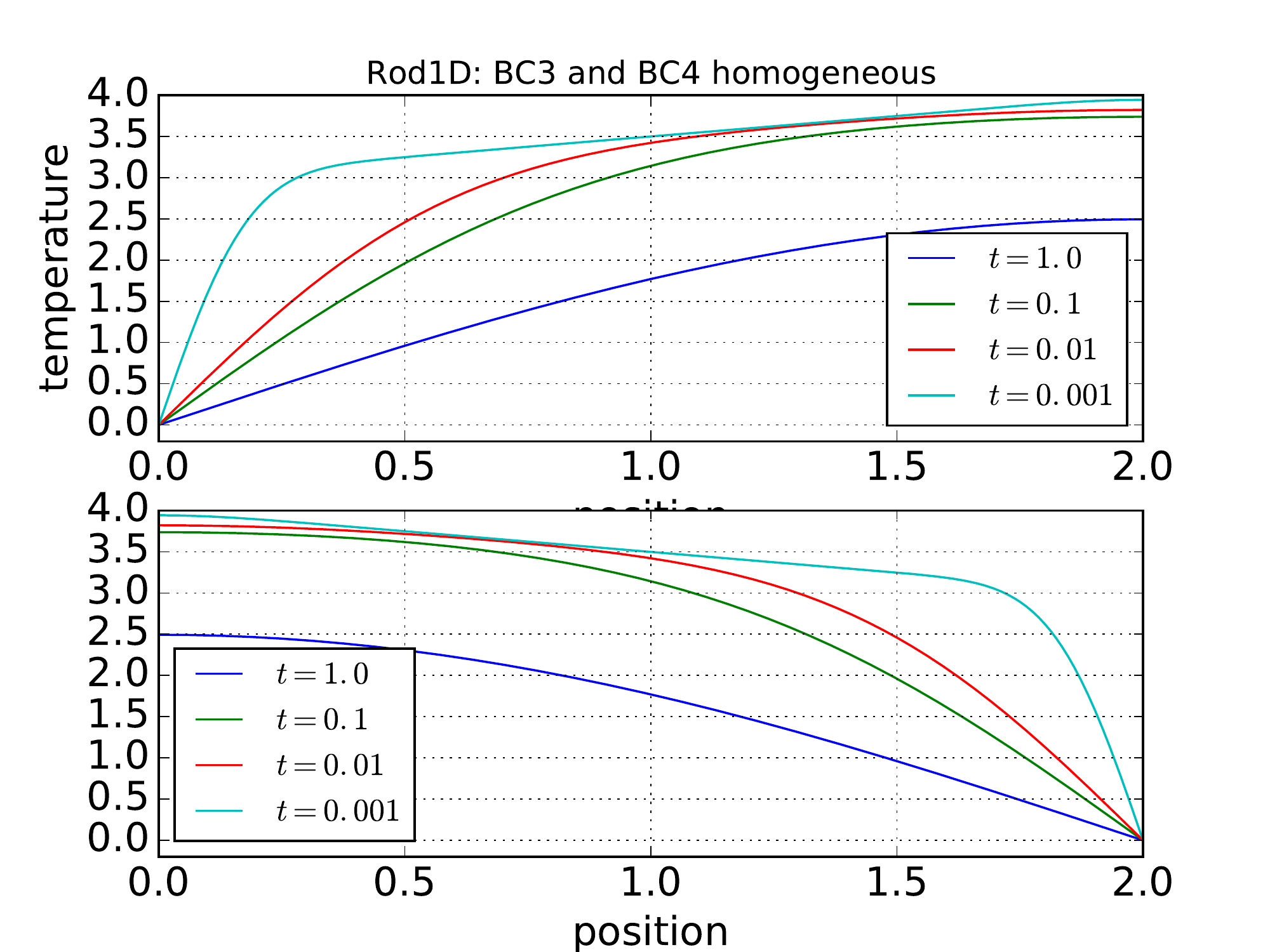} 
\caption{\footnoteskip  
Note that BC3  and BC4 are physically equivalent, and are related by a parity inversion 
across the midpoint. The Figure uses $\kappa=1$, $L=2$, $T_\smA=T_\smB=3$.  
By symmetry principles, 
the two profiles are mirror images of one another. BC3 can be instantiated by 
\texttt{Rod1D(alpha1=1, beta1=0, gamma1=1, alpha2=0, beta2=1, gamma2=0, 
TL=0, TR=0)}, and BC4 by \texttt{Rod1D(alpha1=0, beta1=1, gamma1=0, 
alpha2=0, beta2=1, gamma2=F1, TL=0, TR=0)}.  Note that $T_\smA$ and 
$T_\smB$ are interchanged between BC3 and BC4. The class \texttt{Rod1D} is the parent 
class of \texttt{PlanarSandwich}.
}
\label{fig_BC34_flip} 
\end{figure}

\begin{figure}[t!]
\includegraphics[scale=0.37]{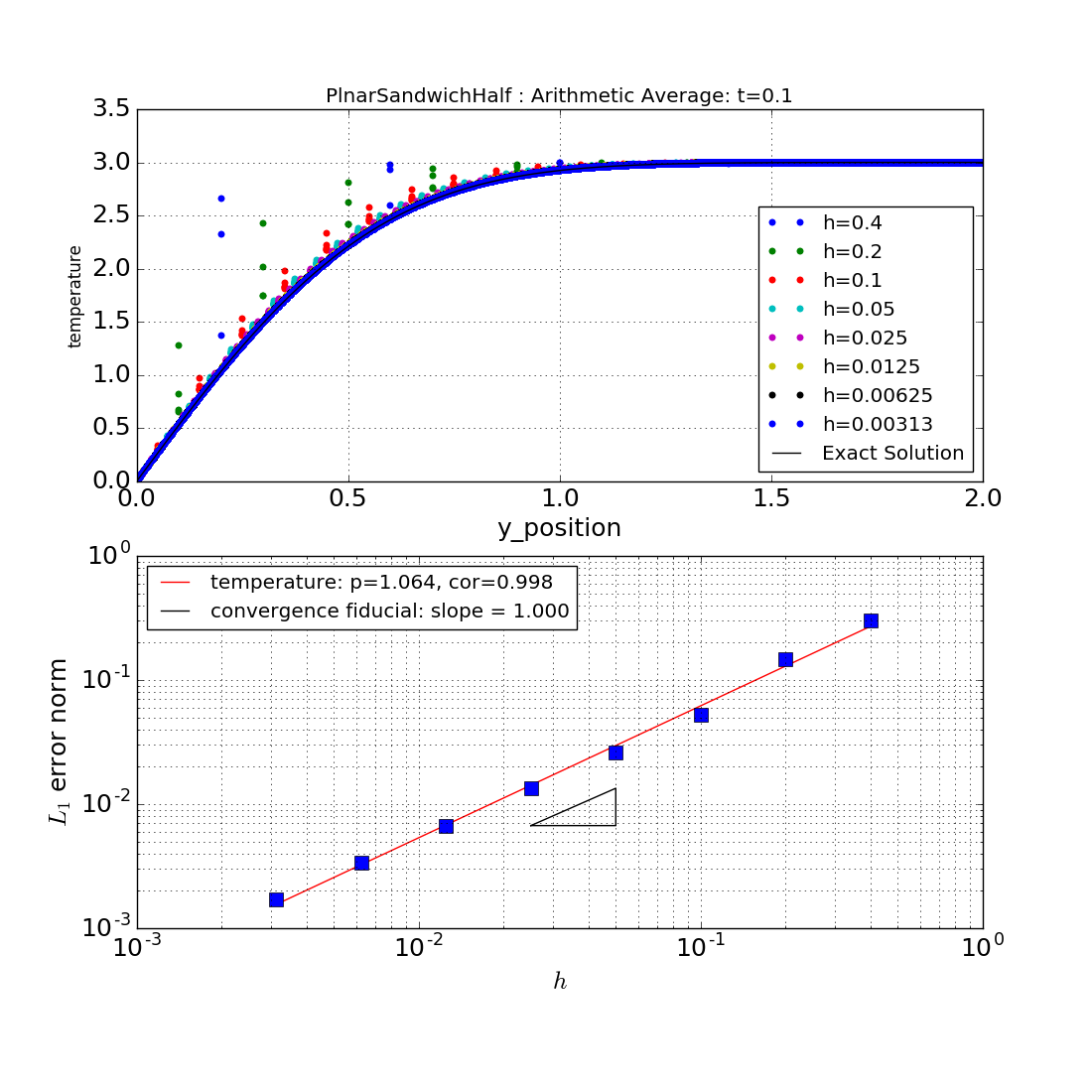}
\vskip-1.0cm
\caption{\footnoteskip  
The parameter setting are the same as in Fig.~\ref{fig_planar_sandwich_c1_res_run9_soln}.
}
\label{fig_planar_sandwich_c1_run11}
\end{figure}

\begin{figure}[h!]
\includegraphics[scale=0.37]{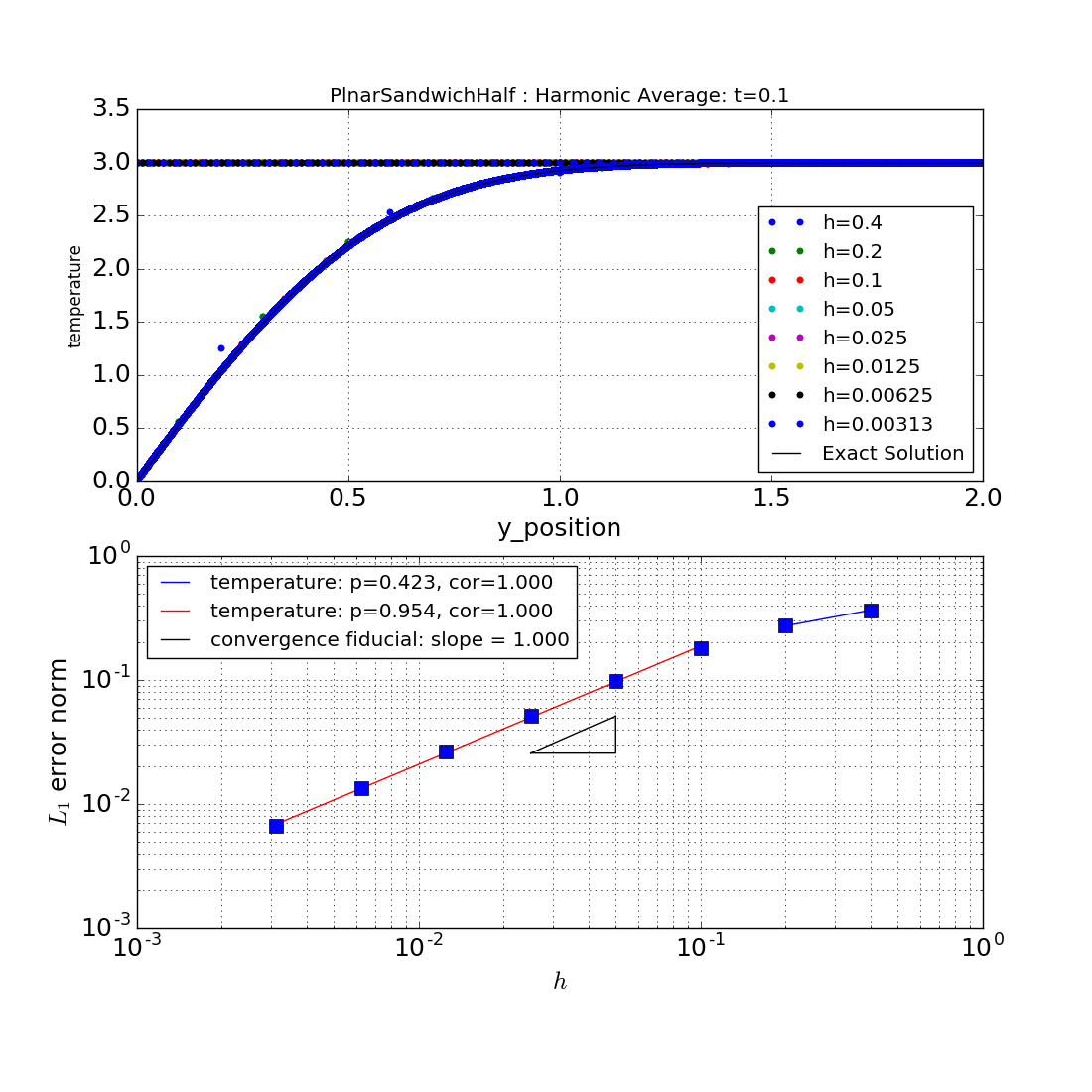}
\vskip-1.0cm
\caption{\footnoteskip  
The parameter setting are the same as in Fig.~\ref{fig_planar_sandwich_c1_res_run9_soln},
except the harmonic average is used for multimaterial cells.
}
\label{fig_planar_sandwich_c2_run11}
\end{figure}

\begin{figure}[h!]
\includegraphics[scale=0.37]{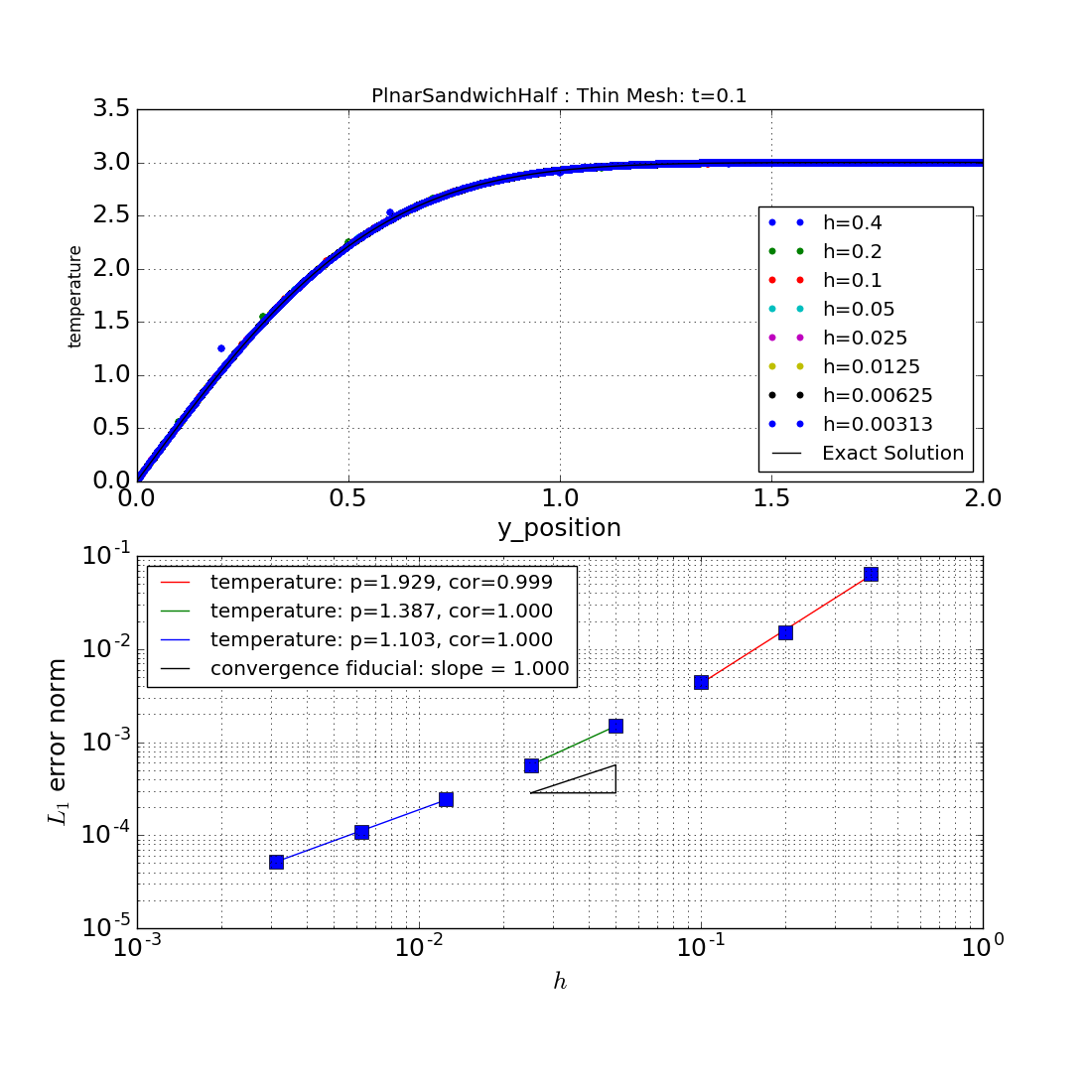}
\vskip-1.0cm
\caption{\footnoteskip  
The parameter setting are the same as in Fig.~\ref{fig_planar_sandwich_c1_res_run9_soln},
except the thin mesh option is used for multimaterial cells.
}
\label{fig_planar_sandwich_tm_run11}
\end{figure}

\begin{figure}[h!]
\includegraphics[scale=0.37]{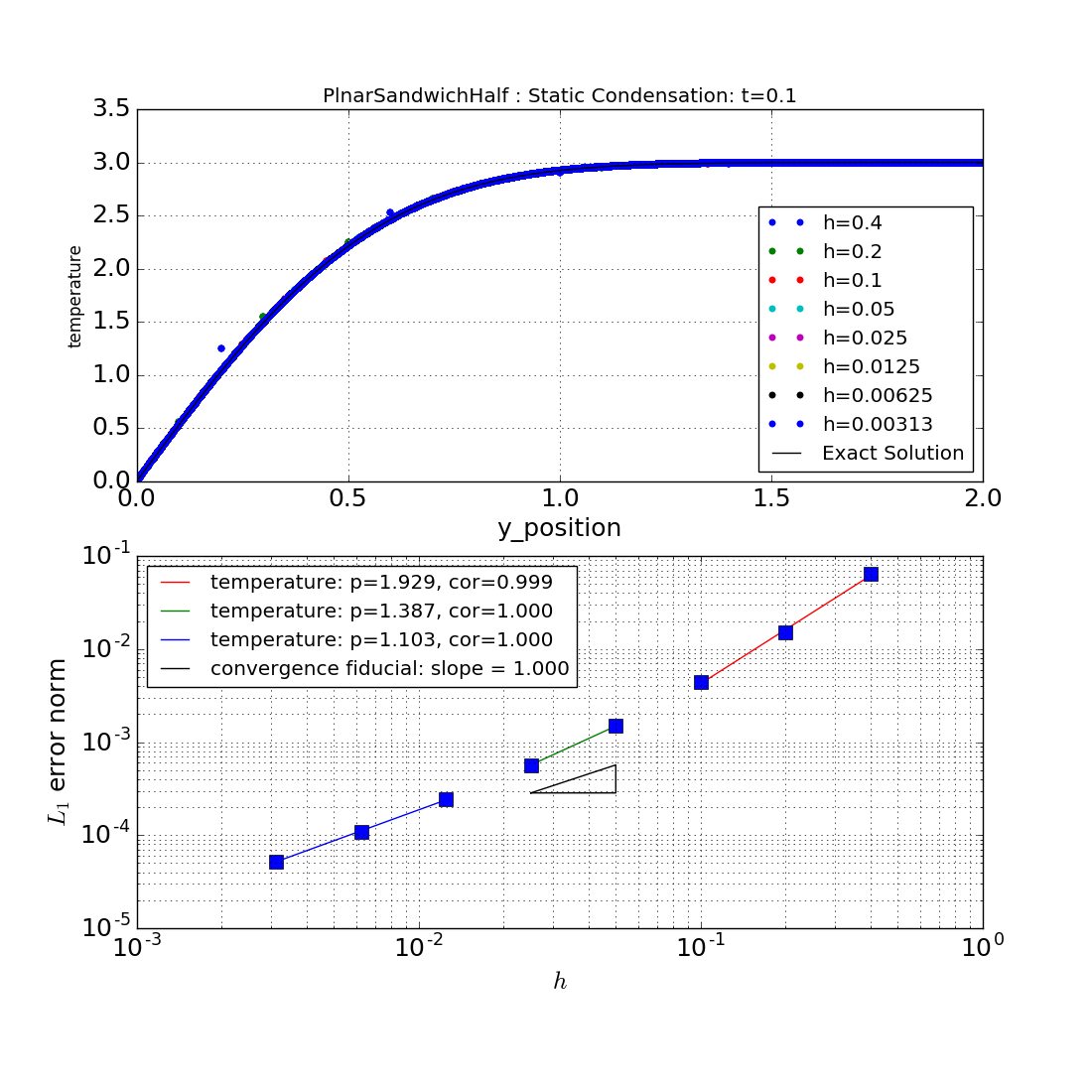}
\vskip-1.0cm
\caption{\footnoteskip  
The parameter setting are the same as in Fig.~\ref{fig_planar_sandwich_c1_res_run9_soln},
except the static condensation option is used for multimaterial cells.
}
\label{fig_planar_sandwich_sc_run11}
\end{figure}

\noindent
If we had chosen $\partial_y T(0)=0$ and $T(L)=0$, as in BC4, then the plot would have 
been reflected about the midpoint of the rod, but is otherwise physically identical, as 
illustrated in Fig.~\ref{fig_BC34_flip}. Figures~\ref{fig_planar_sandwich_c1_run11} and
\ref{fig_planar_sandwich_c2_run11} provide the solution plots and the convergence analyses 
for the arithmetic and harmonic averages of the half planar sandwich. As in the previous 
section, the top panel for the harmonic average in Fig.~\ref{fig_planar_sandwich_c2_run11} 
exhibits near-zeros along the horizontal axis. Similarly, Figs.~\ref{fig_planar_sandwich_tm_run11} 
and \ref{fig_planar_sandwich_sc_run11} illustrate the convergence analyses for the thin 
mesh and static condensation options for the half planar sandwich. The convergence
plots for the half planar sandwich are qualitatively similar to those for the planar sandwich.
In both cases, the arithmetic and harmonic averages are approximately $1^{\rm st}$ order,
while thin mesh and static condensation start out at $2^{\rm nd}$ order and approach
$1^{\rm st}$ order as the mesh is refined.


%

%

\pagebreak

\section{Conclusions and Future Research}

Two dimensional (2D) multimaterial heat diffusion can be a challenging
numerical problem when the material boundaries are misaligned with the
numerical grid. Even when the boundaries start out aligned, they typically 
become misaligned through hydrodynamic motion; therefore, it is
important to perform rigorous verification analyses of the heat transport 
algorithms in any heat-conduction hydrocode.
In this paper we perform  convergence analyses for the four multimaterial heat  flow 
algorithms in the multi-physics hydrodynamics code FLAG: (i) the arithmetic 
average, (ii) the harmonic average, (iii) thin mesh, and (iv) static condensation.  To 
perform the analyses and to produce the corresponding convergence plots, we employ 
the code verification tool ExactPack. We  concentrate on the 2D {\em planar sandwich} test 
problem, along with three generalizations called the {\em half planar sandwich}, the
the {\em hot planar sandwich}, and the {\em warm planar sandwich},
all of which possess simple exact solutions. These test 
problems were designed to exhibit multimaterial cells along a fixed boundary, thereby 
exercising the multimaterial algorithms by the simplest means possible. 

The geometry of the planar sandwich is illustrated in Fig.~\ref{fig_sub_grid}, 
which shows a square numerical grid overlaid on a rectangular physical geometry 
consisting of three parallel sandwich-like regions.  The numerical grid partitions 
the physical geometry into a number of corresponding numerical cells, which need 
not align with the material regions. The outer two regions, called the {\em bread} 
of the sandwich, are composed of an insulating material with zero heat diffusion 
constant $\kappa_1=0$, while the inner region, called the {\em meat} of the 
sandwich, has diffusion coefficient $\kappa_2 = \kappa >0$. 

One of our primary results is that the arithmetic and harmonic averages converge at 
$1^{\rm st}$ order for all three variants of the planar sandwich. We also find that 
both the thin mesh and static condensation algorithms start out converging at $2^{\rm nd}$ 
order, but as the mesh is refined, the convergence rate levels off to $1^{\rm st}$ order.
We conjecture that this is because the error in interface reconstruction algorithms becomes 
less precise at finer resolutions. 

With more work, one can also construct an exact solution for the case in which 
$\kappa_1 \ne 0$ in the bread of the sandwich\,\cite{RLSnotes}, and these solutions 
might be an interesting avenue for future verification work. We are currently adding
these solutions to ExactPack.  We are also exploring Voronoi mesh simulations for the 
planar sandwich. Unlike the square mesh, Voronoi verification must be performed in 
2D. This is because the Voronoi mesh does not align uniformly along $y=$constant, 
and one cannot use the 1D profiles for the exact solutions. Voronoi mesh verification is 
further complicated by the fact that the cells are not necessarily of equal area. 
Ref.~\cite{Kamm99} explores the various choices of norm in such cases. We are adding 
a VTK reader to ExactPack, and this will greatly facilitate verification work with nonuniform 
meshes. 

\begin{figure}
\begin{minipage}[c]{0.4\linewidth}
\includegraphics[scale=0.40]{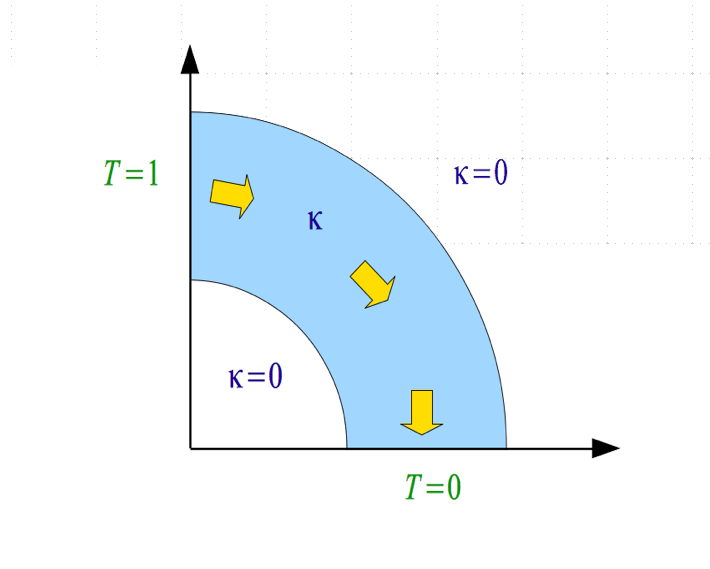}
\end{minipage}
\hfill
\begin{minipage}[c]{0.4\linewidth}
\includegraphics[scale=0.15]{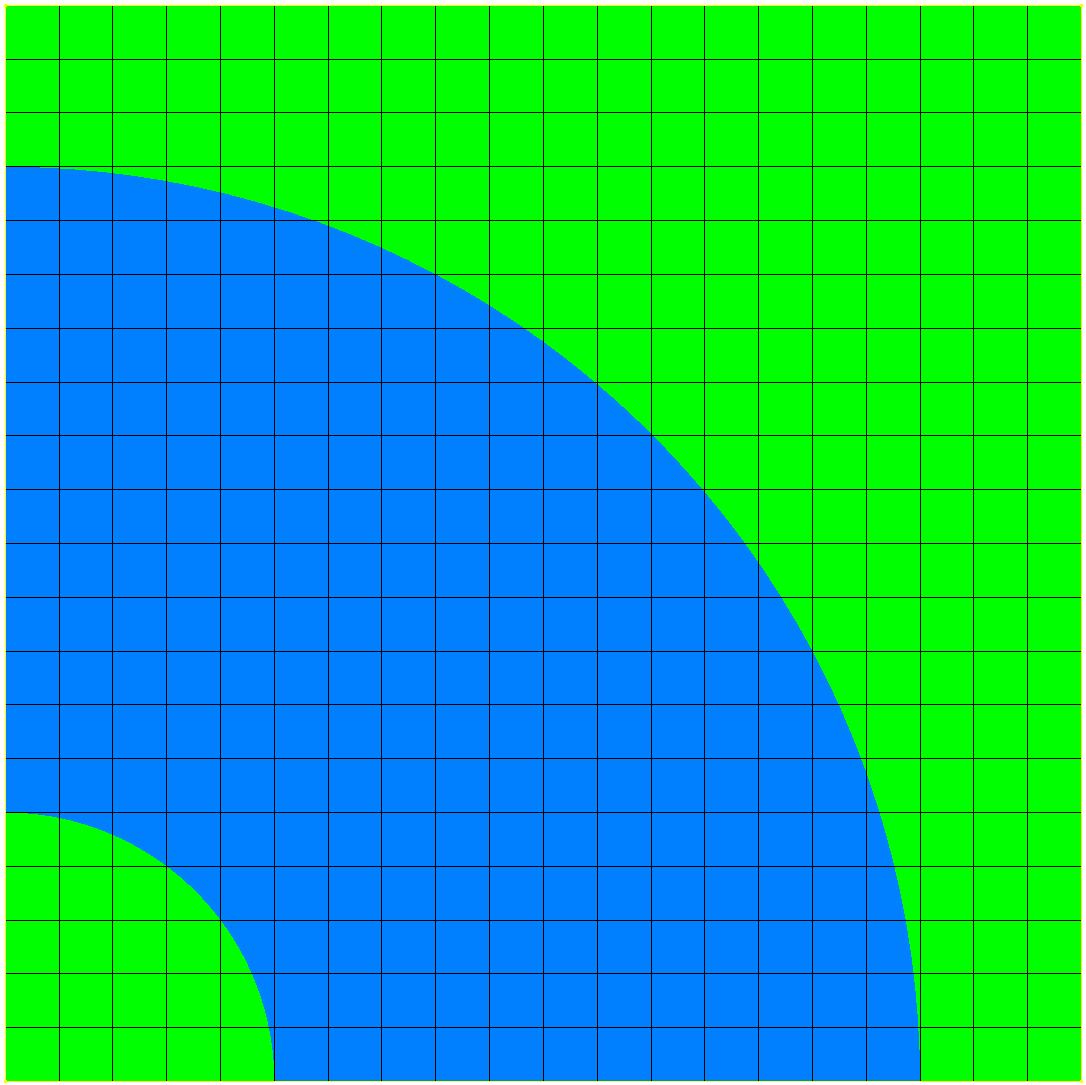}
\end{minipage}
\vskip-0.5cm
\caption{\footnoteskip
The cylindrical sandwich test problem. On the square grid, the 
multimaterial cells never align with the material boundary.
}
\label{fig_cylindrical_sandwich_def}
\end{figure}

We also plan to study problems with more complicated geometry, such as the {\em
cylindrical sandwich}. This test problem was proposed by Alan Dawes in Ref~\cite{Dawes},
and has been analyzed by Dawes,  Shashkov, and Malone\,\cite{DMS}. These authors
did not perform convergence analyses, and they used a highly resolved \lq\lq{}reference
solution\rq\rq{} rather than an exact solution, although the exact solution was presented
in Appendix B of Ref.~\cite{DMS}. 
This test problem is much like the planar sandwich, except that the heat-conducting
material is annular and lies in the first quadrant. The left most region is set to
$T=1$ and the bottom region to $T=0$, and therefore heat moves clockwise along
the annulus. See the left panel of Fig.~\ref{fig_cylindrical_sandwich_def}. 
As illustrated by the right panel of Fig.~\ref{fig_cylindrical_sandwich_def}, the square
mesh will never align with the material boundary. Generalizations of the cylindrical
sandwich exist much like those of Ref.~\cite{PlanarSandwichExactPackDoc}, and we 
plan to investigate these solutions as well.

%


\vfill
\pagebreak
\clearpage
\appendix

\section{Python script for the Planar Sandwich}
\label{sec_python_script}

\footnoteskip
\begin{verbatim}
import numpy as np
import matplotlib.pylab as plt

from exactpack.solvers.heat import PlanarSandwich

L = 2.0
x = np.linspace(0.0, L, 1000)
t1 = 1.0
t2 = 0.2
t3 = 0.1
t4 = 0.05
t5 = 0.01
t6 = 0.001

solver = PlanarSandwich(TB=1, TT=0, L=L, Nsum=1000)
soln1 = solver(x, t1)
soln2 = solver(x, t2)
soln3 = solver(x, t3)
soln4 = solver(x, t4)
soln5 = solver(x, t5)
soln6 = solver(x, t6)
soln1.plot('temperature', label=r'$t=1.000$')
soln2.plot('temperature', label=r'$t=0.200$')
soln3.plot('temperature', label=r'$t=0.100$')
soln4.plot('temperature', label=r'$t=0.050$')
soln5.plot('temperature', label=r'$t=0.010$')
soln6.plot('temperature', label=r'$t=0.001$')

plt.title('PlanarSandwich')
plt.ylabel(r'temperature', fontsize='18')
plt.ylim(0,1)
plt.xlim(0,L)
plt.legend(loc=0)
plt.grid(True)
plt.savefig('planar_sandwich.png')
plt.show()

\end{verbatim}
\bodyskip

\pagebreak
\section{ExactPack  Script for Convergence Analysis}

\footnoteskip
\begin{verbatim}

import numpy as np
import os.path
import glob
import matplotlib.pylab as plt

from exactpack.solvers.heat import PlanarSandwich
from exactpack.analysis import *
from lanl_readers.flag import FlagVarDump

#####################################################################
# problem parameters
L = 2.0
t = 0.1

# solver name
solver_name = 'PlanarSandwich'

# multimaterial alorithm
multimat = 'c1' # arithmetic average
multimat_name = 'Arithmetic Average'

# study resolutions
res_study = [5, 10, 20, 40, 80, 160, 320, 640] # 8

# output tag
problem_out = 'planar_sandwich_{}_run16'.format(multimat)

# variable selection
variables = 'temperature'

# plot parameters
plot_params = {'temperature': {'ymin': 0, 'ymax': 1, 'xmin':0, 'xmax': L, 
  'error_min': 1.e-4, 'error_max': 1.e-0, 'num': [8], 'loc': [1, 2, 1, 2]} 
  }

# run dir and dump files
dumpfiles =/Users/bobs/mygit_repos/heat/runs/planar_sandwich/{}/res*/vardump/
  planar_sandwich_{}_VarDump.00000.1.000000E-01.zx'.format(multimat, multimat)

# create study_parameters
#####################################################################
study_parameters = [L/float(res) for res in res_study]

# creat solver
#####################################################################
solver = PlanarSandwich(TB=1, TT=0, L=L, Nsum=1000)

# study object
#####################################################################
print "*** plot solution and code data ..."

study = Study(sorted(glob.glob(dumpfiles)),
    reference=solver,
    study_parameters=study_parameters,
    time=t,
    reader=FlagVarDump(),
    abscissa='y_position'
    )

#####################################################################
print "*** plot solution and code data ..."

xmin = plot_params[variable]['xmin']
xmax = plot_params[variable]['xmax']
ymin = plot_params[variable]['ymin']
ymax = plot_params[variable]['ymax']
loc = plot_params[variable]['loc'][0]

# plot solution profiles
plt.clf()
study.plot('temperature')
plt.xlim(xmin, xmax)
plt.ylim(ymin, ymax)
plt.ylabel('$T$')
plt.xlabel('$y$')
plt.title(r'{} : {}: t={}'.format(solver_name, multimat_name, t))
plt.legend(loc=loc)
plt.grid(True)
plt.savefig(problem_out+'_'+'temperature_soln.png')
plt.show()

#####################################################################
print "*** convergence analysis ..."

error_max = plot_params[variable]['error_max']
error_min = plot_params[variable]['error_min']
loc = plot_params[variable]['loc'][1]
n0 = plot_params[variable]['num'][0]

domain = (0, L)
fiducials = {'temperature': 1}
fit = RegressionConvergenceRate(study, norm=PointNorm(), domain=domain, fiducials=fiducials)
fit[:n0].plot_fit('temperature', "-", c='r')
fit.norms.plot('temperature', label=None, markersize=10)
fit.plot_fiducial('temperature')
plt.ylim(error_min, error_max)
plt.title(r'{} : {}: t={}'.format(solver_name, multimat_name, t))
plt.xlabel(r'$h$')
plt.ylabel(r'$L_1$ error norm')
plt.legend(loc=loc)
plt.savefig(problem_out+'_'+'temperature_conv.png')
plt.show()

# plot on same axis
#####################################################################

xmin = plot_params[variable]['xmin']
xmax = plot_params[variable]['xmax']
ymin = plot_params[variable]['ymin']
ymax = plot_params[variable]['ymax']
error_max = plot_params[variable]['error_max']
error_min = plot_params[variable]['error_min']
n0 = plot_params[variable]['num'][0]
loc1 = plot_params[variable]['loc'][2]
loc2 = plot_params[variable]['loc'][3]

plt.figure(figsize=(11, 11))
ax = plt.subplot(211)
study.plot('temperature')
plt.xlim(xmin, xmax)
plt.ylim(ymin, ymax)
plt.ylabel('$T$')
plt.xlabel('$y$')
plt.title(r'{} : {}: t={}'.format(solver_name, multimat_name, t))
plt.legend(loc=loc1)
plt.grid(True)

ax = plt.subplot(212)
domain = (0, L)
fiducials = {'temperature': 1}
fit = RegressionConvergenceRate(study, norm=PointNorm(), 
  domain=domain, fiducials=fiducials)
fit[:n0].plot_fit('temperature', "-", c='r')
fit.norms.plot('temperature', label=None, markersize=10)
fit.plot_fiducial('temperature')
plt.ylim(error_min, error_max)
plt.xlabel(r'$h$')
plt.ylabel(r'$L_1$ error norm')
plt.legend(loc=loc2)
plt.savefig(problem_out+'_'+'temperature_soln_conv.png')
plt.show()
\end{verbatim}
\bodyskip

\pagebreak
\clearpage
\section{Solution to the Planar Sandwich}
\label{sec_solving_heat_eq}

In this section we find the general solution to the heat equation  
(\ref{eq_oneDrodAnh})--(\ref{eq_oneDrodCnh}). The differential equation (DE), 
the boundary conditions (BCs), and the initial condition (IC) are  reproduced here 
for convenience,
\begin{eqnarray}
  {\rm DE}: 
  \hskip3.73cm
  \frac{\partial T(y,t)}{\partial t}
  &=&
  \kappa\, \frac{\partial^2 T(y,t)}{\partial y^2}
  \hskip1.2cm 
  0 < y < L ~{\rm and}~ t > 0
\label{eq_oneDrodAnh_ap}
\\[5pt]
  {\rm BCs}:  
  \hskip1.16cm
  \alpha_1 T(0,t) + \beta_1 \partial_y T(0,t) &=& \gamma_1
  \hskip2.8cm t > 0 ~,~ y=0,L
\label{eq_oneDrodBnhA_ap}    
\\[-3pt]
  \alpha_2 T(L,t) + \beta_2 \partial_y T(L,t) &=& \gamma_2
\label{eq_oneDrodBnhB_ap}    
\\[5pt]
  {\rm IC}:  
  \hskip4.05cm
  T(y,0) &=& T_0(y)   
  \hskip2.2cm 
  0 < y < L ~,~ t=0
  \ .
\label{eq_oneDrodCnh_ap}
\end{eqnarray}
The solution is obtained by solving two independent problems: (i) finding a specific 
{\em static nonhomogeneous} solution $\bar T(y)$ and (ii)~finding the general 
{\em homogeneous} solution $\tilde T(y,t)$ satisfying the initial condition
\begin{eqnarray}
  \tilde T(y,0) = T_0(y) - \bar T(y) 
  \ .
  \label{TBC_ab}
\end{eqnarray}
The homogeneous solution $\tilde T(y,t)$ can be represented as a Fourier 
series.
Note that $\bar T$ depends upon the BCs, while $\tilde T$ depends upon the IC and the BCs. 
 Once $\bar T$ and $\tilde T$ have been found, the general solution 
is given by
\begin{eqnarray}
  T(y,t) =  \bar T(y) + \tilde T(y,t)
  \ .
  \label{Tgeneral}
\end{eqnarray}
Note that $T(y,t)$ satisfies the initial condition $T(y,0) = T_0(y)$, and is the unique 
solution because of the maximum principle.

\subsection{The Static Nonhomogeneous Problem}
\label{sec:nonhomogeneous}

Because of its simplicity, we first turn to finding the static nonhomogeneous solution 
$\bar T(y)$. In the static limit, the differential equation (\ref{eq_oneDrodAnh_ap}) 
and boundary conditions (\ref{eq_oneDrodBnhB_ap}) reduce to 
\begin{eqnarray}
  \frac{\partial^2 \bar T(y)}{\partial x^2}
  &=&
  0
\label{eq_oneDrodALP}
\\[5pt]
  \alpha_1 \bar T(0) + \beta_1 \bar T^\prime(0) &=& \gamma_1
\label{eq_oneDrodBLP}  
\\[-3pt]
  \alpha_2 \bar T(L) + \beta_2 \bar T^\prime(L) &=& \gamma_2
  \ .
\label{eq_oneDrodCLP}
\end{eqnarray}
The initial condition (\ref{eq_oneDrodCnh_ap}) of the full time dependent problem 
can be ignored since we are only interested in static solutions. The general solution 
to (\ref{eq_oneDrodALP}) is trivial, and takes the form
\begin{eqnarray}
  \bar T(y) = a + b y
  \ .
  \label{eq_barTab}
\end{eqnarray}
For Dirichlet boundary conditions, we must specify the temperature values $T_1$ and 
$T_2$ at the endpoints, thereby giving the nonhomogenous static solution
\begin{eqnarray}
  \bar T(y) &=& T_1 + \frac{T_2 - T_1}{L}\, y
  \ .
 \label{eq_barTlr}
 \end{eqnarray}
The coefficients $a$ and $b$, or equivalently $T_1$ and $T_2$, are determined by 
$\bar T(0) = a = T_1$ and \hbox{$\bar T(L) = a + b L = T_2$}. For a
Neumann BC specified by flux $F_i$,  we can always rewrite the corresponding homogeneous 
solution in the form  (\ref{eq_barTlr}) by the defining temperature $T_i = F_i L$, where we
can take $i=1,2$. The solution to 

The BCs (\ref{eq_oneDrodBLP}) and (\ref{eq_oneDrodCLP}) can be written as 
a linear equation in terms of $a$ and $b$,
\begin{eqnarray}
   \left(
   \begin{array}{cc}
   \alpha_1  & \beta_1 \\
   \alpha_2  &  ~\beta_2 + \alpha_2L 
  \end{array}
  \right)
  \left(
  \begin{array}{c}
  a \\
  b
  \end{array}
  \right)
  =
    \left(
  \begin{array}{c}
  \gamma_1 \\
  \gamma_2
  \end{array}
  \right)
  \ .
\end{eqnarray}
Upon solving the system of equations we find
\begin{eqnarray}
  a 
 &=& 
 \frac{\beta_2 \gamma_1 - \beta_1 \gamma_2  + L \alpha_2 \gamma_1}
 {\alpha_1 \beta_2 - \alpha_2 \beta_1 + L \alpha_1 \alpha_2 }
  \\[5pt] 
  b 
  &=&  
  \frac{\alpha_1 \gamma_2 - \alpha_2 \gamma_1}
  {\alpha_1 \beta_2 - \alpha_2 \beta_1 + L \alpha_1 \alpha_2}
  \ ,
\end{eqnarray}
or in terms of temperature parameters $T_1=a$ and $T_2=a + b L$,
\begin{eqnarray}
  T_1
 &=& 
 \frac{\beta_2 \gamma_1 -\beta_1 \gamma_2 + L \alpha_2 \gamma_1}
 {\alpha_1 \beta_2 - \alpha_2 \beta_1 + L \alpha_1 \alpha_2 }
 \label{T1genBC}
  \\[5pt] 
  T_2
  &=&  
 \frac{\beta_2 \gamma_1 -\beta_1 \gamma_2 + L \alpha_1 \gamma_2}
 {\alpha_1 \beta_2 - \alpha_2 \beta_1 + L \alpha_1 \alpha_2 }
  \label{T2genBC}
\ .
\end{eqnarray}
Note that the determinant of the linear equations vanishes for BC2, 
and we must handle this case separately.
We can also express the BCs in terms of the fluxes 
\begin{eqnarray}
  F_1 &=&  T_1/L 
  \\[5pt]
  F_2 &=& T_2/L
  \ 
 \label{eq_barFdefT}
 \end{eqnarray}
by writing
\begin{eqnarray}
  \bar T(y) &=& T_1 + (F_1 - F_2) y
  \ .
 \label{eq_barT_flux}
 \end{eqnarray}
We can also express $\bar T(y)$ by combinations of temperature and flux.

\pagebreak

\vskip0.2cm
\noindent
{\bf Special Cases of the Static Problem}

\vskip0.3cm
\noindent
{\em a. BC1}

Let us consider the simple Dirichlet boundary conditions  (\ref{BCTone}) and (\ref{BCTtwo}),
\begin{eqnarray}
  \bar T(0) &=& T_1
  \label{bc_Tone}
  \\
  \bar T(L) &=& T_2
  \label{bc_Ttwo}
  \ ,
\end{eqnarray}
which gives the solution
\begin{eqnarray}
  \bar T(y) &=& T_1 + \frac{T_2 - T_1}{L}\, y
  \ .
 \end{eqnarray}
The temperature coefficients are given by 
\begin{eqnarray}
  T_1 &=& \frac{\gamma_1}{\alpha_1}
\\[5pt]
 T_2 &=& \frac{\gamma_2}{\alpha_2}
  \ ,
 \end{eqnarray}
which follows from Eqs.~(\ref{eq_oneDrodBLP}) and (\ref{eq_oneDrodCLP}), or equivalently 
from Eqs.~(\ref{T1genBC}) and  (\ref{T2genBC}) with $\beta_1=\beta_2=0$. Similarly, the 
coefficients in  (\ref{eq_barTab})  are $a=T_1$ and $b = (T_2 - T_1)/L$.

\vskip0.2cm
\noindent
{\em b. BC2}

Let us now find the nonhomogeneous equilibrium solution for the Neumann
boundary conditions (\ref{eq_BC2A_non}) and (\ref{eq_BC2B_non}),
\begin{eqnarray}
  \partial_y \bar T(0) &=& F_1
  \\
  \partial_y \bar T(L) &=& F_2
  \ ,
\end{eqnarray}
where $F_1$ and $F_2$ are the heat fluxes at $y=0$ and $y=L$, respectively. The 
fluxes and are related to the boundary condition parameters in (\ref{eq_oneDrodBLP})  
and (\ref{eq_oneDrodCLP}) by $F_1 = \gamma_1/\beta_1$ and $F_2 = 
\gamma_2/\beta_2$ with $\alpha_1=\alpha_2=0$. As before, the general solution 
is $\bar T(y)=a + b y$,
and we see that $\bar T^\prime(y)=b$ is independent of $y$.  In other words, 
the heat flux at either end of the rod must be identical, $F_1 = b = F_2$. 
In fact, this result follows from energy conservation, since, in static equilibrium, 
the heat flowing into the rod must be equal the heat flowing out of the rod.
More correctly, we should therefore start with the boundary conditions
\begin{eqnarray}
  \partial_y \bar T(0) &=& F
  \\
  \partial_y \bar T(L) &=& F
  \label{BCFsame}
  \ ,
\end{eqnarray}
where
\begin{eqnarray}
  F = \frac{\gamma_1}{\beta_1} = \frac{\gamma_2}{\beta_2}
  \ .
\end{eqnarray}
The value of the constant term $a$ is not uniquely determined in this case; 
however, we are free to set it to zero, or to combine it with the constant $A_0$ 
term of $\tilde T(y,t)$, thereby giving
\begin{eqnarray}
  \bar T(y) = F y
  \label{barTFsame}
  \ .
\end{eqnarray}
There is nothing wrong with setting $a=0$, since we only need to find
{\em one} nonhomogeneous solution, and (\ref{barTFsame}) fits
the bill. We can write this solution in the form (\ref{eq_barTlr}), with
\begin{eqnarray}
  T_1 &=& 0
  \\
  T_2 &=&  F L 
  \ . 
\end{eqnarray}
%

\vskip0.3cm
\noindent
{\em c. BC3}

We now consider the mixed Dirichlet and Neumann  boundary conditions (\ref{BC3nonhomoA}) 
and (\ref{BC3nonhomoB}),
\begin{eqnarray}
  \bar T(0) &=& T_1
  \\
  \partial_y \bar T(L) &=& F_2
  \ .
\end{eqnarray}
We can express the solution (\ref{eq_barTlr}) in terms of the temperature 
$T_1$, and the effective temperature
\begin{eqnarray}
  T_2 &=& T_1 + F_2 L
  =
  \frac{\gamma_1}{\alpha_1} + \frac{\gamma_2 L}{\beta_2}
  \ ,
\end{eqnarray}
and the solution takes the form
\begin{eqnarray}
  \bar T(y) = T_1 + F_2 y
  \ . 
\end{eqnarray}
%

\vskip0.3cm
\noindent
{\em d. BC4}

The boundary conditions are (\ref{BC4nonhomoA}) and (\ref{BC4nonhomoB}),
\begin{eqnarray}
  \partial_y \bar T(0) &=& F_1
  \\
  \bar T(L) &=& T_2
  \ ,
\end{eqnarray}
and the solution (\ref{eq_barTlr}) can be written
The solution takes the form
\begin{eqnarray}
  \bar T(y) = (T_2 - F_1 L) + F_1 y 
  \  .
\end{eqnarray}
We can define an effective temperature 
\begin{eqnarray}
  T_1 &=& T_2 - F_1 L
  =
  \frac{\gamma_2}{\alpha_2} - \frac{\gamma_1 L}{\beta_1}
  \  ,
\end{eqnarray}

\subsection{The General Homogeneous Problem}

Now that we have constructed the static nonhomogenous solution $\bar T(y)$ 
appropriate to the choice of boundary conditions,  we turn to the slightly more 
involved task of finding the general homogeneous solution $\tilde T(y,t)$. This 
is equivalent to solving a discrete eigenvalue problem, albeit in an infinite number 
of dimensions. We then construct the solution $\tilde T(y,t)$ as a weighted sum 
over the normal modes, where the weights are determined by the choice of BCs 
and the IC. The special cases BC1, BC2, BC3, and BC4 are particularly simple. 
The homogeneous equations of motion, for which $\gamma_1=0$ and $\gamma_2=0$,
take the form
\begin{eqnarray}
  {\rm DE}: 
  \hskip3.73cm
  \frac{\partial \tilde T(y,t)}{\partial t}
  &=&
  \kappa\, \frac{\partial^2 \tilde T(y,t)}{\partial y^2}
  \hskip1.2cm 
  0 < y < L ~{\rm and}~ t > 0
\label{eq_oneDrodA}
\\[5pt]
  {\rm BC}:  
  \hskip1.16cm
  \alpha_1 \tilde T(0,t) + \beta_1 \partial_y \tilde T(0,t) &=& 0
  \hskip2.95cm t > 0
\label{eq_oneDrodB}  
\\[-3pt]
  \alpha_2 \tilde T(L,t) + \beta_2 \partial_y \tilde T(L,t) &=& 0
\nonumber
\\[5pt]
  {\rm IC}:  
  \hskip4.05cm
  \tilde T(y,0) &=& T_0(y)   
  \hskip2.2cm 
  0 < y < L
  \ .
\label{eq_oneDrodC}
\end{eqnarray}
As we have discussed, we shall focus on the linear initial condition
\begin{eqnarray}
  T_0(y) = T_\smA + \frac{T_\smB - T_\smA}{L}\,y
  \hskip4.5cm 
  0 < y < L
 \label{IClinearAgain}
  \ ,
\end{eqnarray}
although, more generally, any continuous function $T_0(y)$ will produce 
a solution. 

The solution technique is by separation of variables, for which we 
assume the solution to be a product of independent functions 
of $y$ and $t$, 
\begin{eqnarray}
  \tilde T(y,t) = Y(y) \, U(t) \ .
\end{eqnarray}
Substituting this {\em Ansatz} into the heat equation gives
\begin{eqnarray}
  \frac{dU(t)}{dt}\, Y(y) = \kappa\, U(t) \,\frac{d^2Y(y)}{dy^2}
  \ ,
\end{eqnarray}
or
\begin{eqnarray}
  \frac{1}{\kappa}\,\frac{U^\prime(t)}{U(t)}
  =
  \frac{Y^{\prime\prime}(y)}{Y(y)} 
  =
  {\rm const} 
  \equiv 
  -k^2
  \ ,
\end{eqnarray}
where we have chosen the constant negative value $-k^2$, and we 
have expressed the derivatives of $U(t)$ and $Y(y)$ by primes. As usual in the 
separation of variables technique, when two functions of different variables are
equated, they must be equal to a constant, independent of $y$ and $t$. The 
variable $U(t)$ satisfies
\begin{eqnarray}
  U^\prime(t)
  =
  - \kappa \, k^2\,U(t)
  \ ,
\end{eqnarray}
which has the solution
\begin{eqnarray}
  U_k(t) &=& U_0 \, e^{-\kappa \, k^2 t}
  \ .
\end{eqnarray}
We have introduced a $k$-subscript in $U_k$ to indicate that the solution
depends upon the value of $k$. Without loss of generality we set
$U_0=1$. We now find that the equation for $Y$ reduces to
\begin{eqnarray}
  Y^{\prime\prime}(y) + k^2 Y(y)
  &=&
  0
\label{eq_oneDrodAX}
\\[5pt]
  \alpha_1 Y(0) + \beta_1 Y^\prime(0) &=& 0
\label{eq_oneDrodBX}  
\\[-3pt]
  \alpha_2 Y(L) + \beta_2 Y^\prime(L) &=& 0
\nonumber
  \ .
\end{eqnarray}
The general solution to (\ref{eq_oneDrodAX}) is 
\begin{eqnarray}
  Y_k(y)  =  A_k \cos k y + B_k \sin k y
  \ .
  \label{Xgen}
\end{eqnarray}
When the BCs are applied, the modes $Y_k$ will be orthogonal, 
\begin{eqnarray}
  \int_0^L dx \, Y_k(y) Y_{k^\prime}(y)
  =
  N_k \, \delta_{k k^\prime}
  \ .
  \label{XkXkprime}
\end{eqnarray}
It is instructive to prove the orthogonality relation  (\ref{XkXkprime})
directly from the differential equation. Given two solutions $Y_k$ and
$Y_{k^\prime}$ to (\ref{eq_oneDrodAX}), we can write the two alternative 
forms,
\begin{eqnarray}
  Y_{k^\prime}\Big[ Y_k^{\prime\prime} + k^2 Y_k \Big] &=& 0
  \\[5pt]
  Y_k \Big[ Y_{k^\prime}^{\prime\prime} + k^{\prime\, 2} Y_{k^\prime} \Big] 
  &=& 0  
 \ .
\end{eqnarray}
These forms differ only in the interchange of $k$ and $k^\prime$.
Upon subtracting these equations, and then integrating over space, 
we find
\begin{eqnarray}
  (k^2 - k^{\prime \, 2})  
  \int_0^L \! dy \,  Y_k \, Y_{k^\prime}
  &=&
  \int_0^L \! dy \, 
  \Big[ Y_k Y_{k^\prime}^{\prime\prime} - 
   Y_{k^\prime} Y_k^{\prime\prime} 
  \Big]
  \\[5pt]
  \nonumber
  &=&
  \int_0^L \! dy \, 
  \Big[ \frac{d}{dy}\,\Big(Y_k Y_{k^\prime}^{\prime}\Big) 
  -
  Y_k^\prime  Y_{k^\prime}^\prime
  - 
  \frac{d}{dy}\Big( Y_{k^\prime} Y_k^{\prime}  \Big)
  +
  Y_{k^\prime}^\prime Y_k^{\prime}
  \Big]
  \\[5pt]
  &=&
  \int_0^L \! dy \, 
  \frac{d}{dy}\,\Big(Y_k Y_{k^\prime}^{\prime}
  - 
   Y_{k^\prime} Y_k^{\prime}  \Big)
   \\[5pt]
   &=&
  \Big(Y_k Y_{k^\prime}^{\prime}
   - 
   Y_{k^\prime} Y_k^{\prime}  \Big)
   \Big\vert_0^L
   =
   0
   \ ,
\end{eqnarray}
where each contribution from $y=0$ and $y=L$ vanishes separately
because of their respective boundary conditions. We therefore arrive at
\begin{eqnarray}
  (k^2 - k^{\prime \, 2})  
  \int_0^L \! dy \,  Y_k \, Y_{k^\prime} = 0
  \ .
  \label{eq_kminuskXX}
\end{eqnarray}
Provided $k \ne k^\prime$, we can divide (\ref{eq_kminuskXX}) by 
$k^2 - k^{\prime \, 2}$ to obtain
\begin{eqnarray}
  \int_0^L \! dy \,  Y_k(y) \, Y_{k^\prime}(y)
   &=&
   0
   ~~~{\rm when}~ k \ne k^\prime
  \ .
\end{eqnarray}
When $k=k^\prime$, (\ref{eq_kminuskXX}) gives no constraint on the 
normalization integral. However, since the differential equation is linear,
and since the BCs are homogeneous and linear, we are free to normalize 
$Y_k$ over $[0,L]$ such that $\int dy \, Y_k^2 = N_k$ for any convenient 
choice of $N_k$. 

We now express the general time dependent solution as a sum over all modes, 
\begin{eqnarray}
  \tilde T(y,t) = {\sum}_k  D_k \, Y_k(y) \, e^{-\kappa \, k^2 t}
  \ .
\end{eqnarray}
The coefficients $D_k$ themselves are chosen so that the initial 
condition is satisfied,
\begin{eqnarray}
  \tilde T(y,0) &=&  {\sum}_k D_k Y_k(y) =  T_0(y) - \bar T(y)
  \label{tildeTyzero}
  \\[5pt]
  ~~~\Rightarrow~~~ 
    D_k &=& \frac{1}{N_k} \int_0^L \! dy\, \Big[ T_0(y) - \bar T(y) \Big]\, Y_k(y) 
  \ .
\end{eqnarray}
Substituting (\ref{IClinearAgain})  and (\ref{eq_barTlr}) into (\ref{tildeTyzero}) gives 
\begin{eqnarray}
  \tilde T(y,0) 
  &=& 
  T_0(y) - \bar T(y) 
  \\[5pt]
  &=&
  T_a + \frac{T_b - T_a}{L}\, y
  \ ,
  \label{barTT0_app}
 \end{eqnarray}
with
\begin{eqnarray}
  T_a &=& T_\smA- T_1
  \label{Ta_def_app}
  \\
  T_b &=& T_\smB - T_2
  \label{Tb_def_app}
    \ .
 \end{eqnarray}
Therefore, the coefficients $D_k=D_k(T_a, T_b)$ are functions of $T_a$ and $T_b$.

\vskip0.75cm
\noindent
{\bf Special Cases of the Homogeneous Problem}

\vskip0.3cm
\noindent
{\em a. BC1}

In the first case we hold the temperature fixed to zero at both ends 
of the rod,
\begin{eqnarray}
  \tilde T(0,t) &=&0
  \label{BC1A}
  \\
  \tilde T(L,t) &=& 0
  \label{eq_BC1B}
  \ . 
\end{eqnarray}
The general solution $Y_k(y)  \!=\!   A_k \cos ky + B_k \sin ky$ reduces to 
$Y_k(y) = B_k\sin k y$ under (\ref{BC1A}), while (\ref{eq_BC1B}) restricts the 
wave numbers to satisfy $\sin k L = 0$, {\em i.e.} $k=k_n =n \pi/L$ for 
$n = 1, 2, 3, \cdots$. Note that $n=0$ does not contribute, since this gives 
the trivial vanishing solution. It is convenient to label the modes by the mode 
number $n$ rather than the wave number $k_n$,  so that the homogeneous 
solution takes the form
\begin{eqnarray}
 \tilde T(y,t)
 &=&
 \sum_{n=1}^\infty 
 B_n \, Y_n(y)\, e^{-\kappa \, k_n^2 t}
 \\[5pt]
  Y_n(y) &=& \sin k_n y 
  \\[5pt]
  k_n &=& 
  \frac{n \pi}{L}
    \hskip1.0cm    n = 1, 2, 3, \cdots  
  \ .
\end{eqnarray}
The tilde over the temperature is meant to explicitly remind us that this 
is the general {\em homogeneous} solution. The orthogonality condition 
on the modes $Y_n$ can be checked by a simple integration,
\begin{eqnarray}
 \int_0^L dy \, Y_n(y) Y_m(y) &=& \frac{L}{2}\, \delta_{nm}
 \ .
\end{eqnarray}
For an initial condition $\tilde T(y,0) =T_0(y)$, we can calculate 
the corresponding coefficients in the Fourier sum,
\begin{eqnarray}
  B_n 
  = 
  \frac{2}{L}
  \int_0^L dy\, T_0(y) \sin k_n y 
  \ .
\end{eqnarray}
For the linear initial condition (\ref{IClinearAgain}), a simple calculation gives
\begin{eqnarray}
  B_n 
  &=& 
  2 T_\smA \, \frac{1 - (-1)^n}{n\pi}
  +
  2(T_\smA - T_\smB) \,\frac{(-1)^n}{n \pi}
  \label{eq_Bnfirstline}
  \\[8pt]
  &=&
  \frac{2 T_\smA - 2 T_\smB (-1)^n}{n\pi}
  \label{eq_Bnsecdondline}
  \ .
\end{eqnarray}

\vskip0.3cm
\noindent
{\em b. BC2}

The second special boundary condition that we consider sets the heat 
flux at both ends of the rod to zero,
\begin{eqnarray}
  \partial_y \tilde T(0,t) &=& 0 
  \label{eq_BC2A}
  \\
  \partial_y \tilde T(L,t) &=& 0 
  \label{eq_BC2B}
  \ .
\end{eqnarray}
This is the hot planar sandwich. The general 
solution $Y_k(y)  \!=\!   A_k \cos ky + B_k \sin ky$ reduces to $Y_k(y) = 
A_k \cos k y$ under (\ref{eq_BC2A}) , while (\ref{eq_BC2B}) restricts the wave 
numbers to $k \sin k L = 0$,  so that $k=k_n=n \pi/L$ for $n=0,1,2 \cdots$. 
In this case, the $n=0$ mode is permitted (and indeed essential), and the 
solution can be written
\begin{eqnarray}
  \tilde T(y,t)  &=& 
  \frac{A_0}{2} + \sum_{n=1}^\infty A_n \, Y_n(y)\,
  e^{-\kappa\, k_n^2 t}
  \label{A0half}
   \\[5pt]
  Y_n(y) &=& \cos k_n y   
  \\[5pt]
    k_n &=& \frac{n \pi}{L}
  \hskip1.0cm
   n = 0, 1, 2, \cdots
  \ ,
\end{eqnarray}
where a conventional factor of $1/2$ has been inserted in the $A_0$ term. This
is because of the  difference in normalization between $n=0$ and $n \ne 0$,
\begin{eqnarray}
  \int_0^L dy \, Y_0^2(y) &=& L
  \\[5pt]
  \int_0^L dy \, Y_n^2(y) &=& \frac{L}{2} ~~~~ n \ne 0
  \ .
\end{eqnarray}
Given the initial condition $\tilde T(y,0) = T_0(y)$, the Fourier modes become
\begin{eqnarray}
  A_n = \frac{2}{L}\, \int_0^L dy \, T_0(y) \cos k_n y
  \label{AnBCTwo}
\end{eqnarray}
for $n=0,1, 2 \cdots$.
This holds for all values of $n$, including $n=0$. This is why we used 
the factor of 1/2 in  the $A_0$ term of (\ref{A0half}).
For simplicity, we will take the linear initial condition (\ref{IClinearAgain})
for $T_0(y)$, in which case,  (\ref{AnBCTwo}) gives the coefficients
\begin{eqnarray}
  \frac{A_0}{2}
  &=& 
  \frac{1}{2}\Big(T_\smA + T_\smB\Big)
  \label{eq_Azeroovertwo}
  \\[5pt]
  A_n 
  &=& 
  2\, \Big(T_\smA - T_\smB \Big) \, \frac{1 - (-1)^n}{n^2 \pi^2}
  \label{eq_Annotzero}
  \ .
\end{eqnarray}
For pedagogical purposes, let us work through 
the algebra for the $A_n$ coefficients, doing the $n=0$ case first:
\begin{eqnarray}
  \frac{A_0}{2}
  &=& 
  \frac{1}{L} \int_0^L \, T_0(y)
  =
  \frac{1}{L} \int_0^L \, \left[
    T_\smA + \frac{T_\smB - T_\smA}{L}\, y
    \right]
    \\[5pt]
    &=&
    T_\sm L + \left[\frac{T_\smB - T_\smA}{2}
    \right]
    =
    \frac{1}{2}\left[ T_\smB + T_\smA \right]
    \ .
\end{eqnarray}
Next, taking $n \ne 0$, we find:
\begin{eqnarray}
  A_n
  &=& 
  \frac{2}{L} \int_0^L dy\, T_0(y) \cos k_n y
  \\[5pt]
  &=& 
    \frac{2}{L} \int_0^L dy\, \left[
    T_\smA + \frac{T_\smB - T_\smA}{L}\, y
    \right] \cos k_n y
    \\[5pt]
      &=& 
    T_\smA  \, \frac{2}{L} \int_0^L dy\, \cos k_n y
    + 
    \Big( T_\smB - T_\smA \Big) \frac{2}{L^2} \int_0^L dy \, y\, \cos k_n y
    \ .
\end{eqnarray}
The first term integrates to zero since
\begin{eqnarray}
  \frac{2}{L} \int_0^L dy \,\cos k_n y
  &=&
  \frac{2}{L} \, \sin k_n y \Big\vert_{y=0}^{y=L}
  =
  0 
  \ ,
\end{eqnarray}
and the second term gives
\begin{eqnarray}
  \frac{2}{L^2} \int_0^L dy \, y\, \cos k_n y
  &=&
  \frac{2}{L^2} \left[ \frac{\cos k_n y}{k_n^2} + \frac{y \sin k_n y}{k_n}
  \right]_{y=0}^{y=L}
  \\[5pt]
  &=&
  \frac{2}{L^2} \, \frac{L^2}{n^2 \pi^2}\,\Big[\cos k_n L - 1 \Big]
  =
  2 \, \frac{(-1)^n - 1}{n^2 \pi^2}
  \ ,
\end{eqnarray}
which leads to (\ref{eq_Annotzero}).

\vskip0.3cm
\noindent
{\em c. BC3}

The next specialized boundary condition is
\begin{eqnarray}
  \tilde T(0,t) &=&0
  \label{eq_BC3A}
  \\
  \partial_y \tilde T(L,t) &=& 0
  \label{eq_BC3B}
  \ .
\end{eqnarray}
This is the half planar sandwich.
The general solution $Y_k(y)  \!=\!   A_k \cos ky + B_k \sin ky$ 
under (\ref{eq_BC3A}) reduces to
$Y_k(y) = B_k \sin k y$, while (\ref{eq_BC3B}) restricts the wave 
numbers to $k \cos k L = 0$, so that $k=k_n=(2 n + 1) \pi/2L$ for 
$n=0,1,2 \cdots$.  The general homogeneous solution is therefore
\begin{eqnarray}
  \tilde T(y,t)  &=& \sum_{n=0}^\infty 
  B_n \, Y_n(y) \, e^{-\kappa \, k_n^2 t}
  \\[5pt]
  Y_n(y) &=& \sin k_n y  
  \\[5pt]
    k_n &=& 
  \frac{(2 n + 1) \pi}{2 L}
  \hskip1.0cm
   n = 0, 1, 2, \cdots
  \ .
\end{eqnarray}
The initial condition $\tilde T(y,0) = T_0(y)$ gives the Fourier modes
\begin{eqnarray}
  B_n = \frac{2}{L}\, \int_0^L dy \, T_0(y) \sin k_n y
  \ ,
  \label{eq_Bnbcthree}  
\end{eqnarray}
and, upon taking the linear initial condition (\ref{IClinearA}),  we find
\begin{eqnarray}
  B_n 
  &=&  
  \frac{4 T_\smA}{(2 n + 1) \pi}
  +
  4\big(T_\smB - T_\smA\big)\left[
  \frac{1}{(2n+1) \pi} 
  -
  \frac{2}{(2n+1)^2 \pi^2} 
  \right]
  \\[5pt]
    &=&  
  \frac{4 T_\smB}{(2 n + 1) \pi}
  -
  \frac{8\big(T_\smB - T_\smA\big) }{(2n+1)^2 \pi^2} 
  \ .
\end{eqnarray}
%

%
\pagebreak
\noindent
{\em d. BC4}

The last special case is the boundary condition
\begin{eqnarray}
  \partial_y \tilde T(0,t) &=& 0 
  \label{eq_BC4A}
  \\
  \tilde T(L,t) &=& 0
  \ .
 \label{eq_BC4B}
\end{eqnarray}
The general solution $Y_k(y) \!=\!  A_k \cos ky + B_k \sin ky$ 
reduces to $Y_k(y) = A_k \cos k y $ under (\ref{eq_BC4A}), 
while (\ref{eq_BC4B}) restricts the wave 
numbers to $\cos k L = 0$, {\em i.e.} $k=k_n=(2 n + 1) \pi/2L$ for 
$n=0,1,2 \cdots$, which gives rise to the homogeneous solution
\begin{eqnarray}
  \tilde T(y,t)  &=& \sum_{n=0}^\infty 
  A_n \, Y_n(y)\, e^{-\kappa \, k_n^2 t}
  \\[5pt]
  Y_n(y) &=& \cos k_n y  
  \\[5pt]
    k_n &=& 
  \frac{(2 n + 1) \pi}{2 L}
  \hskip1.0cm 
  n = 0, 1, 2, \cdots
  \ .
\end{eqnarray}
Similar to (\ref{eq_Bnbcthree}), the mode coefficient is
\begin{eqnarray}
  A_n = \frac{2}{L}\, \int_0^L dy \, T_0(y) \cos k_n y
  \ ,
\end{eqnarray}
and, upon taking the linear initial condition (\ref{IClinearAgain}), we find
\begin{eqnarray}
  A_n 
  &=&
  4 T_\smA \, \frac{(-1)^n }{(2 n + 1)\pi }
  -
  8\Big(T_\smB - T_\smA \Big) \,\frac{1 - (-1)^n}{(2n + 1)^2\, \pi^2}
  \ .
\end{eqnarray}
Note that BC3 and BC4 are in fact equivalent, and represent a rod that has been flipped 
from left to right about its center, as illustrated in Fig.~\ref{fig_BC34_flip}.

\vskip0.4cm
\noindent
{\bf General Boundary Conditions}
\vskip0.3cm

\begin{figure}[t!]
\includegraphics[scale=0.45]{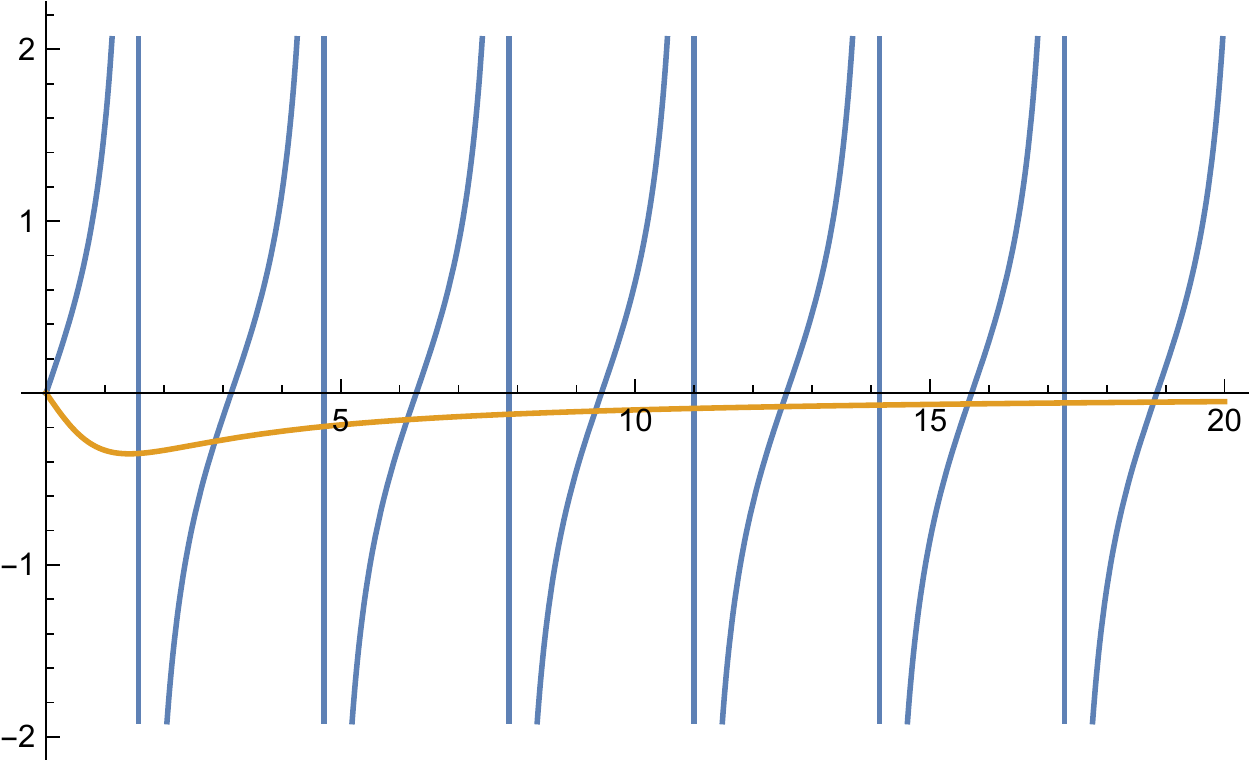} 
\caption{\footnoteskip  
  The roots $\mu_n$ for $\alpha_1=1$, $\bar\beta=1/2$, $\alpha_2=1$, and
  $\bar\beta_2=1$. For $L=2$ this gives $\beta_1=1$ and $\beta_2=2$.
}
\label{fig_tankL}
\end{figure}

We now turn to the general form of the boundary conditions,
which, expressed in terms of $Y$, take the form
\begin{eqnarray}
  \alpha_1 Y_k(0) + \beta_1 Y_k^\prime(0) &=& 0
\label{eq_XBCzero}  
\\
  \alpha_2 Y_k(L) + \beta_2 Y_k^\prime(L)&=& 0
  \ .
\label{eq_XBCL}  
\end{eqnarray}
The solution and its derivative are
\begin{eqnarray}
  Y_k(y) &=& A \cos k y + B \sin k y
  \\
  Y_k^\prime(y) &=& - A k \sin k y  + B k \cos k y
  \ .
\end{eqnarray}
Substituting this into (\ref{eq_XBCzero}) and (\ref{eq_XBCL}) gives
\begin{eqnarray}
  \alpha_1 A + \beta_1 B k &=& 0
  \label{BCa}
  \\
  \alpha_2 \Big[A \cos k L + B \sin k L \Big] 
  + 
  \beta_2\Big[ -A k \sin k L + B k \cos k L \Big] &=& 0
  \label{BCb}
  \ .
\end{eqnarray}
There are two cases, $\cos k L \ne 0$ and $\cos k L = 0$.  We have already
addressed the latter, so let us now consider the former. Upon dividing (\ref{BCb}) 
by $\cos k L$ we find
\begin{eqnarray}
  (\alpha_2 \, B    -\beta_2 \, A k) \tan k L 
  + \alpha_2 A + \beta_2 \, B k  &=& 0
  \ ,
\end{eqnarray}
or
\begin{eqnarray}
   \tan k L 
   &=&
   \frac{\beta_2 \, B k + \alpha_2 A }{ \beta_2 \, A k - \alpha_2 \, B }
   \ .
   \label{BCtankL}
\end{eqnarray}
From (\ref{BCa}) we have $B k = - \alpha_1 A / \beta_1$ (if $\beta_1 \ne 0)$, 
and substituting this into (\ref{BCtankL}) gives 
\begin{eqnarray}
   \tan k L 
  &=&
   \frac{-\alpha_1 \beta_2 \, k + \alpha_2 \beta_1 k }{ \beta_1 \beta_2 \, k^2 
   +  \alpha_2 \, \alpha_1  }      
  \ .
  \label{BCtankLone}
\end{eqnarray}
Similar reasoning provides the same result for the case in which $\beta_1 = 0$.
It is convenient to express this equation in the form 
\begin{eqnarray}
   \tan \mu
   &=&
   \frac{ (\alpha_2 \bar\beta_1 - \alpha_1 \bar\beta_2) \, \mu }{\alpha_1 \alpha_2  +  \bar\beta_1 \bar\beta_2\, \mu^2  }      
   \ ,
  \label{BCtanmu}
\end{eqnarray}
where $\mu \equiv k L$ and $\bar\beta_i \equiv \beta_i/L$.
The solution is illustrated in Fig.~\ref{fig_tankL}.
Equation (\ref{BCtanmu}) will provide the mode numbers $\mu_n$ for $n=0,1,2,\cdots$,
which are used to calculate the wave numbers
\begin{eqnarray}
  k_n = \frac{\mu_n}{L}
  \ .
  \label{knDef}
\end{eqnarray}
Note that $\mu_0=0$, and therefore $k_0=0$. The solution now takes the form
\begin{eqnarray}
  Y_n(y) &=& A_n \cos k_n y + B_n \sin k_n y
  \label{SolAB}
  \\[5pt]
  A_n &=& - \frac{\beta_1 k_n}{\alpha_1}\, B_n
  \ ,
\end{eqnarray}
where $\alpha_1 \ne 0$. The case of $\alpha_1=0$ will be handled 
separately. Setting $B_n=1$ for convenience, the solution (\ref{SolAB}) 
can be expressed as
\begin{eqnarray}
  Y_n(y) &=& \sin k_n y - \frac{\beta_1 k_n}{\alpha_1} \cos k_n y  
  \ ,
  \label{SolABtwo}
\end{eqnarray}
while the general solution takes the form
\begin{eqnarray}
  Y(y) &=& \sum_{n=1}^\infty D_n Y_n(y)
  \ .
\end{eqnarray}
Note that the $n=0$ term does not contribute, and the Fourier coefficients are given by
\begin{eqnarray}
  D_n = \frac{1}{N_n}\int_0^L dy\, T_0(y) Y_n(y)
  \ .
\end{eqnarray}
Also note that the modes are orthogonal,
\begin{eqnarray}
  \int_0^L dy \, Y_n(y) Y_m(y) = 0 ~~~{\rm for}~ n \ne m 
  \ ,
  \label{SolABtwo}
\end{eqnarray}
with the normalization factor $N_k$ determined by 
\begin{eqnarray}
  \int_0^L dy \, Y_n^2(y)  
  &=&
  \frac{1}{4 k_n \alpha_1^2}\Bigg[
  -2 \alpha_1 \beta_1 k_n  + 2 (\beta_1^2 k_n^2 + \alpha_1^2 ) k_nL +
  \\
  \nonumber && \hskip1.5cm
  2\alpha_1 \beta_1 k_n \cos 2 k_n L + 
  (\beta_1^2 k_n^2 - \alpha_1^2 ) \sin 2 k_n L
  \Bigg]
  \ . 
  \label{SolABthree}
\end{eqnarray}
That is to say,
\begin{eqnarray}
 &&
  \int_0^L dy \, Y_n(y) Y_m(y) 
  =
  N_n \, \delta_{nm} \ ,
  \\
  &&N_n
  =
  \frac{1}{4 k_n \alpha_1^2}\Big[
  -2 \alpha_1 \beta_1 k_n  + 2 (\beta_1^2 k_n^2 + \alpha_1^2 ) k_nL +
  2\alpha_1 \beta_1 k_n \cos 2 k_n L + 
  (\beta_1^2 k_n^2 - \alpha_1^2 ) \sin 2 k_n L
  \Big] \ .
  \nonumber
  \\
  \label{SolABfour}
\end{eqnarray}

It is convenient for numerical work to express this in
terms of $A_n$ and $B_n$ coefficients:
\begin{eqnarray}
  Y(y) 
  &=&
  \sum_{n=1}^\infty D_n\, \Big[ 
  - \frac{\beta_1 k_n}{\alpha_1}\,\cos k_n y + \sin k_n y 
  \Big]
  \\[5pt]
  &=&
  \sum_{n=1}^\infty\Big[ 
  A_n \cos k_n y + B_n \sin k_n y 
  \Big]
  {\rm ~~~with}
  \\[5pt]
  \nonumber
  A_n &=& - \frac{\beta_1 k_n}{\alpha_1}\, D_n
  ~~~{\rm and}~~~
  B_n = D_n
  \ .
\end{eqnarray}
The temperature $\tilde T(y,t)$ is therefore,
\begin{eqnarray}
  \tilde T(y,t) 
  &=&
  \sum_{n=1}^\infty\Big[ 
  A_n \cos k_n y + B_n \sin k_n y 
  \Big] \, e^{-\kappa \, k_n^2 t}
  \\[5pt]
    B_n &=& \frac{1}{N_n}\int_0^L dy\, T_0(y) Y_n(y)
  \\[5pt]
  A_n &=& - \frac{\beta_1 k_n}{\alpha_1}\, B_n
  \ .
\end{eqnarray}
For $T_0^a(y)=T_1$ we have
\begin{eqnarray}
  B_n^a &=& \frac{T_1}{N_n} \, \left[
  \frac{1 - \cos k_n L}{k_n}  - \frac{\beta_1 \sin k_n L}{\alpha_1}\right]
  \ .
\end{eqnarray}
For $T_0^b(y) = (T_2 - T_1) \, y/L$ we have
\begin{eqnarray}
  B_n^b &=& \frac{T_2 - T_1}{N_n\, L}\,\frac{1}{\alpha_1 k_n^2} \, \Big[
  \beta_1 k_n  - (\alpha_1 k_n L + \beta_1 k_n) \cos k_n L + (\alpha_1 - 
  \beta_1 k_n^2 L) \sin k_n L 
  \Big]
  \ ,
\end{eqnarray}
with $B_n = B_n^a + B_n^b$.

\begin{figure}[h!]
\includegraphics[scale=0.45]{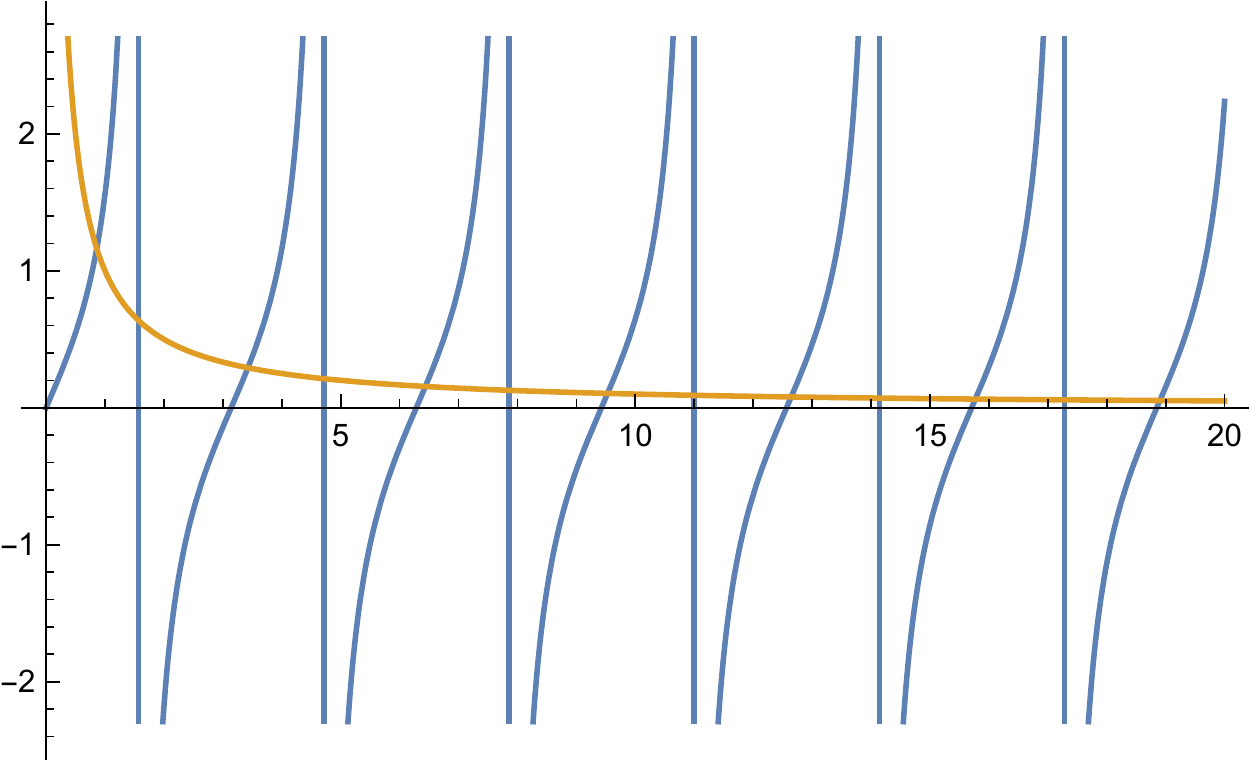} 
\caption{\footnoteskip  
  The roots $\mu_n$ for $\alpha_1=0$, $\alpha_2=1$, and
  $\bar\beta_2=1$. For $L=2$ we have $\beta_2=2$.
}
\label{fig_tankLzero}
\end{figure}

Let us now consider the case of $\alpha_1=0$, so that (\ref{BCtanmu}) becomes
\begin{eqnarray}
   \tan \mu
   &=&
   \frac{a}{\mu}
   ~~~~{\rm with} ~~    a = \alpha_2 / \bar\beta_2
   \ .
  \label{BCtanmuzero}
\end{eqnarray}
We can find an approximate solution for large values of $\mu$: since
the RHS is very small for \hbox{$\mu \gg 1$}, we must solve $\tan \mu = 0$,
and therefore $\mu_n^{(0)} = n \pi$. The exact solution can be expressed
as $\mu_n = n \pi + h$, where $0 < h \ll 1$, and we find ${\rm LHS} =
\tan(n\pi + h) = \tan(h) = h + {\cal O}(h^2)$. Similarly, ${\rm RHS}=
a/(n\pi + h) = (a/n\pi)\big(1 + h /n\pi \big)^{-1} =  (a/n\pi)\big(1 - h /n\pi
\big) + {\cal O}([h/n]^2) = a/n\pi - a h + {\cal O}(h^2/n^2)$, thus
\begin{eqnarray}
  h = \frac{a}{n\pi} - a h 
  ~~~\Rightarrow~~~
  h = \frac{a}{1 + a}\,\frac{1}{n\pi}
  \ ,
\end{eqnarray}
and the first order solution becomes
\begin{eqnarray}
  \mu_n^{(1)}
  =
  n \pi + \frac{a}{1 + a}\,\frac{1}{n\pi} + {\cal O}(1/n^2)
  \ .
\end{eqnarray}
This can be used as an initial guess when using an iteration
method to find the $\mu_n$. The solution is
\begin{eqnarray}
  T(y,t) &=& \sum_{n=1}^\infty A_n Y_n(y)\,e^{-\kappa \, k_n^2 t}
  \\
  Y_n(y) &=& \cos k_n L
  \\
  \int_0^L dy \, Y_n(y) Y_m(y) &=& N_n \, \delta_{nm}
 \\
  N_n &=& \frac{1}{4 k_n}\,\Big[2 k_n L + \sin 2 k_n L \Big]
  \ ,
\end{eqnarray}
and
\begin{eqnarray}
  A_n 
  &=&
  \frac{1}{N_n}\int_0^L dy\, T(y,0) Y_n(y)
  \\[5pt]
  &=&
  \frac{T_1}{k_n} \,\sin k_n L + \frac{T_2 - T_1}{k_n^2 L}\,
  \Big[ -1 + \cos k_n L + k_n L \sin k_n L \Big]
  \ .
\end{eqnarray}
%

%
\subsection{The General Solution}

\noindent
{\em a. BC1: Given Temperatures $T_1$ and $T_2$}

\begin{eqnarray}
 T(y,t)
 &=&
 T_1 + \frac{T_2 - T_1}{L}\,y
 +
 \sum_{n=1}^\infty 
 B_n \, \sin k_n y \, e^{-\kappa \, k_n^2 t}
  \\[5pt]
  B_n 
  &=&
  \frac{2 T_\smA - 2 T_\smB (-1)^n}{n\pi}
 \\[5pt]
  k_n &=&   \frac{n \pi}{L}
  \ .
\end{eqnarray}

\noindent
{\em b. BC2: Given Flux $F$}

\begin{eqnarray}
 T(y,t)
 &=&
  F y
 +
  \frac{A_0}{2} + \sum_{n=1}^\infty A_n \, \cos k_n y   \,
  e^{-\kappa\, k_n^2 t}
  \\[5pt]
  A_0
  &=& 
  T_\smA + T_\smB
  \\[5pt]
  A_n 
  &=& 
  2\, \Big(T_\smA - T_\smB \Big) \, \frac{1 - (-1)^n}{n^2 \pi^2}
 \\[5pt]
  k_n &=& \frac{n \pi}{L}
  \ .
\end{eqnarray}

\noindent
{\em c. BC3: Given $T_1$ and $F_2$}

\begin{eqnarray}
  T(y,t)  
  &=& 
  T_1 + F_2\,y
  +
  \sum_{n=0}^\infty 
  B_n \, \sin k_n y  \, e^{-\kappa \, k_n^2 t}
  \\[5pt]
  B_n &=& 
  \frac{4 T_\smB}{(2 n + 1) \pi}
  -
  \frac{8\big(T_\smB - T_\smA\big) }{(2n+1)^2 \pi^2} 
    \\[5pt]
    k_n &=& 
  \frac{(2 n + 1) \pi}{2 L}
  \ .
\end{eqnarray}

\noindent
{\em d. BC4: Given $T_2$ and $F_1$}

\begin{eqnarray}
  T(y,t)  &=& 
  (T_2 - F_1 L) + F_1\, y
  +
  \sum_{n=0}^\infty 
  A_n \, \cos k_n y  \, e^{-\kappa \, k_n^2 t}
  \\[5pt]
  A_n 
  &=&
  4 T_\smA \, \frac{(-1)^n }{(2 n + 1)\pi }
  -
  8\Big(T_\smB - T_\smA \Big) \,\frac{1 - (-1)^n}{(2n + 1)^2\, \pi^2}
  \\[5pt]
  k_n &=& 
  \frac{(2 n + 1) \pi}{2 L}
  \ .
\end{eqnarray}
%

\pagebreak
\clearpage

\end{document}